\documentclass[10pt]{iopart}
\usepackage{mathrsfs}
\expandafter\let\csname equation*\endcsname=\relax 
\expandafter\let\csname endequation*\endcsname=\relax 
\usepackage{amsmath,amssymb,amsthm}
\usepackage{physics}
\usepackage{hyperref}
\usepackage{color}
\newcommand{\lket}[1]{\left. \left| #1 \right> \! \right>} 
\newcommand{\lbra}[1]{\left< \! \left<  #1 \right|\right.} 
\newcommand{\lbraket}[2]{\left< \! {\left< #1 \vphantom{#2} \right|
 \left. #2 \vphantom{#1} \right>}\! \right>} 
\newcommand{\lmatrixel}[3]{\left< \!{\left< #1 \vphantom{#2#3} \right|
 #2 \left| #3 \vphantom{#1#2} \right>} \! \right>} 
\begin{document}

\title{Fundamentals of Quantum Mechanics in Liouville Space}
\author{Jerryman A. Gyamfi}
\address{Scuola Normale Superiore di Pisa, Piazza dei Cavalieri 7, 56126 Pisa, Italy.}
\ead{jerryman.gyamfi@sns.it}
\vspace{10pt}
\begin{indented}
\item[]June 2020
\end{indented}

\begin{abstract}
The purpose of this paper is to articulate a coherent and easy-to-understand way of doing quantum mechanics in any finite-dimensional Liouville space, based on the use of Kronecker product and what we have termed the `bra-flipper' operator. One of the greater strengths of the formalism expatiated on here is the striking similarities it bears with Dirac's bra-ket notation. For the purpose of illustrating how the formalism can be effectively employed, we use it to solve a quantum optical master equation for a two-level quantum system and find its Kraus operator sum representation. The paper is addressed to students and researchers with some basic knowledge of linear algebra who want to  acquire a deeper understanding of the Liouville space formalism. The concepts are conveyed so as to make the application of the formalism to more complex problems in quantum physics straightforward and unencumbered.
\end{abstract}


\submitto{European Journal of Physics. [For final peer review record, please see: {\color{blue}Jerryman A. Gyamfi 2020 \textit{Eur. J. Phys.} \textbf{41} 063002}. DOI: \url{https://doi.org/10.1088/1361-6404/ab9fdd}]}

\section{Introduction} 
For many complex problems in quantum mechanics, the Liouville space formalism turns out to be very effective in finding solutions or mathematically characterizing the problem. This is true, for example, in solving master equations in the theory of open quantum systems
\cite{inbook:Fano-1964, art:Albert-2014}. To see why this is the case, we first have to recognize that the density matrix
\cite{art:Fano-1957, book:Blum-2012} (Sec. \ref{subsec:pure_mixed}) is a more pragmatic way of describing the quantum state of a system compared to vectors in the state space (Sec. \ref{subsec:Postulate_I}) because it offers a general, compact and elegant way of describing quantum states. This makes it easier (with respect to the use of state vectors) to derive general formulas for probabilities and observable averages. Perhaps, nothing makes one appreciate more the invention of density matrix than the quantum theory of relaxation
\cite{book:Blum-2012}. The vast majority of quantum relaxation processes (from nuclear and electron spin magnetic resonance studies to relaxation processes in quantum optics) studied in physics, chemistry and biology become cumbersome and more likely to be fraught with unnecessary complications without the use of density matrices. However, though the equations of motion for the density matrix for such relaxation processes are often linear in the former, they are often not amenable to easy resolution. The Gorini-Kossakowski-Sudarshan--Lindblad master equation
\cite{art:Manzano-2020, misc:Lidar-2019, art:Pearle-2012, book:Breuer-2007} is an excellent example. The Liouville space formalism offers a way around the problem. This is because, in this linear space, operators defined on the state space become column vectors (called supervectors, Sec. \ref{subsec:bra-flipper}) and given that we are adept at solving equations of motion for vectors in state space (think of the Schr\"odinger equation, \eqref{eq:Schrodinger}), we can easily apply those same techniques also in Liouville space
\cite{inbook:Fano-1964, art:Fano-1959}. 
\par The literature already provides a number of important introductions to the subject
\cite{inbook:Fano-1964, art:Jeener-1982, book:Mukamel-1995, book:Ernst_Bodenhausen-1990, inbook:Mayne-2007, inbook:Petrosky_Prigogine-1996} but for students and researchers new to it, grasping the basics is often challenging and confusing. A careful analysis of the literature points to one important source of this confusion. It has to do with the operational meaning an author assigns to the symbol `$\otimes$' (sometimes indicated as `$\times$'
\cite{art:Jeener-1982}). In some expositions (which we may call the `symbolic approach'), the common operational symbol `$\otimes$' is deprived of its common meaning and operations involving said symbol are assigned customized results which cannot be traced to any basic mathematical operations. This is the approach used, for example, in 
\cite{book:Gamliel-Levanon-1995}. In fact, in referring to the symbol `$\otimes$' the authors clarify that it ``is \textit{not} a tensor product, but a related operation" 
\cite[chap. 1, pg. 20]{book:Gamliel-Levanon-1995}. 
\par In addition, there are those expositions (which we may call the `literal approach') in which the symbol `$\otimes$' retains its normal operational meaning  ---  namely, the tensor product. The literal approach is used, for example, in 
\cite{book:Ernst_Bodenhausen-1990}. There are also those introductions where neither `$\otimes$' nor `$\times$'  are explicitly used but either the symbolic approach or the literal one are implicitly implied in the operations (see for example 
\cite{inbook:Fano-1964, art:Barnett-1987}).  
\par The rationale behind this paper is threefold: 1) give a coherent account of quantum mechanics in Liouville space, 2) impact the readership with a simple and unified view of the subject so as to render the study of quantum systems of diverse nature in Liouville space conceptually and mathematically easy, and 3) strengthen further the literal approach. We emphasize that the present account is restricted to separable (see \ref{appendix:A1} for definition) and finite Hilbert spaces (thus, separable and finite Liouville spaces as well). We shall rely primarily on a notation very similar to the familiar Dirac bra-ket notation
\cite{book:Dirac-1967} and only introduce here  ---  as a matter of formality  ---  a new operator, which we call the `bra-flipper' operator, $\mho$ (see Sec. \ref{subsec:bra-flipper}). As we shall soon see, this is all that we need (together with the tensor product operation and its properties) to make doing quantum mechanics in Liouville space easy and straightforward. One important thing the Reader will notice in the course of our discussion is the striking similarities between the Liouville space formalism presented here and the formalism taught in undergraduate courses on quantum theory. 
\par The rest of the paper is organized into three main sections. In the first section (Sec. \ref{sec:Overview_state space}), we give a quick overview of standard quantum mechanics in state space. We then gradually construct the notion of Liouville space and its related formalism in the second section (Sec. \ref{sec:LS}) --- where, among other things, we introduce the bra-flipper operator (Sec. \ref{sec:bra-flipper}). This is followed by the third section (Sec. \ref{sec:Applications}) where we apply the mathematical apparatus and concepts developed in the preceding section to formulate equations of motion in Liouville space according to Schr\"odinger, Heisenberg and interaction pictures. We also discuss in this section solutions to a specific class of master equations (for isolated and open quantum systems) to which the Lindblad master equation belongs.
\par To show a concrete application of the formalism, we consider in Sec. \ref{subsec:optics} a quantum optical master equation for a two-level system. There, we show how one can aptly use the formalism for many ends: from solving the master equation to determining the Kraus operator sum representation of the solution. 
\par Readers who may be less familiar with linear algebra and functional analysis may consult \ref{appendix:A1} for some basic notions and definitions. Readers already well-versed in quantum mechanics may proceed directly to Sec. \ref{sec:LS}.

\section{Overview of quantum mechanics in state space}\label{sec:Overview_state space}
The postulates of quantum mechanics are listed and formulated differently from author to author, even though the contents fundamentally do not disagree. The presentation offered here follows closely Ref. \cite{book:Nielsen-2011}.
\subsection{Postulate I}\label{subsec:Postulate_I}
The first postulate defines the complex linear space in which one can describe \emph{isolated} (or \emph{closed}) quantum systems -- \emph{i.e.} systems not participating in any kind of interaction with other systems. (Some authors, for example \cite{book:Breuer-2007}, make a distinction between closed and isolated systems. We shall use the two terms interchangeably.)
\begin{quotation}
\item \textbf{Postulate I}: Let $S$ be an arbitrary isolated physical system.  Associated to $S$ is a Hilbert space referred to as its \emph{state space}.  The state of $S$ can be completely described by a unit vector $\ket{\psi}$ of its state space. $\ket{\psi}$ is called a \emph{state vector}.
\end{quotation}  
We draw the Reader's attention to the fact that, in the literature, the term `Hilbert space' is often taken as synonymous with (or used as a shorthand for) `state space'. This can sometimes be misleading because the notion of Hilbert space is a very broad one. Even the so-called Liouville space  ---  the subject of this article --- is in itself a Hilbert space (more on this later). In fact, any linear space which is complete and endowed with a norm is a Hilbert space (see \ref{appendix:A1}). For the sake of clarity, we shall keep this distinction in our discussion. For a solid introduction to Hilbert spaces, we recommend  
\cite{book:Jordan-2006, book:Moretti-2018, book:Conway-2007, book:Riesz-1955, book:Tarasov-2008, book:Dennery-1996}  ---  to name a few excellent references. 
\par Let us focus now on the state space. As mentioned in the introduction, we shall be concerned with only finite-dimensional state spaces. So, let $\mathcal{H}_d$ be a finite-dimensional state space of dimension $d$.  Let $\ket{x}$ be a vector in $\mathcal{H}_d$.  $\ket{x}$ is, thus, a $(d \times 1)$ column vector. In linear algebra, we would have indicated such a $(d\times 1)$ vector with the bold symbol $\pmb{x}$ or $\overrightarrow{x}$. But in quantum mechanics, we use what is called the Dirac bra-ket notation
\cite{book:Dirac-1967}, and indicate such a column vector with the symbol `$\ket{x}$', called a \emph{ket}. The advantage of Dirac's bra-ket notation is that it works fine even for infinite dimensional Hilbert spaces and allows us to do many interesting manipulations without much effort, as we shall soon see.
\par Associated with $\mathcal{H}_d$ is an \emph{adjoint space} (or \emph{dual space}), indicated as $\mathcal{H}^*_d$. There is a one-to-one correspondence between the elements of $\mathcal{H}_d$ and $\mathcal{H}^*_d$: if $\ket{x} \in \mathcal{H}_d$, then there exists its corresponding element, denoted as `$\bra{x}$' (called \emph{bra}) in $\mathcal{H}^*_d$. The relation between the two is
				\begin{equation}
				\label{eq:bra_ket_relation}
				\bra{x} =  \ket{x}^\dagger
				\end{equation}				 
where, for any matrix (vector or operator) $A$, `$A^\dagger$' indicates the conjugate transpose of $A$. We therefore see that $\bra{x}$ must be a row vector of dimension $(1 \times d)$. $\bra{x}$ is also said to be the dual vector of $\ket{x}$. 
\par For $\bra{y} \in \mathcal{H}^*_d$ and $\ket{x} \in \mathcal{H}_d$, $\bra{y}$ maps $\ket{x}$ to a scalar through the matrix product 
				\begin{equation}
				\label{eq:inner_prod}
				\big( \bra{y} \big) \big( \ket{x}\big) \equiv \braket{y}{x} \ .
				\end{equation}
Equation \eqref{eq:inner_prod} also defines an inner product (or scalar product) for $\mathcal{H}_d$
\cite{ book:Jordan-2006, book:Conway-2007, book:Riesz-1955}(\ref{appendix:A1}). That is, given two vectors $\ket{y},\ket{x} \in \mathcal{H}_d$, we can choose their inner product to be defined as
				\begin{equation}
				\label{eq:inner_prod_Hilbert}
				\big< \ket{y}, \ket{x} \big> := \braket{y}{x} \ .
				\end{equation}				  
For separable Hilbert spaces, \eqref{eq:inner_prod_Hilbert} is the commonly used inner product, and so shall we in this paper. Note that for $\ket{x}$ to be a state vector, according to Postulate I of quantum mechanics, $\ket{x}$ must be a unit vector -- meaning $\braket{x}{x}=1$. The \emph{null vector} of $\mathcal{H}_d$, which we may indicate as $\ket{\emptyset}$, is the vector whose inner product with any given element of $\mathcal{H}_d$ is zero, \emph{i.e.} $\braket{\emptyset}{x}=\braket{x}{\emptyset}=0 \ , \forall \ket{x} \in \mathcal{H}_d$. $\ket{\emptyset}$ is simply the $(d \times 1)$ column vector whose entries are all zero.
\par The vector $\ket{x}$ can be expanded in any orthonormal basis $\{\ket{\phi_n}\} \  (n= 1, \ldots, d)$ of the Hilbert space $\mathcal{H}_d$. That is,
				\begin{equation}
				\label{eq:x}
				\ket{x} = \sum^d_{n=1} c_n \ket{\phi_n} \ , \ \qquad \ \braket{\phi_n}{\phi_{n'}}=\delta_{n,n'} 
				\end{equation} 
where the coefficients $\{c_n\}$ are complex scalars. Multiplying $\ket{x}$ in \eqref{eq:x} from the left by $\bra{\phi_{m}}$, we get
				\begin{equation}
				\label{eq:c_n}
				\braket{\phi_{m}}{x} = \sum^d_{n=1} c_n \braket{\phi_{m}}{\phi_n} = \sum^d_{n=1} c_n \ \delta_{n,m} = c_{m} \ .
				\end{equation}
Equation \eqref{eq:c_n} therefore gives a prescription on how to determine the coefficients $\{c_n\}$ given the generic vector $\ket{x}$ and an orthonormal basis $\{\ket{\phi_n}\}$. For any given $\mathcal{H}_d$, there is an infinite number of possible orthornormal basis  $\{\ket{\phi_n}\}$ -- which means there is also an infinite number of ways of expressing the same vector $\ket{x}$ of $\mathcal{H}_d$.
\par If $\ket{x}$ is a unit vector, then \eqref{eq:x} represents the state of an isolated system expressed as the linear combination of other state vectors $\{\ket{\phi_n}\}$  (remember each $\ket{\phi_n}$ is a unit vector, thus a potential state vector). In this case, $\sum^d_{n=1} \abs{c_n}^2 = 1$ (\emph{normalization condition}). Furthermore, it must be emphasized that for a state vector expressed by the sum in \eqref{eq:x}, the state of the system is \emph{simultaneously} $\ket{\phi_1}, \ldots, \ket{\phi_d}$, each to some degree; for each $\ket{\phi_n}$, this degree is quantified by the respective coefficient $c_n$ and we say there is \emph{quantum coherence} between the components $\{\ket{\phi_n}\}$
 \cite{book:Schlosshauer-2007,book:Zeh-2003}. And $\ket{x}$ in \eqref{eq:x} is said to be a \emph{coherent superposition} of the state vectors $\ket{\phi_1}, \ldots, \ket{\phi_d}$. We observe that the phenomenon of quantum superposition is a basis-dependent one. This basis dependency plays an important role in ongoing research aimed at explaining how the classical world we are used to emerges from the quantum world
\cite{book:Schlosshauer-2007, book:Zeh-2003, art:Schlosshauer-2019}. 
\subsubsection{The Hilbert space of linear operators $\mathcal{O}_d$ and the extended Hilbert-Schmidt inner product.}
\par A vector $\ket{x}$ of $\mathcal{H}_d$ may be transformed into another vector $\ket{x'}$ of the same Hilbert space by means of what is called a \emph{linear operator} $A$, \emph{i.e.} $\ket{x'}=A\ket{x}$. A linear operator $A$ on $\mathcal{H}_d$ is such that if $\ket{x}, \ket{y} \in \mathcal{H}_d$, and $c$ is a scalar, then
			\begin{subequations}
			\begin{align}
			A(\ket{x} + \ket{y}) & = A \ket{x} + A\ket{y}\\
			A \big(c \ket{x}\big) & = c A \ket{x} \ .
			\end{align}
			\end{subequations}
One important operator we shall often deal with is the identity operator on $\mathcal{H}_d$, denoted as $\mathbb{I}_d$. When applied to an arbitrary vector $\ket{x} \in \mathcal{H}_d$, $\mathbb{I}_d$ leaves the vector unchanged: \emph{i.e.} $\ket{x} = \mathbb{I}_d \ket{x} , \  \forall \ket{x}\in \mathcal{H}_d$. If we go back to \eqref{eq:x}, use the result in \eqref{eq:c_n} -- and the fact that as a scalar, $c_n$ commutes with the vector $\ket{\phi_n}$, \emph{i.e.} $c_n\ket{\phi_n}=\ket{\phi_n}c_n$--, we see that
				\begin{equation}
				\label{eq:x_2}
				\ket{x} = \sum^d_{n=1} \ket{\phi_n} c_n = \sum^d_{n=1} \ket{\phi_n} \braket{\phi_n}{x}\ .
				\end{equation}
Based on \eqref{eq:inner_prod}, we may rewrite \eqref{eq:x_2} as
				\begin{equation}
				\ket{x} = \left(\sum^d_{n=1} \ket{\phi_n} \!\bra{\phi_n}\right) \ket{x} \ .
				\end{equation}
Since $\ket{x}$ is an arbitrary vector of $\mathcal{H}_d$, it follows that
				\begin{equation}
				\label{eq:identity_I}
				\mathbb{I}_d = \sum^d_{n=1} \ket{\phi_n}\!\bra{\phi_n} \ .
				\end{equation}
This identity is commonly referred to as the \emph{closure} (or \emph{completeness}) \emph{relation}. Equation \eqref{eq:identity_I} is valid for any arbitrary orthonormal basis of $\mathcal{H}_d$. 
\par Certainly, for a given $\mathcal{H}_d$ there is an infinite number of linear operators one can define on it. These operators also form a finite-dimensional complex linear space. Let it be denoted as $\mathcal{O}_d$. In reality, there is a one-to-one correspondence between elements of $\mathcal{O}_d$ and $(d \times d)$ square matrices
\cite{book:Jordan-2006}, so we shall speak of the elements of $\mathcal{O}_d$ as operators or square matrices, interchangeably. 

\par Just as we could expand any vector $\ket{x}\in \mathcal{H}_d$ in any basis of of the same linear space, we can also expand any operator $X \in \mathcal{O}_d$ in any given basis of $\mathcal{H}_d$. What makes it easy to do so is the closure relation in \eqref{eq:identity_I} and the fact that for any $X\in \mathcal{O}_d$ and integers $n,m$, the relation: $\mathbb{I}^m_d X \mathbb{I}^n_d=X$ always holds. For example, say we want to expand the operator $X \in \mathcal{O}_d$ in the basis $\{\ket{\phi_n}\}$. Then, the following identities follows:
				\begin{equation}
				\label{eq:X_expansion}
				\begin{split}
				X & = \mathbb{I}_d X \mathbb{I}_d = \left( \sum^d_{n=1} \ket{\phi_n}\!\bra{\phi_n} \right) X\left( \sum^d_{{n'}=1} \ket{\phi_{n'}}\!\bra{\phi_{n'}} \right)\\
				& = \sum^d_{n=1}\sum^d_{{n'}=1} \ket{\phi_n}\!\matrixel{\phi_n}{ X } {\phi_{n'}}\bra{\phi_{n'}} 
				\end{split}
				\end{equation}
where $\matrixel{\phi_n}{ X } {\phi_{n'}}$ is a shorthand for the matrix product $\big(\bra{\phi_n} \big)\big(X \big)\big(\ket{\phi_{n'}} \big)$. Since $\bra{\phi_n}$ is  a $(1 \times d)$ matrix, $X$ is a $(d\times d )$ matrix and  $\ket{\phi_{n'}}$ is a $(d\times 1)$ matrix, we note that the product $\matrixel{\phi_n}{ X } {\phi_{n'}}$ must therefore be a scalar, and hence, commutes with both $\ket{\phi_n}$ and $\bra{\phi_{n'}}$. Thus, 
				\begin{equation}
				\label{eq:X_expanded_phi_n_n'}
				X = \sum^d_{n=1}\sum^d_{{n'}=1} \matrixel{\phi_n}{ X } {\phi_{n'}} \ket{\phi_n}\!\bra{\phi_{n'}} = \sum^d_{n=1}\sum^d_{{n'}=1} X_{nn'} \ket{\phi_n}\!\bra{\phi_{n'}}
				\end{equation}
where $X_{nn'} \equiv \matrixel{\phi_n}{ X } {\phi_{n'}}$. The matrix product $\ket{\phi_n}\!\bra{\phi_{n'}}$ is between a $(d\times 1)$ matrix (on the left) and a $(1 \times d)$ matrix (on the right), so $\ket{\phi_n}\!\bra{\phi_{n'}}$ is actually a $(d\times d)$ matrix just like $X$. In fact, $\ket{\phi_n}\!\bra{\phi_{n'}}$ is an element of $\mathcal{O}_d$. If we should write down $X$ as a $(d\times d)$ matrix in the basis $\{\ket{\phi_n}\}$, we note that $X_{nn'}$ corresponds to the matrix element at the intersection between the $n-$th row and $n'-$th column. The sum of all the diagonal elements of $X$ in the basis $\{\ket{\phi_n}\}$ is called its \emph{trace} with respect to $\{\ket{\phi_n}\}$, denoted $\Tr_\phi[X]$:
				\begin{equation}
				\label{eq:trace_def}
				\Tr_\phi[X] = \sum^d_{n=1} \matrixel{\phi_n}{X}{\phi_n} \ .
				\end{equation}
One important property of the trace is that it is independent of the basis. That is, if $\{\ket{\phi_n}\}$ and $\{\ket{\vartheta_n}\}$ are two independent basis of $\mathcal{H}_d$, the trace of $X\in \mathcal{O}_d$ remains the same with respect to both bases. In fact, from  \eqref{eq:trace_def} we have
				\begin{equation}
				\label{eq:trace_inv_def}
				\begin{split}
				\Tr_\phi[X] & = \sum^d_{n=1} \matrixel{\phi_n}{X}{\phi_n} = \sum^d_{n=1} \matrixel{\phi_n}{ \mathbb{I}_d X }{\phi_n}
				 = \sum^d_{n=1} \bra{\phi_n}\left(\sum^d_{n'=1}\ket{\vartheta_{n'}}\!\bra{\vartheta_{n'}} \right)X\ket{\phi_n}\\
				 & = \sum^d_{n=1} \sum^d_{n'=1} \braket{\phi_n}{\vartheta_{n'}} \! 
\matrixel{\vartheta_{n'}}{X}{\phi_n} = \sum^d_{n=1} \sum^d_{n'=1}  \matrixel{\vartheta_{n'}}{X}{\phi_n} \braket{\phi_n}{\vartheta_{n'}}
				\end{split}
				\end{equation}
where in the last step we have used that fact that $\braket{\phi_n}{\vartheta_{n'}}$ and $\matrixel{\vartheta_{n'}}{X}{\phi_n}$ are scalars and therefore commute. Proceeding, we have
				\begin{equation}
				\label{eq:trace_inv_def_2}
				\begin{split}
				\Tr_\phi[X]
				 & =  \sum^d_{n'=1}  \bra{\vartheta_{n'}}{X}\left(\sum^d_{n=1} \ket{\phi_n} \bra{\phi_n}\right)\ket{\vartheta_{n'}} = \sum^d_{n'=1}  \matrixel{\vartheta_{n'}}{X \mathbb{I}_d}{\vartheta_{n'}} = \sum^d_{n'=1}  \matrixel{\vartheta_{n'}}{X}{\vartheta_{n'}} \\
				 & = \Tr_\vartheta [X] \ .
				\end{split}
				\end{equation}
Given that the trace of $X$ is independent of the basis, it is commonly indicated as $\Tr[X]$, without specifying the basis.

\par  The linear space $\mathcal{O}_d$ also has its corresponding adjoint space, which we denote as $\mathcal{O}^*_d$. As usual, there is a one-to-one correspondence between the elements of $\mathcal{O}_d$ and $\mathcal{O}^*_d$: if $A \in \mathcal{O}_d$, then its dual is $A^\dagger \left(\in \mathcal{O}^*_d\right)$. Also, since it is a linear space, we would expect to be able to define an inner product on $\mathcal{O}_d$. The commonly used one here is the Hilbert-Schmidt  inner product. If $A$ and $B$ are two elements of $\mathcal{O}_d$, then their Hilbert-Schmidt inner product is
				\begin{equation}
				\label{eq:inner_prod_Hilbert-Schmidt}
				\big< A, B \big> = \mbox{Tr} \big[  B A^\dagger \big]\ . 
				\end{equation}			 
$A$ and $B$ are square matrices so $\big< A, B \big> = \mbox{Tr} \big[   A^\dagger B\big]$, due to a property of the trace functional which can easily be proved. But on close examination, one observes that if we define an \emph{ordered}  inner product (meaning, the position of the elements $A,B$ are to be strictly maintained) as given in \eqref{eq:inner_prod_Hilbert-Schmidt} (which me may call `\emph{extended} Hilbert-Schmidt' to differentiate it from the conventional one), then the inner product we defined for $\mathcal{H}_d$, \eqref{eq:inner_prod_Hilbert}, is also of the same kind. Indeed, from   \eqref{eq:inner_prod_Hilbert} and \eqref{eq:identity_I}, and the fact that $\ket{x}=\mathbb{I}_d \ket{x}$, it follows that
				\begin{equation}
				\begin{split}
				\label{eq:inner_prod_2}
				\big<\ket{y} , \ket{x}\big> & = \braket{y}{x} = \matrixel{y}{\mathbb{I}_d}{x} 
				 = \sum^d_{n=1} \braket{y}{\phi_n}\!\braket{\phi_n}{x} = \sum^d_{n=1} \braket{\phi_n}{x}\!\braket{y}{\phi_n} = \Tr\big[\ket{x}\!\bra{y}\big] \ .
				\end{split}
				\end{equation}
Thus, for both $\mathcal{H}_d$ and $\mathcal{O}_d$, we have the extended Hilbert-Schmidt inner product, \eqref{eq:inner_prod_Hilbert-Schmidt}, as the chosen inner product. Moreover, if $\{\ket{\phi_n}\}$ is an orthonormal basis for $\mathcal{H}_d$, then the set of operators $\{\ket{\phi_n}\!\bra{\phi_{n'}}\}$ --- where $n,n'=1,2,\ldots, d$ --- constitute an orthonormal basis for $\mathcal{O}_d$: That is, any element of $\mathcal{O}_d$ can be expressed as linear combination of the elements of the set $\{\ket{\phi_n}\!\bra{\phi_{n'}}\}$. This is what we achieved, for example, in \eqref{eq:X_expanded_phi_n_n'}. The orthonormality of the elements of $\{\ket{\phi_n}\!\bra{\phi_{n'}}\}$ can be verified with the extended Hilbert-Schmidt inner product, \eqref{eq:inner_prod_Hilbert-Schmidt}. The linear space $\mathcal{O}_d$ is therefore complete and normed, which makes it a \emph{(complex) Hilbert space}. The dimension of the set $\{\ket{\phi_n}\!\bra{\phi_{n'}}\}$ ---  and, therefore, of $\mathcal{O}_d$ --- is easily seen to be $d^2$.
\par An important class of elements of $\mathcal{O}_d$ are the so-called \emph{Hermitian operators}. An operator $A \in \mathcal{O}_d$ is said to be Hermitian if it coincides with its conjugate transpose, \emph{i.e.} $A = A^\dagger$. It turns out that physical dynamical variables of a system (like its total energy, magnetization vector, etc.) are represented by Hermitian operators. The \emph{spectral decomposition theorem}
\cite{book:Nielsen-2011, book:Jordan-2006, book:Moretti-2018, book:Dennery-1996} of quantum mechanics asserts that given a Hermitian operator $A \in \mathcal{O}_d$ there exists an orthonormal basis $\{\ket{a_n}\!\bra{a_{n'}}\}$ of $\mathcal{O}_d$ such that
				\begin{equation}
				\label{eq:spectral_decomposition_thm}
				A = \sum^d_{n=1} \lambda_n \ket{a_n}\!\bra{a_n} \ \qquad \braket{a_n}{a_{n'}} = \delta_{n,n'} 
				\end{equation}
where the scalars $\{\lambda_n\}$ are real. The expansion in \eqref{eq:spectral_decomposition_thm} is said to be the \emph{diagonal representation} of the operator $A$. The vectors $\{\ket{a_n}\}$ are the eigenvectors of $A$ and $\{\lambda_n\}$ are the corresponding eigenvalues; indeed, if we multiply \eqref{eq:spectral_decomposition_thm} from the left by $\ket{a_m}$, we get $A \ket{a_m}=\lambda_m \ket{a_m}$ -- which is a normal eigenvalue/eigenvector equation.
\subsection{Postulate II}\label{subsec:Postulate_II}
The second postulate asserts how the state vector of a closed quantum system evolves in time.
	\begin{quotation}
\item \textbf{Postulate II}
The state vector $\ket{\psi}$ of a closed quantum system evolves in time according to the \emph{Schr\"odinger equation}:
				\begin{equation}
				\label{eq:Schrodinger}
				\frac{d}{dt}\ket{\psi} = -\frac{i}{\hbar} H \ket{\psi}
				\end{equation}
-- where $\hbar$ is the reduced Planck constant;  $H$ is a linear operator called the \emph{Hamiltonian} of the quantum system and it represents the total energy of the system.
	\end{quotation} 
Because $\ket{\psi}$ in \eqref{eq:Schrodinger} clearly depends on time, it is customary to indicate its time-dependence by writing $\ket{\psi}$ as $\ket{\psi(t)}$. Suppose $\ket{\psi(t)}$ is a vector of $\mathcal{H}_d$. If we choose to expand $\ket{\psi(t)}$ in the basis $\{\ket{\phi_n}\}$, then
				\begin{equation}
				\ket{\psi(t)} = \mathbb{I}_d \ket{\psi(t)} = \sum^d_{n=1} \ket{\phi_n}\braket{\phi_n}{\psi(t)} = \sum^d_{n=1} c_n(t) \ket{\phi_n}
				\end{equation}
where the coefficients $c_n(t) \equiv \braket{\phi_n}{\psi(t)}$ are now time-dependent.
\par At first glance, this postulate may seem very limiting because no system in the universe can be truly closed. In one way or the other, every system interacts with another system. Nonetheless, Postulate II is the basis on which \emph{open} quantum systems (\emph{i.e.} quantum systems engaged in some form of interaction with other systems) are effectively described. In fact, many approaches have been devised for describing how open quantum systems evolve in time and they all start with Postulate II by assuming the relevant system we wish to describe and the other systems with which it interacts (collectively called the \emph{environment} or \emph{reservoir}) form a closed system. These approaches collectively go under what is called theory of open quantum systems. 
\subsection{Postulate III}\label{subsec:Postulate_III}
To determine properties like energy or spin magnetic moment of a quantum system, we need to perform some measurements on the system. These actions ultimately constitute some form of interaction with the quantum system. In this sense, interaction and measurement are almost synonymous. It is thus no wonder this third postulate is important to the theory of open quantum systems. Postulate III of quantum mechanics is the quantum outlook on the process of measurements, and also introduces the concept of `quantum measurement operators'.
	\begin{quotation}
	\item \textbf{Postulate III} Let the outcomes of a (quantum) measurement be a  countable set indexed  $m$, \emph{i.e.} $\{r_m\}$. To each outcome $r_m$ is associated an operator $M_m$ called \emph{measurement operator}. 
	Let $\ket{\psi}$ be the state of the quantum system at the instant immediately prior to the measurement. Then, the probability $p(m)$ that the outcome is $r_m$ is 
				\begin{equation}
				\label{eq:p(m)}
				p(m) = \matrixel{\psi}{M^\dagger_m M_m}{\psi} 
				\end{equation}
and the new state $\ket{\psi'}$ of the system immediately after the measurement is
				\begin{equation}
				\label{eq:new_state_psi'}
				\ket{\psi'} = \frac{M_m \ket{\psi}}{\sqrt{p(m)}}
				\end{equation}
If $\ket{\psi} \in \mathcal{H}_d$, then the measurement operators $\{M_m\}$ are such that
				\begin{equation}
				\label{eq:completeness_M_m}
				\sum_m M^\dagger_m M_m =  \mathbb{I}_d \ .
				\end{equation}
	\end{quotation}
Equation \eqref{eq:completeness_M_m} is the \emph{completeness equation} for the measurement operators $\{M_m\}$. It is important because it indirectly instills the requirement that the probabilities $p(m)$ summed over all $m$ adds up to the value $1$. In fact, from \eqref{eq:p(m)}, we have $\sum_m p(m) = \sum_m \matrixel{\psi}{M^\dagger_m M_m}{\psi} =  \matrixel{\psi}{\sum_m M^\dagger_m M_m}{\psi} = \braket{\psi}{\psi}=1$.  It should be quite clear that the observable being measured and the nature of the quantum system being probed determine the set $\{M_m\}$.
\par Another concept central to quantum mechanics is that of \emph{expectation value} (or \emph{mean value}) of a measurement. The expectation value of a measurement (related to an observable) is the average outcome of the measurements. If the outcomes of the observable represented by the operator $B$ are $\{\lambda_m\}$ with corresponding probabilities $\{p(m)\}$, it is clear that the expectation value of the observable, indicated as $\big<B\big>$, is the weighted sum
				\begin{equation}
				\label{eq:expectation_value_0}
				\big< B \big> = \sum_m \lambda_m \ p(m) = \sum_m \lambda_m  \matrixel{\psi}{M^\dagger_m M_m}{\psi} = \matrixel{\psi}{B}{\psi}
				\end{equation}		
where we have made use of \eqref{eq:p(m)}, and have defined the operator $B$ as
				\begin{equation}
				\label{eq:observable_gen}
				B = \sum_m \lambda_m M^\dagger_m M_m \ , \qquad \text{with }\sum_m  M^\dagger_m M_m = \mathbb{I}_d  \ .
				\end{equation}
\Eref{eq:observable_gen} is a more general operator representation of an observable in quantum mechanics (for finite-dimensional state space). It is evident from \eqref{eq:expectation_value_0} that the value $\big< B \big>$ is always real. Note that  the measurement operators $M_m$ may be Hermitian or not. 
As an example, consider the spin angular momentum operator $S_z$ of an electron. We know the spin state space of the electron is a two-dimensional Hilbert space and 
				\begin{equation}
				S_z = \hbar\begin{pmatrix}
				\frac{1}{2} & 0 \\
				0 & -\frac{1}{2} \ .
				\end{pmatrix}
				\end{equation}
which is a Hermitian operator. Note that we if we take $\{\ket{0},\ket{1}\}$ as the basis of the electron's spin state space, where
				\begin{equation}
				\label{eq:matrix_rep_ket_0_1}
				\ket{1} = \begin{pmatrix}
				1 \\
				0
				\end{pmatrix}
				\qquad 
				\ket{0} = \begin{pmatrix}
				0 \\
				1
				\end{pmatrix} \ .
				\end{equation}
we may rewrite $S_z$ as
				\begin{equation}
				\label{eq:S_z_expanded}
				S_z= \frac{\hbar}{2} \ket{1}\!\bra{1} - \frac{\hbar}{2} \ket{0}\!\bra{0} 
				\end{equation}
From \eqref{eq:S_z_expanded}, we can infer the outcomes $\{\lambda_m\}$ of the measurement of $S_z$, and their corresponding measurement operators $\{M_m\}$. Indeed, we notice that \eqref{eq:S_z_expanded} may be rewritten as
				\begin{equation}
				\label{eq:S_z_measurement_operators}
				S_z = \lambda_0 \ M^\dagger_0 M_0 + \lambda_1\  M^\dagger_1 M_1
				\end{equation}
where 		
				\begin{subequations}
				\label{eq:S_z_M_m_lambda_m}
				\begin{align}
				\lambda_0 = -\frac{\hbar}{2} \ , \quad & M_0 = \ket{0}\!\bra{0}	\\
				\lambda_1 = +\frac{\hbar}{2} \ , \quad & M_1 = \ket{1}\!\bra{1}	
				\end{align}
				\end{subequations}									 
and the completeness relation $M^\dagger_0M_0 + M^\dagger_1M_1=\mathbb{I}_2$ is satisfied. For these particular measurement operators, it is observed that $M_m M_{m'}=\delta_{m,m'} M_m$. Such measurement operators give rise to what is called \emph{projective measurements}
\cite{book:Nielsen-2011}. Also, note that $M_0$ and $M_1$ may be expanded in any basis of our choice, but the fundamental structure of \eqref{eq:S_z_measurement_operators} remain unchanged. With the help of \eqref{eq:S_z_M_m_lambda_m}, one can determine, for example, through \eqref{eq:new_state_psi'} the probability of the outcomes $\lambda_0,\lambda_1$ given any initial spin state $\ket{\psi}$ of the electron. 
\par As another example, consider an observable $F$ of the electron spin given by the outcomes $\{\lambda_{\pm}\}$ and their corresponding measurement operators $\{\sigma_\pm\}$, \emph{i.e.}
					\begin{equation}
					F = \lambda_+ \sigma^\dagger_+\sigma_+ + \lambda_- \sigma^\dagger_- \sigma_-
					\end{equation}
-- where $\sigma_+=\ket{1}\!\bra{0}$ and $\sigma_-=\ket{0}\!\bra{1}$. We note that in this case the measurement operators are not Hermitian (\emph{i.e.} $\sigma_\pm \neq \sigma^\dagger_\pm$) but, nonetheless, they satisfy the completeness relation for measurement operators, \eqref{eq:completeness_M_m}, \emph{i.e.} $\sigma^\dagger_+\sigma_+ + \sigma^\dagger_- \sigma_- = \mathbb{I}_2$.

\subsection{Pure states, mixed states and density matrices}\label{subsec:pure_mixed}
The state of a quantum system (closed or open) may be \emph{pure} or \emph{mixed}. It is \emph{pure} when we have complete information on it
\cite{book:Blum-2012, book:Schlosshauer-2007, art:Luo-2005}. `Complete information' in the sense that there is no classical uncertainty as to what the quantum state is
\cite{art:Luo-2005}, so we can assign a single state vector $\ket{\psi}$ to the system. A classic way of illustrating this concept is through the Stern-Gerlach experiment, where a beam of silver atoms is passed through an inhomogeneous magnetic field. The spin quantum number of a normal silver atom is $\frac{1}{2}$, so in trasversing the inhomogeneous magnetic field the beam is split into two: one in which each silver atom has a spin magnetic moment of $+\frac{1}{2}\hbar$ (let us call it Beam 1) and another in which each atom has a spin magnetic moment of $-\frac{1}{2}\hbar$ (Beam 2). Beam 1 can be collected. The spin state of all the silver atoms in Beam 1 is completely known so the beam is in a pure (spin) state. And we can assign a single spin state vector $\ket{\psi}$ to Beam 1. The same applies to Beam 2.
\par When we cannot assign a single state vector $\ket{\psi}$ to the system because there is classical uncertainty on the system's quantum state, we resort to  (classical) statistics to describe the state. The result is what we call a  \emph{mixed state}
\cite{book:Blum-2012, book:Schlosshauer-2007, art:Luo-2005}. This is a collection of positive real numbers $0 \leq P_k \leq 1$ and their corresponding vectors $\ket{\psi_k} \in \mathcal{H}_d$, where $P_k$ is the probability that the state of the quantum system is given by the state vector $\ket{\psi_k}$. Mixed states are therefore typically expressed in the form of the collection $\{P_k, \ket{\psi_k}\} $ and they are symptom of the observer's lack of complete information on the quantum state of the system under study. Either $P_k$ or $\ket{\psi_k}$ (or both) may depend on time. Nonetheless, the condition $\sum_k P_k =1$ always holds.
\par Given the mixed state $\{P_k, \ket{\psi_k}\}$, where the state vectors $\ket{\psi_k}$ form an orthonormal basis of $\mathcal{H}_d$, we may introduce an operator $\rho \in \mathcal{O}_d$, the density matrix (or density operator), defined on $\mathcal{H}_d$ such that
				\begin{equation}
				\matrixel{\psi_k}{\rho}{\psi_k}=P_k \ , \  \forall k \ .
				\end{equation}
It is not difficult to realize that the operator $\rho$ must be of the form
				\begin{equation}
				\label{eq:rho}
				\rho = \sum^d_{k=1} P_k \ket{\psi_k}\!\bra{\psi_k} \ .
				\end{equation}	
Unlike the sum in \eqref{eq:x}, state of the quantum system as expressed in \eqref{eq:rho} is \emph{not} one which is simultaneously the $\{\ket{\psi_k}\}$. Rather, \eqref{eq:rho} expresses a classical mixture of states. \Eref{eq:rho} is also reminiscent of \eqref{eq:spectral_decomposition_thm}. Indeed, $\rho$ is Hermitian and \eqref{eq:rho} is its diagonal representation. If we choose to represent $\rho$ in a different orthonormal basis --- say $\{\ket{\phi_n}\}$ --- , then, from   \eqref{eq:rho} and \eqref{eq:identity_I}, we have
				\begin{equation}
				\begin{split}
				\label{eq:rho_2}
				\rho & = \sum^d_{k=1} P_k \ \mathbb{I}_d\ket{\psi_k}\!\bra{\psi_k}\mathbb{I}_d 
				 = \sum^d_{k=1} P_k \sum^d_{n=1}\sum^d_{n'=1} \ket{\phi_n}\braket{\phi_n}{\psi_k}\braket{\psi_k}{\phi_{n'}}\bra{\phi_{n'}}\\
				 & =  \sum^d_{n=1}\sum^d_{n'=1} \bigg( \sum^d_{k=1} P_k \braket{\phi_n}{\psi_k}\braket{\psi_k}{\phi_{n'}} \bigg) \ket{\phi_n}\!\bra{\phi_{n'}}\\
				 & = \sum^d_{n=1}\sum^d_{n'=1} \rho_{n n'} \ket{\phi_n}\!\bra{\phi_{n'}} \ .
				\end{split} 
				\end{equation}				
where we can see $\rho_{nn'}\equiv \matrixel{\phi_n}{\rho}{\phi_{n'}}$. Thus, in a different orthonormal basis, $\rho$ ceases to be diagonal even though the statistics of measurement outcomes remain the same. Such different representations of the same density matrix $\rho$ are related to each other through a unitary transformation.
    If the system is isolated, with Hamiltonian $H$, and we take the time derivative of $\rho(t)$, then, from   \eqref{eq:Schrodinger} and \eqref{eq:rho}, we obtain
    			\begin{equation}
    			\label{eq:LvN}
    			\frac{d}{dt}\rho(t) = -\frac{i}{\hbar}\left[H,\rho(t) \right]
    			\end{equation}
where $\rho$ is now written as $\rho(t)$ to show its time-dependence and $\ket{\psi_k}\!\bra{\psi_k} \to \ket{\psi_k(t)}\!\bra{\psi_k(t)}$ in \eqref{eq:rho}. (For any pair of operators $A,B \in \mathcal{O}_d$, their commutator $\left[ A,B\right]$ is defined as $\left[ A,B\right]\equiv AB - BA$. And their anti-commutator, $\left[ A,B\right]_+$, is defined as $\left[ A,B\right]_+ \equiv AB + BA$. We note that $[A,B]=-[B,A]$ and  $[A,B]_+=[B,A]_+$.) \Eref{eq:LvN} is referred to as the \emph{Liouville-von Neumann equation}.
\par Note that a pure state may also be expressed in the form of a density matrix. If the quantum system is in the pure state $\ket{\psi(t)}$, then $\{P_k, \ket{\psi_k}\} \to \{ 1,\ket{\psi(t)} \}$, and so from \eqref{eq:rho}, it follows that
				\begin{equation}
				\label{eq:rho_pure_0}
				\rho(t)= \ket{\psi(t)}\!\bra{\psi(t)} \ .
				\end{equation}
If we express this pure state density matrix in the basis $\{\ket{\phi_n}\}$, for example, we get
				\begin{equation}
				\label{eq:rho_pure}
				\begin{split}
				\rho(t)& = \mathbb{I}_d \ket{\psi(t)}\bra{\psi(t)} \mathbb{I}_d = \sum^d_{n=1}\sum^d_{n'=1} \ket{\phi_n}\braket{\phi_n}{\psi(t)} \braket{\psi(t)}{\phi_{n'}}\bra{\phi_{n'}}\\
				& = \sum^d_{n=1}\sum^d_{n'=1} c_{nn'}(t) \ket{\phi_n}\!\bra{\phi_{n'}} 
				\end{split}
				\end{equation}
with $c_{nn'}(t) \equiv \braket{\phi_n}{\psi(t)}\! \braket{\psi(t)}{\phi_{n'}}= \matrixel{\phi_n}{\rho(t)}{\phi_{n'}}$. 
We recognize the similarity between the final forms of $\rho$ in \eqref{eq:rho_2} and \eqref{eq:rho_pure}. This tells us that the fact that a density matrix $\rho$ has nonzero off-diagonal elements (\emph{i.e.} coherence) in a basis, does not necessarily mean it represents a mixed state. To check whether a density matrix represents a mixed or pure state, one has to put it in its diagonal form, \eqref{eq:rho}: if there are more than one nonzero $P_k$, then the state is mixed, otherwise it is pure. Put more elegantly, $\rho$ represents a pure state if its rank (\emph{i.e.} the number of nonzero eigenvalues) is $1$; if the rank is greater than $1$ then the state is mixed. Consider, for example, the following qubit density matrices:
				\begin{equation}
				\label{eq:rho_1_rho_2}
				\rho_1=\begin{pmatrix}
				 \frac{1}{2} & -\frac{e^{i\theta}}{2}\\
				-\frac{e^{-i\theta}}{2} & \frac{1}{2}
				\end{pmatrix} \ \qquad \ 
				\rho_2=\begin{pmatrix}
				 \frac{1}{2} & -i\frac{\sqrt{2}}{3}\sin{\theta}\\
				i\frac{\sqrt{2}}{3}\sin{\theta} & \frac{1}{2} 
				\end{pmatrix} 				\ .
				\end{equation}
where $\theta$ is some parameter of the system. Note that both density matrices have the same populations (\emph{i.e.} diagonal elements), but $\rho_1$ is a pure state and $\rho_2$ is a mixed state.
\par The fact that the rank of a pure state's density matrix is $1$ also leads to the assertion that $\Tr[\rho^2] = 1$ if $\rho$ represents a pure state; while for a mixed state, $\Tr[\rho^2] < 1$. For $\rho_1$ and $\rho_2$ in \eqref{eq:rho_1_rho_2}, for example, one can easily verify that $\Tr[\rho^2_1]=1$ and $\Tr[\rho_2^2]= \frac{1}{2}+\frac{4}{9}\sin^2\theta < 1$.
\par The final problem we wish to attend to before moving on to the Liouville space formalism is how Postulate III turns out when dealing with a mixed state $\{P_k, \ket{\psi_k}\}$ instead of a state vector $\ket{\psi}$.  Suppose we have an observable $B$ --  described by a set of quantum measurement operators $\{M_m \mid \sum_m M^\dagger_m M_m = \mathbb{I}_d\}$ and their corresponding outcome values $\{\lambda_m\}$. We want to determine the probability $p(m)$ that the outcome of measuring $B$ is $\lambda_m$, given that the state of the quantum system immediately prior to the measurement is given by the mixed state $\rho = \{P_k, \ket{\psi_k}\}$ -- where the state vectors $\{\ket{\psi_k}\}$ are assumed orthogonal to each other. Naturally, we have to apply Postulate III. The only complication is that, contrary to what Postulate III originally describes, we are dealing here with a collection of states $\{P_k, \ket{\psi_k}\}$.  But this is no complication because if immediately prior to the measurement, the state of the quantum system is $\{P_k, \ket{\psi_k}\}$, then the probability $p(m)$ that we get $\lambda_m$ as the outcome is 
				\begin{equation}
				\label{eq:p(m)_mixed}
				p(m) = \sum_k p(m|k) \ P_k
				\end{equation}
where $p(m|k)$ is the probability that the outcome $\lambda_m$ is a consequence of the quantum system being in the state $\ket{\psi_k}$ immediately prior to the measurement. That is, following \eqref{eq:p(m)},
				\begin{equation}
				\label{eq:p(m|k)}
				p(m|k) = \matrixel{\psi_k}{M^\dagger_mM_m}{\psi_k}.
				\end{equation}
Note, however, that
				\begin{equation}
				\begin{split}
				\matrixel{\psi_k}{M^\dagger_mM_m}{\psi_k} & = \matrixel{\psi_k}{\mathbb{I}_d M^\dagger_m  M_m}{\psi_k}  = \sum^d_{n=1} \braket{\psi_k}{\phi_n}\!\matrixel{\phi_n}{ M^\dagger_m  M_m}{\psi_k}\\
				& = \sum^d_{n=1} \matrixel{\phi_n}{ M^\dagger_m  M_m}{\psi_k}\!\braket{\psi_k}{\phi_n}
				 = \Tr \bigg[ M^\dagger_m  M_m\ket{\psi_k}\!\bra{\psi_k}\bigg]\ .
				\end{split}
				\end{equation}
Thus, $p(m|k)= \Tr \bigg[ M^\dagger_m  M_m\ket{\psi_k}\!\bra{\psi_k}\bigg]$ and \eqref{eq:p(m)_mixed} may be written as
				\begin{equation}
				\begin{split}
				\label{eq:p(m)_mixed_2}
				p(m)  = & \sum_k \Tr \bigg[ M^\dagger_m  M_m\ket{\psi_k}\!\bra{\psi_k}\bigg] \ P_k 
				= 	 \Tr \bigg[ M^\dagger_m  M_m \left(\sum_k\ket{\psi_k}\!\bra{\psi_k}P_k \right)\bigg] \\
				= &  \Tr\bigg[ M^\dagger_m  M_m \ \rho \bigg] \ .
				\end{split} 	
				\end{equation}				 
The expectation value $\big< B \big>$ is still given by the weighted sum $\big< B \big> = \sum_m \lambda_m p(m)$, \eqref{eq:expectation_value_0}, as we saw above. But now, introducing the expression for $p(m)$, \eqref{eq:p(m)_mixed_2}, we get
				\begin{equation}
				\label{eq:B_avg_general}
				\begin{split}
				\big< B \big> 
				= &  \sum_m \lambda_m \ p(m) 
				= \sum_m \lambda_m \Tr\bigg[ M^\dagger_m  M_m \ \rho \bigg] 
				=  \Tr\bigg[ \left( \sum_m \lambda_mM^\dagger_m  M_m\right)  \rho \bigg]\\
				= & \Tr \big[ B \ \rho \big] \ .
				\end{split}
				\end{equation}	 	
Furthermore, we may be interested in what the mixed state $\rho=\{P_k, \ket{\psi_k}\}$ transforms into, immediately after the outcome of the measurement of $B$, if the outcome is recorded to be $\lambda_m$. Let us indicate this new mixed state as $\rho_m$. It can be shown that
\cite{book:Nielsen-2011}
				\begin{equation}
				\rho_m = \frac{M_m \rho M^\dagger_m}{p(m)}
				\end{equation}	
where $p(m)$ is given by \eqref{eq:p(m)_mixed_2}. (This may be verified by taking $\rho$ to be a pure state, $\rho=\ket{\psi}\!\bra{\psi}$, and using \eqref{eq:new_state_psi'}.) Suppose we do not record the outcome, so that we are ignorant of what the state of the quantum system is immediately after the measurement. Then, it is easy to see that an effective way to describe the new mixed state  $\rho'$ is as a \emph{mixture} of the states $\{\rho_m\}$ with probabilities $\{p(m)\}$. That is,
				\begin{equation}
				\label{eq:non-selective_measurement}
				\rho' = \sum_m p(m) \ \rho_m = \sum_m M_m \rho M^\dagger_m \ .
				\end{equation}		
This is called \emph{non-selective measurement}
\cite{misc:Lidar-2019}.	For a comprehensive introduction to the quantum theory of measurement see, for example, \cite{book:Holevo-2011}.	 				 
\section{Mathematical foundations of Quantum Mechanics in Liouville space}\label{sec:LS}
\subsection{Some initial considerations}\label{subsec:initial_consid}
For finite-dimensional separable Hilbert spaces (\ref{appendix:A1}) like $\mathcal{H}_ d$ (the Hilbert space of state vectors) and $\mathcal{O}_d$ (the Hilbert space of linear operators on $\mathcal{H}_d$), we have seen in the previous section that it is always possible to represent their elements as column vectors and square matrices, respectively. Indeed, quantum mechanics in separable Hilbert spaces turns out to be essentially matrix analysis
\cite{book:Horn-2012, book:Lax-2007}. Thus, if we consider again Schr\"odinger's equation, \eqref{eq:Schrodinger}, we may conveniently view it as the classic matrix calculus problem 
				\begin{equation}
				\label{eq:x_A_diff_eq}
				\frac{d}{dt}x(t) = A\ x(t) \ , \qquad \ x(0) = c
				\end{equation}
where $x(t)$ and $c$ are $(d \times 1)$ complex column matrices and $A$ is a $(d \times d)$ complex square matrix. We know from matrix analysis (via the theory of matrix functions) that the differential equation in \eqref{eq:x_A_diff_eq} has as solution
\cite{book:Lax-2007, book:Michal-2008, book:Higham-2008}
				\begin{equation}
				\label{eq:x_A_sol}
				x(t)=e^{tA}c
				\end{equation}
where $e^{tA}$ (sometimes also indicated as $\exp[tA]$) is a $(d \times d)$ complex square matrix defined as
				\begin{equation}
				\label{eq:e^At}
				e^{tA} := \mathbb{I}_d + t A + \frac{t^2}{2!} A^2 + \frac{t^3}{3!} A^3 + \ldots = \sum^{\infty}_{n=0} \frac{t^n}{n!} A^n
				\end{equation}
where we define $A^n \equiv \mathbb{I}_d$ ($\mathbb{I}_d$ is the $d \times d$ identity matrix) for $n=0$. For a review on how $e^{tA}$ may be effectively computed, we recommend the updated version of Moler and Van Loan's celebrated ``Nineteen dubious ways'' paper
\cite{art:Moler-2003}. If we set $x(t)=\ket{\psi(t)}$ and $A=-\frac{i}{h} H$, \eqref{eq:x_A_sol} solves the Schr\"odinger equation, \eqref{eq:Schrodinger}.
\par Had $x(t)$ and $A$ been scalars, the solution to the differential equation in \eqref{eq:x_A_diff_eq} would still be of the form given in \eqref{eq:x_A_sol}. This equivalence does not hold true in general. If we consider the Liouville-von Neumann equation --- \eqref{eq:LvN} --- 
				\begin{equation}
				\frac{d}{dt} \rho(t) = -\frac{i}{\hbar}\big(H \rho(t) - \rho(t) H \big)
				\end{equation}
for example, we cannot solve it assuming $\rho(t)$ and $H$ were scalars. What makes matrix differential equations of this sort challenging to solve is the fact that given two matrices $A$ and $B$, in general, $AB \neq BA$. Solving the linear equation
				\begin{equation}
				\label{eq:AX+XB}
				A X + X B = C
				\end{equation}
for $X$ (where $X,A,B$ and $C$ are square matrices of the same dimension $n$) also poses similar problems. The conundrum can be put to rest by resorting to \emph{Kronecker product}. The point is that solving a problem like \eqref{eq:AX+XB} for $X$ is, in ultimate analysis, solving for a countable set of functions (\emph{i.e.} the elements of $X$): however these functions may be organized --- in the form of a square matrix or otherwise --- is of secondary importance. With the help of the Kronecker product, a problem like \eqref{eq:AX+XB} may be solved by factoring out $X$ in the form of a column vector from the l.h.s. turning the equation into (see for example \cite{book:Graham-1981, book:Turkington-2002})
				\begin{equation}
				\label{eq:vec_AX+XB}
				\widetilde{D} \ \mbox{vec}[X] = \mbox{vec}[C]
				\end{equation}
where $\widetilde{D}$ is a square matrix that depends on $A$ and $B$, while $\mbox{vec}[X]$ and $\mbox{vec}[C]$ are the column vector representation of $X$ and $C$, respectively. Note that while $\mbox{vec}[X]$ and $\mbox{vec}[C]$ are column vectors of dimension $(n^2 \times 1)$, $\widetilde{D}$ is a square matrix of dimension $(n^2 \times n^2)$. Moreover, while the form in \eqref{eq:vec_AX+XB} can \emph{always} be achieved from \eqref{eq:AX+XB}, care must be taken when solving the former for $\mbox{vec}[X]$. The solution set for $\mbox{vec}[X]$ may consist of a unique element, infinite elements or be empty (\emph{i.e.} no solution), depending on the nature of $\widetilde{D}$ and $\mbox{vec}[C]$. For example, if $\widetilde{D}$ is invertible and $\mbox{vec}[C]$ is not a null column vector, then
				\begin{equation}
				\label{eq:vec_AX+XB_sol}
				\mbox{vec}[X] = \widetilde{D}^{-1} \mbox{vec}[C] \ .
				\end{equation}
\par If the transformation $A X + X B = C \mapsto \widetilde{D} \ \mbox{vec}[X] = \mbox{vec}[C]$ is always possible, then it is easy to imagine also the following transformation of the Liouville-von Neumann equation
				\begin{equation}
				\label{eq:vec_LvN}
				\frac{d}{dt} \rho(t) = -\frac{i}{\hbar}\big(H \rho(t) - \rho(t) H \big) \quad \mapsto \quad \frac{d}{dt} \mbox{vec}[\rho(t)] = \mathfrak{L} \ \mbox{vec}[\rho(t)]
				\end{equation}
from which follows the general solution
				\begin{equation}
				\mbox{vec}[\rho(t)] = e^{t \mathfrak{L}} \ \mbox{vec}[\rho(0)]
				\end{equation} 
where $\mathfrak{L}$ and $e^{t\mathfrak{L}}$ are $(d^2 \times d^2)$ matrices which depend on the Hamiltonian $H$, while $\mbox{vec}[\rho(t)]$ and $\mbox{vec}[\rho(0)]$ are $(d^2 \times 1)$ column vectors. The square matrix $\mathfrak{L}$ is called the \emph{Liouvillian} (more on this in Sec. \ref{subsec:solving_EMs}). Moreover, the square matrix $e^{t\mathfrak{L}}$ still has a series expression similar to \eqref{eq:e^At}, namely,
				\begin{equation}
				\label{eq:exp(tmahfrakL)}
				e^{t\mathfrak{L}} = \mathbb{I}_{d^2} + t \mathfrak{L} + \frac{t^2}{2!} \mathfrak{L}^2 + \frac{t^3}{3!} \mathfrak{L}^3 + \ldots = \sum^{\infty}_{n=0} \frac{t^n}{n!} \mathfrak{L}^n \ .
				\end{equation}
The transformation in \eqref{eq:vec_LvN} is the quintessence of the Liouville space formalism: \emph{viz}. the linear space where state space linear operators (\emph{i.e.} elements of $\mathcal{O}_d$) become column vectors.
\par Before going any further, it is important we introduce the Kronecker product and its properties in the next subsection. In the subsequent subsections, we employ these properties to develop the Liouville space formalism. Readers familiar with the Kronecker product may proceed directly to Sec. \ref{subsec:bra-flipper}.
\subsubsection{Kronecker product and properties.} \label{sec:Kronecker-product}
Let $X=[x_{ij}]$ be a matrix of dimension $(m \times n)$ and $Y=[y_{ij}]$ a matrix of dimension $(m' \times n')$, 
				\begin{equation}
				\label{eq:X_Y_matrices}
				X  = 
				\begin{pmatrix}
				x_{11} & x_{12} & \ldots & x_{1n}\\
				x_{21} & x_{22} & \ldots & x_{2n}\\
				\vdots & \vdots & \ldots & \vdots \\
				x_{m1} & x_{m2} & \ldots & x_{mn} 
				\end{pmatrix}
				\qquad  
				Y= \begin{pmatrix}
				y_{11} & y_{12} & \ldots & y_{1n'}\\
				y_{21} & y_{22} & \ldots & y_{2n'}\\
				\vdots & \vdots & \ldots & \vdots \\
				y_{m'1} & y_{m'2} & \ldots & y_{m'n'} 				
				\end{pmatrix} \ .
				\end{equation}
Then, the Kronecker product (also called `tensor product' or `direct product') $X \otimes Y$ is a $(m m' \times n  n')$ matrix defined as
				\begin{equation}
				X \otimes Y := 
				\begin{pmatrix}
				x_{11}Y & x_{12}Y & \ldots & x_{1n}Y\\
				x_{21}Y & x_{22}Y & \ldots & x_{2n}Y\\
				\vdots & \vdots & \ldots & \vdots \\
				x_{m1}Y & x_{m2}Y & \ldots & x_{mn}Y 
				\end{pmatrix} \ .
				\end{equation}
For example, if 
				\begin{equation}
				X = \begin{pmatrix}
				x_{11} & x_{12}\\
				x_{21} & x_{22}
				\end{pmatrix} \qquad
				Y = \begin{pmatrix}
				y_{11} & y_{12}\\
 				y_{21} & y_{22}\\
				y_{31} & y_{32}
				\end{pmatrix}
				\end{equation}
then
				\begin{equation}
				X \otimes Y = 
				\begin{pmatrix}
				x_{11} Y & x_{12}Y\\
				x_{21}Y & x_{22}Y
				\end{pmatrix} = 
				\begin{pmatrix}
				x_{11}y_{11} & x_{11}y_{12} & x_{12}y_{11} & x_{12}y_{12}\\
 				x_{11}y_{21} & x_{11}y_{22} & x_{12}y_{21} & x_{12}y_{22}\\
				x_{11}y_{31} & x_{11}y_{32} & x_{12}y_{31} & x_{12}y_{32}\\
				x_{21}y_{11} & x_{21}y_{12} & x_{22}y_{11} & x_{22}y_{12}\\
 				x_{21}y_{21} & x_{21}y_{22} & x_{22}y_{21} & x_{22}y_{22}\\
				x_{21}y_{31} & x_{21}y_{32} & x_{22}y_{31} & x_{22}y_{32}
				\end{pmatrix} \ .
				\end{equation}
We note that, in general, $X \otimes Y \neq Y \otimes X$. Here are some useful properties and identities involving the Kronecker product ($V,X,Y,Z$ are matrices) (proofs can be found in
\cite{book:Graham-1981}):
\begin{enumerate}
\item \emph{Multiplication by scalar}: If $c$ is complex scalar, then
				\begin{equation}
				X \otimes (c Y) = c (X \otimes Y) 
				\end{equation}			
\item \emph{Distributive property with respect to addition}:
				\begin{subequations}
				\begin{align}
				(X + Y) \otimes Z & = X \otimes Z + Y \otimes Z \\
				X \otimes (Y + Z) & = X \otimes Y + X \otimes Z
				\end{align}	
				\end{subequations}	
\item \emph{Associative property}:
				\begin{equation}
				X \otimes (Y \otimes Z) = (X \otimes Y) \otimes Z 
				\end{equation}		
\item \emph{Mixed product rule}: If $X,Y,Z,V$ are matrices of dimension $(m \times n), (n \times m), (m' \times n')$ and $(n' \times m')$, respectively, then
				\begin{equation}
				\label{eq:mixed_product_rule}
				(X \otimes Z)(Y \otimes V) = (X Y) \otimes (Z  V) 
				\end{equation}
\item \emph{The inverse of a Kronecker product}: If $X$ and $Y$ are invertible matrices, then
				\begin{equation}
				\label{eq:inverse_Kronecker}
				(X \otimes Y)^{-1} = X^{-1} \otimes Y^{-1}
				\end{equation}	
\item \emph{The conjugate transpose of a Kronecker product}:
				\begin{equation}
				(X \otimes Y)^\dagger = X^{\dagger} \otimes Y^{\dagger}
				\end{equation}	
\item \emph{Eigenvalues and eigenvectors}: Let $X$ and $Y$ be diagonalizable matrices of dimension $(n \times n)$ and $(m \times m)$, respectively. Let $\{\lambda_j\}$ and $\{x_j\}$ $(j =1, 2, \ldots, n)$ be the eigenvalues and eigenvectors of $X$, respectively. Similarly, let $\{\mu_k\}$ and $\{y_k\}$ $(k=1,2,\ldots, m)$ be the eigenvalues and eigenvectors of $Y$, respectively. Then, $\{\lambda_j \mu_k\}$ and $\{x_j \otimes y_k\}$ are eigenvalues and eigenvectors of $X \otimes Y$, respectively. That is
				\begin{equation}
				(X \otimes Y)(x_j \otimes y_k) = \lambda_j \mu_k \ (x_j \otimes y_k)
				\end{equation}
\item \emph{Trace function of a Kronecker product}:
				\begin{equation}
				\mbox{Tr}[X \otimes Y] = \mbox{Tr}[X]\cdot \mbox{Tr}[Y]
				\end{equation}	
\item \emph{Determinant of a Kronecker product}: If $X$ is a $(n \times n)$ matrix, and $Y$ is a $(m \times m)$ matrix then 
				\begin{equation}
				\det[(X \otimes Y)] = \left(\det[X]\right)^m \cdot \left( \det[Y]\right)^n
				\end{equation}																	\item \emph{Analytic function of a Kronecker product involving an identity matrix}: Let $X$ be a $(n \times n)$ square matrix, and $\mathbb{I}_m$ the $(m \times m)$ identity matrix. If $f$ is an analytic function
\cite{book:Higham-2008} defined for both the Kronecker product (between $X$ and $I_m$) and for $X$, then
				\begin{subequations}
				\label{eq:function_Kronecker_prod}
				\begin{align}
				f(X \otimes \mathbb{I}_m) & = f(X) \otimes \mathbb{I}_m \\
				f(\mathbb{I}_m \otimes X) & =  \mathbb{I}_m \otimes f(X) \ .
				\end{align}
				\end{subequations}		
Examples of matrix analytic functions include $\exp[\bullet]$ and the trigonometric functions $\sin(\bullet)$ and $\cos(\bullet)$. So, we have, for example, that
				\begin{equation}
				e^{X \otimes \mathbb{I}_m} = e^X \otimes \mathbb{I}_m
				\end{equation}
				
				\begin{equation}
				\sin(\mathbb{I}_m \otimes X) = \mathbb{I}_m \otimes \sin(X)
				\end{equation}			

				\begin{equation}
				\cos( X \otimes \mathbb{I}_m ) = \cos(X) \otimes \mathbb{I}_m
				\end{equation}	
where
\cite{book:Higham-2008}
				\begin{subequations}
				\begin{align}
				\cos(X) & := \mathbb{I}_n - \frac{X^2}{2!} + \frac{X^4}{4!} - \frac{X^6}{6!} + \ldots = \sum^\infty_{k=0} (-1)^k\frac{X^{2k}}{(2k)!}\\
				\sin(X) & := X - \frac{X^3}{3!} + \frac{X^5}{5!} - \frac{X^7}{7!} + \ldots = \sum^\infty_{k=0} (-1)^k\frac{X^{2k+1}}{(2k+1)!} \ .
				\end{align}
				\end{subequations}											 
\end{enumerate}

\subsection{The bra-flipper operator, $\mho$}\label{subsec:bra-flipper}\label{sec:bra-flipper}
We mentioned above that in the Liouville space formalism, operators like the density matrix become column vectors. These `vectorized' operators are commonly referred to as `supervectors' or `superkets'. To understand the nuances in the Liouville space formalism, consider, for example, the qubit pure state
				\begin{equation}
				\label{eq:gen_qubit_ps}
				\ket{\psi(t)} = a(t) \ket{1} + b(t) \ket{0}
				\end{equation}
where $a(t)$ and $b(t)$ are complex scalars which satisfy the normalization condition $\abs{a(t)}^2 + \abs{b(t)}^2 = 1$. The matrix representations of the elements of the basis $\{\ket{0} , \ket{1}\}$ are still chosen to be those in \eqref{eq:matrix_rep_ket_0_1}. 
Consequently, 
				\begin{equation}
				\bra{1} = \ket{1}^\dagger = \begin{pmatrix}
				1 & 0
				\end{pmatrix} \qquad 
				\bra{0} = \ket{0}^\dagger = \begin{pmatrix}
				0 & 1
				\end{pmatrix} 
				\end{equation}
and the matrix representation of the generic qubit pure state $\ket{\psi(t)}$ in \eqref{eq:gen_qubit_ps} is
				\begin{equation}
				\ket{\psi(t)} = a(t) \begin{pmatrix}
				1 \\
				0
				\end{pmatrix} + b(t) \begin{pmatrix}
				0 \\
				1
				\end{pmatrix} =
				\begin{pmatrix}
				a(t) \\
				b(t)
				\end{pmatrix} \ .
				\end{equation}
The density matrix $\rho(t)$ corresponding to this pure state is of the form 
				\begin{subequations}
				\begin{align}
				\rho(t) & = \ket{\psi(t)}\!\bra{\psi(t)} \label{eq:rho_LS_1_a}\\
				& = \abs{a(t)}^2 \ket{1}\!\bra{1} + a(t)b^*(t)\ket{1}\!\bra{0} + a^*(t)b(t)\ket{0}\!\bra{1}  + \abs{b(t)}^2\ket{0}\!\bra{0} \label{eq:rho_LS_1_b} \ .
				\end{align}				
				\end{subequations}
The matrix representation of $\rho(t)$ can be obtained either from \eqref{eq:rho_LS_1_a} or \eqref{eq:rho_LS_1_b}. If we use \eqref{eq:rho_LS_1_a}, then
				\begin{equation}
				\label{eq:rho_matrix}
				\rho(t)  = \ket{\psi(t)}\!\bra{\psi(t)}  = 				
				\begin{pmatrix}
				a(t) \\
				b(t)
				\end{pmatrix} 				
				\begin{pmatrix}
				a^*(t) & b^*(t)
				\end{pmatrix} = 
				\begin{pmatrix}
				\abs{a(t)}^2 & a(t)b^*(t)\\
				a^*(t)b(t)  & \abs{b(t)}^2
				\end{pmatrix} \ .
				\end{equation}
Now, if we want to represent $\rho(t)$ in the form of a column vector, we have different choices as to how to rearrange the elements of $\rho(t)$ in the vector. The following, for example, are some of the possible choices:
				\begin{equation}
				\label{eq:possible_rho_vectors}
				\begin{pmatrix}
				\abs{a(t)}^2 \\
				a^*(t)b(t)\\
				a(t)b^*(t)\\
				\abs{b(t)}^2
				\end{pmatrix}\quad
				\begin{pmatrix}
				\abs{a(t)}^2 \\
				a(t)b^*(t)\\
				a^*(t)b(t)\\
				\abs{b(t)}^2
				\end{pmatrix}\quad
				\begin{pmatrix}
				\abs{a(t)}^2 \\
				\abs{b(t)}^2\\
				a(t)b^*(t)\\
				a^*(t)b(t)
				\end{pmatrix}\quad
				\begin{pmatrix}
				\abs{a(t)}^2 \\
				\abs{b(t)}^2\\
				a^*(t)b(t)\\
				a(t)b^*(t)\\
				\end{pmatrix} \quad
				\begin{pmatrix}
				\abs{b(t)}^2\\
				a^*(t)b(t)\\
				\abs{a(t)}^2 \\
				a(t)b^*(t)
				\end{pmatrix}	\ .		
				\end{equation}
In matrix analysis
\cite{book:Graham-1981,book:Turkington-2002}, given a $(m \times n)$ matrix $X$, \eqref{eq:X_Y_matrices}, the column vector $\mbox{vec}[X]$ is meant to correspond to the following arrangement
				\begin{equation}
				\label{eq:X_arrange_matrix_an}
				\mbox{vec}[X] \equiv \begin{pmatrix}
				X_{\bullet 1}\\
				X_{\bullet 2}\\
				\vdots \\
				X_{\bullet n-1}\\
				X_{\bullet n}
				\end{pmatrix} \ , \ \quad \ \text{where}  \ 
				X_{\bullet i} = \begin{pmatrix}
				x_{1i}\\
				x_{2i}\\
				\vdots \\
				x_{mi}
				\end{pmatrix} \ .
				\end{equation}
$X_{\bullet i}$ is the $i-$th column of the matrix (counting from the left). So if we take $\rho(t)$ in \eqref{eq:rho_matrix}, for example,
				\begin{equation}
				\mbox{vec}[\rho(t)] = \begin{pmatrix}
				\abs{a(t)}^2 \\
				a^*(t)b(t)\\
				a(t)b^*(t)\\
				\abs{b(t)}^2				
				\end{pmatrix}
				\end{equation}
which is the first vector we wrote in \eqref{eq:possible_rho_vectors}. In quantum mechanics, different choices may be made. Some authors, for example, prefer the arrangement whereby the diagonal elements are entered first, followed by the off-diagonals (see, for example,
\cite{art:Jeener-1982, book:Mukamel-1995}).  Another popular choice, but rarely explicitly stated by authors, is the arrangement --- which we indicate for now as $\widetilde{\mbox{vec}}[X]$ --- (see, for example, \cite{inbook:Mayne-2007})
				\begin{equation}
				\label{eq:X_arrange_QM}
				\widetilde{\mbox{vec}}[X] \equiv \begin{pmatrix}
				X_{1 \bullet }\\
				X_{2 \bullet }\\
				\vdots \\
				X_{m-1 \bullet }\\
				X_{m \bullet}
				\end{pmatrix} \ , \ \quad \ \text{where}  \ 
				X_{k \bullet } = \begin{pmatrix}
				x_{k 1}\\
				x_{k 2}\\
				\vdots \\
				x_{k n}
				\end{pmatrix} \ .
				\end{equation}
$X_{k \bullet}$ is the $k-$th row of the matrix $X$. One major advantage of this choice is that, unlike \eqref{eq:X_arrange_matrix_an} used in matrix analysis, it can be easily created using the Kronecker product (without any further rearrangements). As a result, the mathematics become greatly simplified. There is, indeed, a direct correlation between the arrangement chosen and the straightforwardness of the mathematics which come into play as a consequence. \par We choose the arrangement in \eqref{eq:X_arrange_QM} in our discussion. To see how it can be easily implemented, we introduce the `bra-flipper' operator $\mho$:
\begin{description}
\item \textit{Let $\ket{a}\!\bra{b}$ be an operator defined on the Hilbert space $\mathcal{H}_d$ of dimension $d$. (Thus, $\ket{a}\!\bra{b} \in \mathcal{O}_d$.) The bra-flipper operator $\mho$, defined on the same $\mathcal{H}_d$, is a superoperator which acts on  $\ket{a}\!\bra{b}$ and transforms it into a superket according to the transformation
				\begin{equation}
				\label{eq:def_mho}
				\mho \big[ \ket{a}\!\bra{b} \big] = \ket{a} \otimes \ket{b}^* \equiv \lket{a,b}
				\end{equation}
where $\ket{b}^*$ is the complex conjugate of $\ket{b}$.
}
\end{description}
In other others, when operating on $\ket{a}\!\bra{b}$, the bra-flipper operator $\mho$ turns the simple matrix product into Kronecker product and changes the bra $\bra{b}$ into its corresponding complex conjugated ket, $\ket{b}^*$. Moreover, $\mho$ is a linear superoperator. That is,
\begin{enumerate}
\item If $\lambda$ is a complex scalar, then
				\begin{equation}
				\mho \big[ \lambda \ \ket{a}\!\bra{b}\big] = \lambda \ \mho \big[ \ket{a}\!\bra{b}\big]
				\end{equation}
\item Let $\ket{a'}\!\bra{b'}$ also be an operator defined on $\mathcal{H}_d$ and $\lambda'$ a complex scalar. Then
				 \begin{equation}
				 \label{eq:mho_linear}
				\mho \big[ \lambda \ \ket{a}\!\bra{b} + \lambda' \ \ket{a'}\!\bra{b'}\big] = \lambda \ \mho \big[ \ket{a}\!\bra{b}\big] + \lambda' \ \mho \big[ \ket{a'}\!\bra{b'}\big] \ .
				\end{equation}
\end{enumerate}
Since the set of operators defined on $\mathcal{H}_d$ form a linear space, it follows from the  above properties of $\mho$ that the superkets $\{ \lket{a,b} \}$ also form a linear space.  (We leave it to the Reader to prove.) This linear space is none other but the finite-dimensional Liouville space associated with $\mathcal{H}_d$, which we denote as $\mathcal{L}_d$.
\par 
\par As an example, observe that if we apply $\mho$ to $\rho(t)$ in \eqref{eq:rho_matrix}, we get
				\begin{equation}
				\lket{\rho(t)} \equiv \mho\big[\rho(t)\big] = \mho\big[ \ket{\psi(t)}\! \bra{\psi(t)} \big] = \ket{\psi(t)} \otimes \ket{\psi(t)}^* 
				\end{equation}
which in matrix representation reads
				\begin{equation}
				\lket{\rho(t)} = \ket{\psi(t)} \otimes \ket{\psi(t)}^* = 										\begin{pmatrix}
				a(t) \\
				b(t)
				\end{pmatrix} \otimes 
				\begin{pmatrix}
				a^*(t) \\
				b^*(t)
				\end{pmatrix}=
				\begin{pmatrix}
				\abs{a(t)}^2 \\
				a(t)b^*(t)\\
				a^*(t) b(t)\\
				\abs{b(t)}^2
				\end{pmatrix}  
				\end{equation}
-- which we observe coincides with the second vector in \eqref{eq:possible_rho_vectors}.
\par With the help of the bra-flipper operator, we can also transform any mixed state density operator into a superket. If we take $\rho$ in \eqref{eq:rho_2}, for example, we have
				\begin{equation}
				\label{eq:mixed_rho_vectorized}
				\begin{split}
				\lket{\rho} = \mho[\rho] & = \sum^d_{n=1}\sum^d_{n'=1} \rho_{nn'} \ \mho \big[ \ket{\phi_n}\!\bra{\phi_{n'}}\big] = \sum^d_{n=1}\sum^d_{n'=1} \rho_{nn'} \  \ket{\phi_n}\otimes \ket{\phi_{n'}}^* \\
				& = \sum^d_{n=1}\sum^d_{n'=1} \rho_{nn'} \  \lket{\phi_n,\phi_{n'}} \ .
				\end{split}
				\end{equation}
This also applies to \emph{any} operator $A$ defined on $\mathcal{H}_d$. Indeed, since
				\begin{equation}
				A = \sum^d_{\nu=1}\sum^d_{\nu'=1} A_{\nu \nu'} \ket{\nu}\!\bra{\nu'} \ , \ \qquad \ A_{\nu \nu'} \equiv \matrixel{\nu}{A}{\nu'}
				\end{equation}
for an arbitrary orthonormal basis $\{\ket{\nu}\}$ of $\mathcal{H}_d$, it readily follows that
				\begin{equation}
				\label{eq:A_vectorized_0}
				\lket{A} = \mho \big[ A \big] = \sum^d_{\nu=1}\sum^d_{\nu'=1} A_{\nu \nu'} \ \mho \big[ \ket{\nu}\!\bra{\nu'} \big] = \sum^d_{\nu=1}\sum^d_{\nu'=1} A_{\nu \nu'}   \lket{\nu,\nu'} \ .
				\end{equation}
If we indicate each superket $\lket{\nu ,\nu'}$ with a distinct single character so that $\lket{\nu ,\nu'} \to \lket{\alpha}$, then
				\begin{equation}
				\label{eq:A_vectorized}
				\lket{A} = \sum^{d^2}_{\alpha =1 } A_\alpha \lket{\alpha} \ .
				\end{equation}
We recognize \eqref{eq:A_vectorized} as the expansion of the supervector $\lket{A}$ in the basis (or `superbasis') $\{\lket{\alpha}\}$ (more on this in the next subsection). The expansions in \eqref{eq:mixed_rho_vectorized}, \eqref{eq:A_vectorized_0} and \eqref{eq:A_vectorized} are formally similar to the expansion in \eqref{eq:x}. In fact, the sets $\{\lket{\phi_n,\phi_{n'}}\}$ and $\{\lket{\alpha}\}$ are both orthonormal superbases for $\mathcal{L}_d$. And since the cardinality of both $\{\lket{\phi_n,\phi_{n'}}\}$ and $\{\lket{\alpha}\}$ is $d^2$, it follows that $\mathcal{L}_d$ is a $d^2-$dimensional linear space (\ref{appendix:A1}). We prove the orthonormality of the set $\{\lket{\alpha}\}$ in the next subsection when we define the adjoint space for $\mathcal{L}_d$. 

\subsection{The Liouville adjoint space $\mathcal{L}^*_d$ and inner product}
To every superket $\lket{a,b}$ of $\mathcal{L}_d$ is associated a unique dual vector $\lbra{a,b}$, called `superbra'. The superbras form the adjoint space $\mathcal{L}^*_d$. Just as the vector $\ket{x}$ and its dual $\bra{x}$ are related through the operation of conjugate transpose, \eqref{eq:bra_ket_relation}, so is the superket $\lket{a,b}$ and its dual $\lbra{a,b}$. That is,
				\begin{equation}
				\label{eq:def_lbra}
				\lbra{a,b} = \lket{a,b}^\dagger = \big( \ket{a} \otimes \ket{b}^* \big)^\dagger =\bra{a} \otimes \bra{b}^* \ .
				\end{equation}
Here too, the matrix product between a superbra and superket constitute an inner product on $\mathcal{L}_d$. Specifically, if $\lket{a',b'} , \lket{a,b} \in \mathcal{L}_d$, then their inner product is defined as
				\begin{equation}
				\label{eq:inner_product_Liouville_space}
				\big< \lket{a',b'} , \lket{a,b} \big> : = \lbraket{a',b'}{a,b} \ .
				\end{equation}
Thus, $\mathcal{L}_d$ is a normed linear space. Given that every finite-dimensional complex linear space is complete (\ref{appendix:A1}), $\mathcal{L}_d$ must also be complete
\cite{book:Jordan-2006, book:Conway-2007, book:Riesz-1955}. Furthermore, since every normed and complete linear space is a Hilbert space, we also conclude that $\mathcal{L}_d$ \emph{is also a Hilbert space} (\ref{appendix:A1}). Specifically, it is a complex and separable Hilbert space. 
\par Note that, according to \eqref{eq:def_mho} and \eqref{eq:def_lbra}, 
				\begin{subequations}
				\begin{align}
				\lbraket{a',b'}{a,b} & = \big(\bra{a'} \otimes \bra{b'}^* \big)\big(\ket{a} \otimes \ket{b}^* \big)\\
				& = \braket{a'}{a} \otimes \braket{b'}{b}^*\\
				& = \braket{a'}{a} \cdot  \braket{b'}{b}^* \label{eq:LS_inner_prod_2c}
				\end{align}
				\end{subequations}
where we have made use of the mixed product rule of the Kronecker product, \eqref{eq:mixed_product_rule}. If the vectors $\ket{a}, \ket{a'}, \ket{b}, \ket{b'}$ belong to the same orthonormal basis of $\mathcal{H}_d$, then it follows from \eqref{eq:LS_inner_prod_2c} that 
				\begin{equation}
				\label{eq:LS_orthonormality}
				\lbraket{a',b'}{a,b} = \delta_{a',a} \delta_{b',b}
				\end{equation}	
which means the set of superkets $\{\lket{a,b}\}$ form an orthonormal (super)basis for the Liouville space $\mathcal{L}_d$.
\par From \eqref{eq:def_lbra}, it follows that to every superket $\lket{X}$ in $\mathcal{L}_d$, is associated a superbra $\lbra{X}$, and $\lbra{X}=\lket{X}^\dagger$. If we take $\lket{A}$ in \eqref{eq:A_vectorized}, for example, its dual $\lbra{A}$ is
				\begin{equation}
				\label{eq:def_lbra_A}
				\lbra{A} = \lket{A}^\dagger = \sum^{d^2}_{\alpha=1} A^*_\alpha \lket{\alpha}^\dagger = \sum^{d^2}_{\alpha=1} A^*_\alpha \lbra{\alpha} \ .
\end{equation}	

\subsection{The superoperator $\mathbb{I}_{d^2}$ and change of superbasis}			 
\par If we  take the inner product between $\lket{\nu'',\nu'''} \in \mathcal{L}_d$ and $\lket{A}$ (defined in \eqref{eq:A_vectorized_0}), it yields the expression
				\begin{equation}
				\label{eq:68}
				\lbraket{\nu'',\nu'''}{A} = \sum^d_{\nu=1} \sum^d_{\nu'=1} A_{\nu \nu'}  \lbraket{\nu'',\nu'''}{\nu,\nu'}\ .
				\end{equation}
If $\ket{\nu}, \ket{\nu'},\ket{\nu''}, \ket{\nu'''}$ are elements of the same set of orthonormal basis of $\mathcal{H}_d$, then it follows from \eqref{eq:LS_orthonormality} and \eqref{eq:68} that
				\begin{equation}
				\label{eq:lmatrix_element}
				\lbraket{\nu'',\nu'''}{A} = A_{\nu'' \nu'''} = \matrixel{\nu''}{A}{\nu'''}\ .
				\end{equation}
This is an important identity because it allows us to rewrite \eqref{eq:A_vectorized_0} as
				\begin{equation}
				\lket{A} = \sum^d_{\nu=1} \sum^d_{\nu'=1}   \lket{\nu,\nu'}A_{\nu \nu'} = \sum^d_{\nu=1} \sum^d_{\nu'=1}   \lket{\nu,\nu'}\lbraket{\nu,\nu'}{A} 
				\end{equation}
from which we deduce that
				\begin{equation}
				\label{eq:I_d^2_a}
				\sum^d_{\nu=1} \sum^d_{\nu'=1}   \lket{\nu,\nu'}\!\lbra{\nu,\nu'} = \mathbb{I}_{d^2}
				\end{equation}
where $\mathbb{I}_{d^2}$ is the identity superoperator defined on $\mathcal{L}_d$. This is very similar to how we derived \eqref{eq:identity_I} in state space. In a single index notation, we may write
				\begin{equation}
				\label{eq:I_d^2}
				\sum^{d^2}_{\alpha = 1 } \lket{\alpha}\! \lbra{\alpha} = \mathbb{I}_{d^2} \ .
				\end{equation}
\par Now, suppose the expansion of $\lket{A}$ in the orthonormal superbasis $\{ \lket{\alpha} \}$, \eqref{eq:A_vectorized_0}, holds. Suppose $\{\lket{\eta}\}$ is another orthonormal superbasis of $\mathcal{L}_d$. We want to change \eqref{eq:A_vectorized_0} from the superbasis $\{ \lket{\alpha} \}$ to  $\{\lket{\eta}\}$. This can be easily achieved in light of \eqref{eq:I_d^2}. Indeed, from \eqref{eq:A_vectorized}, we may write
				\begin{equation}
				\begin{split}
				\lket{A} & = \sum^{d^2}_{\alpha=1} A_\alpha \ \mathbb{I}_{d^2} \lket{\alpha} 
				 = \sum^{d^2}_{\alpha=1} A_\alpha \sum^{d^2}_{\eta=1} \lket{\eta}\!\lbraket{\eta}{\alpha}
				= \sum^{d^2}_{\eta=1} A_{\eta} \lket{\eta}
				\end{split}
				\end{equation}
with
				\begin{equation}
				 A_{\eta} =  \sum^{d^2}_{\alpha=1} A_\alpha  \lbraket{\eta}{\alpha} \ .
				\end{equation}
\subsection{$\mho$ is a bijection and has an inverse, $\mho^{-1}$.}\label{sec:inverse_bra-flipper}
We have seen above that $\mho$ transforms elements of $\mathcal{O}_d$ into superkets, \emph{i.e.} elements of $\mathcal{L}_d$. We show here that the mapping  $\mho: \mathcal{O}_d \to \mathcal{L}_d$ is bijective (see \ref{appendix:inj_sur_bij} for definition). That is, given two operators $A, B \in \mathcal{O}_d$, 1) if $A \neq B$, then $\mho[A] \neq \mho[B]$, 2) $\mho[A]=\mho[B]$ if and only if $A=B$. Another requirement for the map $\mho: \mathcal{O}_d \to \mathcal{L}_d$ to be bijective is that $\mathcal{O}_d$ and $\mathcal{L}_d$ be of the same dimension -- which we have already shown above to be $d^2$. So to show that $\mho$ is a bijection we need to prove the points 1) and 2).
\par We do so by \emph{reductio ad absurdum}. We assume the proposition
\begin{quotation}\label{prop:1}
\textbf{P}: \textit{$\mho$ is \textbf{not} a bijection}
\end{quotation}
is true. Suppose then that we take two non-null operators $A,B \in \mathcal{O}_d$, where $A \neq B$ but $\mho[A] = \mho[B]$. Let $\{\ket{\nu}\!\bra{\nu'}\}\ (\nu,\nu'=1,\ldots,d)$ be an orthonormal basis of $\mathcal{O}_d$. Thus, we have the expansions $A=\sum_\nu \sum_{\nu'} A_{\nu \nu'} \ket{\nu}\!\bra{\nu'}$ and $B=\sum_\nu \sum_{\nu'} B_{\nu \nu'} \ket{\nu}\!\bra{\nu'}$, from which  ---  upon applying $\mho$  ---  we obtain
			\begin{equation}
			\mho[A] = \lket{A} = \sum_{\nu,\nu'} A_{\nu \nu'} \lket{\nu,\nu'} \  , \qquad \ \mho[B] = \lket{B} = \sum_{\nu,\nu'} B_{\nu \nu'} \lket{\nu,\nu'} \ .
			\end{equation}
Since $\mho[A] = \mho[B]$, it follows that
			\begin{equation}
			\begin{split}
			0 & = \mho[A] - \mho[B]
			 = \sum_{\nu,\nu'} \left( A_{\nu \nu'} - B_{\nu \nu'} \right) \lket{\nu,\nu'}
			\end{split}
			\end{equation}
which can only be true if $\left( A_{\nu \nu'} - B_{\nu \nu'} \right)=0$ for any arbitrary $\nu,\nu'$  ---  given that the superkets $\{\lket{\nu,\nu'}\}$ are linearly independent (\ref{appendix:A1}). But for non-null operators $A$ and $B$, $\left( A_{\nu \nu'} - B_{\nu \nu'} \right)=\matrixel{\nu}{(A-B)}{\nu'}=0, \ \forall \nu, \nu'$ can be true iff $A=B$. This contradicts our initial assertion that $A \neq B$. Thus, the proposition \textbf{P} is false.
\par Since $\mho: \mathcal{O}_d \to \mathcal{L}_d$ is a bijection, we know it must have an inverse $\mho^{-1}$ (\ref{appendix:inj_sur_bij}). The superoperator $\mho^{-1}$ transforms superkets in $\mathcal{L}_d$ into operators in $\mathcal{O}_d$:
\begin{description}
\item \textit{Let $\ket{a} \otimes \ket{b}^*=\lket{a,b}$ be a superket of the Liouville space $\mathcal{L}_d$. The inverse bra-flipper operator $\mho^{-1}$, defined on the same $\mathcal{L}_d$, is a superoperator which acts on  $\lket{a,b}$ and transforms it into an element of $\mathcal{O}_d$ according to the transformation
				\begin{equation}
				\label{eq:def_mho_inverse}
				\mho^{-1} \big[ \ket{a} \otimes \ket{b}^*\big] = \ket{a}\!\bra{b} \ .
				\end{equation}
}
\end{description}			 
Like $\mho$, $\mho^{-1}$ is also a linear superoperator. Namely, for $\lket{a,b}, \lket{a',b'} \in \mathcal{L}_d$, and complex scalars $\lambda, \lambda'$, 
				\begin{subequations}
				\begin{align}
				\mho^{-1} \big[ \lambda \ \ket{a} \otimes \ket{b}^*\big] & = \lambda \ \mho^{-1} \big[ \ket{a} \otimes \ket{b}^*\big]\\
				\mho^{-1} \big[ \lambda \ \ket{a} \otimes \ket{b}^* + \lambda' \ \ket{a'} \otimes \ket{b'}^*\big] & = \lambda \ \mho^{-1} \big[\ket{a} \otimes \ket{b}^*\big] + \lambda' \ \mho^{-1} \big[ \ket{a'} \otimes \ket{b'}^*\big] \ .
				\end{align}
				\end{subequations}
Note also that if we take \eqref{eq:def_mho_inverse} and make use of \eqref{eq:def_mho}, we may rewrite the former as:
				\begin{equation}
				\label{eq:mho_inverse}
				\begin{split}
				\mho^{-1} \bigg[\mho \big[\ket{a}\!\bra{b} \big] \bigg] & = \ket{a}\!\bra{b}\\
				\mho^{-1} \mho \big[\ket{a}\!\bra{b} \big]    & = \ket{a}\!\bra{b}
				\end{split}
				\end{equation}
from which we conclude that $\mho^{-1} \mho = \mathbb{I}_d$. Similarly, one can easily prove -- using again \eqref{eq:def_mho_inverse} -- that $\mho \mho^{-1}  = \mathbb{I}_d$.
\subsection{$\mho$ is an isomorphism}
We concluded above that $\mathcal{H}_d$ (the linear space of state vectors), $\mathcal{O}_d$ (the linear space of operators on $\mathcal{H}_d$) and $\mathcal{L}_d$ (the Liouville space) are all complex separable Hilbert spaces; but while $\mathcal{H}_d$ is $d-$dimensional, $\mathcal{O}_d$ and $\mathcal{L}_d$ are $d^2-$dimensional. We prove here that $\mho$ is an \emph{isomorphism} between $\mathcal{O}_d$ and $\mathcal{L}_d$. That is, $\mho$ preserves the inner product between these two spaces.
\par In general, a linear surjective map (\ref{appendix:inj_sur_bij}) $f:\mathcal{X} \to \mathcal{Y}$ (where $\mathcal{X}$ and $\mathcal{Y}$ are Hilbert spaces) is said to be an isomorphism between the two spaces if for $x_1, x_2 \in \mathcal{X}$
\cite{book:Conway-2007}
				\begin{equation}
				\label{eq:def_isomorphism}
				\big< f(x_1),f(x_2) \big> = \big< x_1, x_2\big> \ 
				\end{equation}
-- that is, the inner product of the input elements $x_1,x_2$ always coincides with the inner product of their images under $f$. Therefore, to prove $\mho$ is an isomorphism, we need to prove first of all that it is surjective, and then prove that it preserves the inner product. But we showed in the previous subection that $\mho$ is bijective, which naturally means it is also surjective (\ref{appendix:inj_sur_bij}). Thus, we just need to prove at this point that $\mho$ preserves the extended Hilbert-Schmidt inner product defined in \eqref{eq:inner_prod_Hilbert-Schmidt}.
\par In this respect, let $A$ and $B$ be any two operators which belong to $\mathcal{O}_d$. If $\mho$ is an isomorphism, then according to \eqref{eq:def_isomorphism}, we must have
				\begin{equation}
				\big< \mho \big[ A \big], \mho \big[ B \big]\big> = \big< A, B\big> \ .
				\end{equation}
But we know from \eqref{eq:inner_prod_Hilbert-Schmidt} that $\big< A, B\big> = \Tr \big[  B A^\dagger \big]$. Hence, proving $\mho$ is an isomorphism now reduces to showing that $\big< \mho \big[ A \big], \mho \big[ B \big]\big> = \Tr \big[  B A^\dagger \big]$ for arbitrary $A$ and $B$ -- which is what we are going to do.
\par From  \eqref{eq:inner_product_Liouville_space},
				\begin{equation}
				\label{eq:lbraket_A_B_0}
				\big< \mho \big[ A \big], \mho \big[ B \big]\big> = \big< \lket{A}, \lket{B}\big> = \lbraket{A}{B} \ .
				\end{equation}
Let us consider $\lbraket{A}{B}$ for a moment. We know, in general, we may write $\lket{A}=\sum^d_{\nu=1}\sum^d_{\nu'=1} A_{\nu \nu'} \lket{\nu, \nu'}$ and $\lket{B}=\sum^d_{\nu=1}\sum^d_{\nu'=1} B_{\nu \nu'} \lket{\nu, \nu'}$, where $\{\lket{\nu,\nu'}\}$ constitute an orthonormal superbasis of $\mathcal{L}_d$. So, from \eqref{eq:def_lbra_A}, $\lbra{A}=\lket{A}^\dagger= \sum^d_{\nu=1}\sum^d_{\nu'=1} A^*_{\nu \nu'} \lbra{\nu, \nu'}$, and 
				\begin{equation}
				\begin{split}
				\label{eq:lbraket_A_B}
				\lbraket{A}{B} & = \left(\sum^d_{\nu=1}\sum^d_{\nu'=1} A^*_{\nu \nu'} \lbra{\nu, \nu'}  \right)\left(\sum^d_{\nu''=1}\sum^d_{\nu'''=1} B_{\nu'' \nu'''} \lket{\nu'',\nu'''} \right)\\
				& =  \sum^d_{\nu=1}\sum^d_{\nu'=1} \sum^d_{\nu''=1}\sum^d_{\nu'''=1}  A^*_{\nu \nu'}   B_{\nu'' \nu'''} \lbraket{\nu, \nu'}{\nu'',\nu'''} \\
				& = \sum^d_{\nu=1}\sum^d_{\nu'=1} \sum^d_{\nu''=1}\sum^d_{\nu'''=1}  A^*_{\nu \nu'}   B_{\nu'' \nu'''} \ \delta_{\nu, \nu''} \ \delta_{\nu',\nu'''} \\
				& = \sum^d_{\nu=1}\sum^d_{\nu'=1} A^*_{\nu \nu'} B_{\nu \nu'} = \sum^d_{\nu=1}\sum^d_{\nu'=1}  B_{\nu \nu'}A^*_{\nu \nu'} \ .
				\end{split}
				\end{equation}	
But since 	$A^*_{\nu \nu'}$ is scalar, it is invariant under transpose -- \emph{i.e.} $ A^*_{\nu \nu'} = (A^*_{\nu \nu'})^T$. So, we may write
				\begin{equation}
				A^*_{\nu \nu'} = (A^*_{\nu \nu'})^T =(A_{\nu \nu'})^\dagger = \big(\matrixel{\nu}{A}{\nu'}\big)^\dagger= \matrixel{\nu'}{A^\dagger}{\nu} \ .
				\end{equation}						 
Thus, making the substitutions $A^*_{\nu \nu'} \to \matrixel{\nu'}{A^\dagger}{\nu}$, $B_{\nu \nu'} \to \matrixel{\nu}{B}{\nu'}$ in the last equation of \eqref{eq:lbraket_A_B}, we get
				\begin{equation}
				\label{eq:lbraket_A_B_1}
				\lbraket{A}{B} = \sum^d_{\nu=1}\sum^d_{\nu'=1}  \matrixel{\nu}{B}{\nu'}\matrixel{\nu'}{A^\dagger}{\nu} = \sum^d_{\nu=1} \matrixel{\nu}{B A^\dagger}{\nu} = \Tr\big[ B A^\dagger \big]
				\end{equation}
which is the extended Hilbert-Schmidt inner product, \eqref{eq:inner_prod_Hilbert-Schmidt}. We have thus shown that
				\begin{equation}
				\label{eq:lbraket_A_B_final}
				\lbraket{A}{B}  = \Tr\big[ B A^\dagger \big] \ .
				\end{equation}
If we now go back to \eqref{eq:lbraket_A_B_0}, we may then write
				\begin{equation}
				\label{eq:lbraket_A_B_extremus}
				\big< \mho \big[ A \big], \mho \big[ B \big]\big> = \lbraket{A}{B} = \Tr\big[ B A^\dagger \big] = \big< A,B \big>
				\end{equation}
-- where we have made use of \eqref{eq:lbraket_A_B_final} and \eqref{eq:inner_prod_Hilbert-Schmidt}. This proves that $\mho$ is an isomorphism, and the Hilbert spaces $\mathcal{O}_d, \mathcal{L}_d$ are truly isomorphic.
\par Furthermore, it is worth noting the significance of \eqref{eq:lbraket_A_B}: it allows us to write the trace of any product of two operators in $\mathcal{O}_d$ as the scalar product of two superkets in $\mathcal{L}_d$. So, for example, if $B \in \mathcal{O}_d$, then  since $\Tr[B] = \Tr[B \mathbb{I}^\dagger_d]$, it follows from \eqref{eq:lbraket_A_B_final} that
				\begin{equation}
				\Tr[B] = \Tr[B\  \mathbb{I}^\dagger_d] = \lbraket{\mathbb{I}_d}{B}\ .
				\end{equation}
Note that we could have also written $\Tr[B] = \Tr[\mathbb{I}_d B ]=\Tr[\mathbb{I}_d \left(B^\dagger \right)^\dagger ]$ -- in which case we derive from \eqref{eq:lbraket_A_B_final} that
				\begin{equation}
				\Tr[B] = \Tr[\mathbb{I}_d \left(B^\dagger \right)^\dagger ] = \lbraket{B^\dagger}{\mathbb{I}_d}\ .
				\end{equation}
But there is a more interesting application of \eqref{eq:lbraket_A_B_final}, and it has to do with expectation value $\big< B \big>$ of an observable represented by the operator $B\in \mathcal{O}_d$. We saw in \eqref{eq:B_avg_general} that $\big< B \big> = \Tr[B \rho]= \Tr[\rho B]$. Applying \eqref{eq:lbraket_A_B_final}, we see that
				\begin{subequations}
				\label{eq:LS_expectation_value}
				\begin{align}
				\big< B \big> & = \Tr[B \rho^\dagger] = \lbraket{\rho}{B}\\
				& = \Tr[ \rho B^\dagger] = \lbraket{B}{\rho}
				\end{align}
				\end{subequations}
where we have exploited the fact that density matrix $\rho$ and $B$, \eqref{eq:observable_gen}, are both Hermitian.
Thus, we have from \eqref{eq:LS_expectation_value} that the expectation value $\big<B\big>$ is simply a scalar product in Liouville space between two superkets: $\lket{\rho}$ and $\lket{B}$.

\subsection{Superoperators and the trace functional on $\mathcal{L}_d$}
\par Like linear operators on $\mathcal{H}_d$, we can also think of linear operators acting on $\mathcal{L}_d$. These operators are called \emph{superoperators}. For a given $\mathcal{L}_d$, it is easy to see that its superoperators are matrices of dimension $(d^2 \times d^2)$ and they form a linear space of dimension $d^4$ --- which we indicate as $\mathcal{S}_d$. If $\{\lket{\nu, \nu'}\} \equiv \{\lket{\alpha}\}$ is an orthonormal superbasis of $\mathcal{L}_d$, then the set $\{\lket{\alpha}\!\lbra{\alpha'}\}$ spans $\mathcal{S}_d$ and is called a superoperator basis of $\mathcal{S}_d$. $\mathbb{I}_{d^2}$, \eqref{eq:I_d^2}, is the identity superoperator on $\mathcal{L}_d$.
\par Before we define the inner product on $\mathcal{S}_d$, we need to define the trace function on $\mathcal{L}_d$. Similar to the trace function on $\mathcal{O}_d$, \eqref{eq:trace_def}, if $\mathfrak{B}$ is a superoperator, element of $\mathcal{S}_d$, then its trace $\Tr( \mathfrak{B} ) $ is defined as
				\begin{equation}
				\Tr ( \mathfrak{B} ) = \sum^{d^2}_{\alpha=1} \lmatrixel{\alpha}{\mathfrak{B}}{\alpha} \ .
				\end{equation}	
for an arbitrary basis $\{\lket{\alpha}\}$ of $\mathcal{L}_d$. Not surprisingly, we choose the extended Hilbert-Schmidt inner product as the scalar product on $\mathcal{S}_d$. In particular, given the two superoperators $\mathfrak{A},\mathfrak{B} \in \mathcal{S}_d$, we define their inner product as
				\begin{equation}
				\big< \mathfrak{A} , \mathfrak{B} \big> := \Tr ( \mathfrak{B} \mathfrak{A}^\dagger ) \ .
				\end{equation}
Note that in analogy to \eqref{eq:inner_prod_2}, we observe that $\lbraket{A}{B}$, \eqref{eq:lbraket_A_B}, may also be written as
				\begin{equation}
				\lbraket{A}{B} = \sum^{d^2}_{\alpha=1} \lbraket{A}{\alpha}\!\lbraket{\alpha}{B} = \sum^{d^2}_{\alpha=1} \lbraket{\alpha}{B}\!\lbraket{A}{\alpha} = \Tr \big( \lket{B}\lbra{A} \big) \ .
				\end{equation}
Thus, the linear spaces $\mathcal{H}_d, \mathcal{O}_d, \mathcal{L}_d$ and $\mathcal{S}_d$ share the same type of inner product. Also, since $\mathcal{S}_d$ is normed and finite-dimensional (thus, complete), it is also a Hilbert space like the other three. These concepts are summarized in table \ref{tab:Hilbert_spaces}.
\begin{table}
\caption{\label{tab:Hilbert_spaces}Some characteristics of the linear spaces $\mathcal{H}_d, \mathcal{O}_d, \mathcal{L}_d, \mathcal{S}_d$. (HS=Hilbert space; LO=Linear operators)}
\begin{indented}
\item[]\begin{tabular}{@{}llll}
\br
Symbol   &  Name  &   Dimension  &   Elements\\
\mr
$\mathcal{H}_d$      & HS of state vectors      &     $d$      &   $(d \times 1)$ column vectors\\
$\mathcal{O}_d$      & HS of LO on $\mathcal{H}_d$       &     $d^2$      &   $(d \times d)$ matrices\\
$\mathcal{L}_d$      & HS of `vectorized' LO on $\mathcal{H}_d$       &     $d^2$      &   $(d^2 \times 1)$ column vectors\\
$\mathcal{S}_d$      & HS of LO on $\mathcal{L}_d$       &     $d^4$      &   $(d^2 \times d^2)$ matrices\\
\br
\end{tabular}
\end{indented}
\end{table}
\par In complete analogy to \eqref{eq:X_expansion}, we can also expand any given superoperator $\mathfrak{B} \in \mathcal{S}_d$ in any given orthonormal superbasis $\{\lket{\alpha}\!\lbra{\alpha}\}$ of $\mathcal{S}_d$:
				\begin{equation}
				\mathfrak{B} = \mathbb{I}_{d^2} \mathfrak{B} \mathbb{I}_{d^2} = \sum^{d^2}_{\alpha=1} \sum^{d^2}_{\alpha'=1} \lket{\alpha}\! \lmatrixel{\alpha}{\mathfrak{B}}{\alpha'}\!\lbra{\alpha'} = \sum^{d^2}_{\alpha=1} \sum^{d^2}_{\alpha'=1} \mathfrak{B}_{\alpha, \alpha'}\lket{\alpha}\! \lbra{\alpha'}
				\end{equation}	
where
				\begin{equation}
				\mathfrak{B}_{\alpha, \alpha'} = \lmatrixel{\alpha}{\mathfrak{B}}{\alpha'} \ .
				\end{equation}							 
If instead of the single index $\alpha$ we use the two index representation (see   \eqref{eq:I_d^2_a} and \eqref{eq:I_d^2}), the expansion becomes
				\begin{equation}
				\mathfrak{B} = \sum^{d}_{\nu=1} \sum^{d}_{\nu'=1} \sum^{d}_{\nu''=1} \sum^{d}_{\nu'''=1} \mathfrak{B}_{\nu \nu', \nu'' \nu'''}\lket{\nu, \nu'} \lbra{\nu'',\nu'''}
				\end{equation}
with
				\begin{equation}
				\mathfrak{B}_{\nu \nu', \nu'' \nu'''} = \lmatrixel{\nu, \nu'}{\mathfrak{B}}{\nu'',\nu'''}
				\end{equation}
where the set $\{\ket{\nu}\}$ is an orthonormal basis of $\mathcal{H}_d$.
\par In Sec. \ref{subsec:Postulate_I}, \eqref{eq:spectral_decomposition_thm}, we discussed the spectral decomposition theorem, which applied to Hermitian elements of $\mathcal{O}_d$. Similarly, elements of $\mathcal{S}_d$ can be Hermitian. Like Hermitian elements of $\mathcal{O}_d$, a superoperator $\mathfrak{B}\in \mathcal{S}_d$ is Hermitian if $\mathfrak{B}=\mathfrak{B}^\dagger$. For such Hermitian superoperators, the spectral decomposition theorem still applies. That is, if $\mathfrak{B}=\mathfrak{B}^\dagger$, then, there exists an orthonormal superbasis $\{\lket{\beta}\!\lbra{\beta}\}$ of $\mathcal{S}_d$ such that
				\begin{equation}
				\label{eq:B_spectral_decomp_0}
				\mathfrak{B} = \sum^{d^2}_{\beta=1} \mathfrak{B}_{\beta, \beta} \lket{\beta}\!\lbra{\beta} \qquad \lbraket{\beta}{\beta'} = \delta_{\beta, \beta'}
				\end{equation}
or, in the two-indexed representation,
				\begin{equation}
				\label{eq:B_spectral_decomp}
				\mathfrak{B} = \sum^d_{\mu=1}\sum^d_{\mu'=1} \mathfrak{B}_{\mu \mu', \mu \mu'} \lket{\mu, \mu'}\lbra{\mu, \mu'} \qquad \lbraket{\mu,\mu'}{\mu'',\mu'''}=\delta_{\mu,\mu''}\delta_{\mu',\mu'''}\ .
				\end{equation}
Here too, the expansion coefficients $\{\mathfrak{B}_{\beta, \beta}\} \equiv \{\mathfrak{B}_{\mu \mu', \mu \mu'}\}$ are the eignevalues of $\mathfrak{B}$ and are real. The vectors $\{\lket{\beta}\} \equiv \{\lket{\mu,\mu'}\}$ are the eigenvectors (or `eigen-superkets') of $\mathfrak{B}$; indeed, multiplying \eqref{eq:B_spectral_decomp_0} from the right by $\lket{\beta'}$ yields the eigenvector equation: $\mathfrak{B}\lket{\beta'}=\mathfrak{B}_{\beta', \beta'} \lket{\beta'}$.  The superkets $\{\lket{\beta}\} \equiv \{\lket{\mu,\mu'}\}$ also constitute a complete orthonormal basis for $\mathcal{L}_d$. 
\subsection{The superket triple product identity}
\par The results of this subsection bring to light the full glory of the Liouville space formalism. Say $A,B,C$ are linear operators acting on $\mathcal{H}_d$; thus, they are also elements of $\mathcal{O}_d$. The product $ABC$ is still an operator on $\mathcal{H}_d$. The superket triple product identity states that
				\begin{equation}
				\label{eq:triple_product}
				\lket{ABC} = \left(A \otimes C^T\right) \lket{B}
				\end{equation}
where $C^T$ is the transpose of $C$. In other words, the superket corresponding to the product $ABC \in \mathcal{O}_d$, \emph{i.e.} $\lket{ABC}\in \mathcal{L}_d$, can be written as the result of the superoperator $\left(A \otimes C^T\right) \in \mathcal{S}_d$ acting on the superket $\lket{B}\in \mathcal{L}_d$. As a corollary, the following identities also follow from \eqref{eq:triple_product}:
				\begin{subequations}
				\begin{align}
				\lket{ABC} & = \left(AB \otimes \mathbb{I}_d\right) \lket{C}\\
				\lket{ABC} & = \left(\mathbb{I}_d \otimes C^T B^T\right) \lket{A} \ .
				\end{align}
				\end{subequations}
Let $\{\ket{\nu}\}$ be an orthonormal basis for $\mathcal{H}_d$. 
To prove \eqref{eq:triple_product}, we need to observe that the product of matrix elements $A_{\nu \nu'}B_{\nu''\nu'''}=\matrixel{\nu}{A}{\nu'}\matrixel{\nu''}{B}{\nu'''}$ may be written as an element of the superoperator $A \otimes B^T$. Namely,
				\begin{equation}
				\label{eq:A_otimes_B^T}
				A_{\nu \nu'}B_{\nu''\nu'''} = \lmatrixel{\nu,\nu'''}{A \otimes B^T}{\nu',\nu''} \ .
				\end{equation}
Equation \eqref{eq:A_otimes_B^T} can be easily verified using the mixed product rule, \eqref{eq:mixed_product_rule}. Since the expansion of $ABC$ in the basis $\{\ket{\nu}\}$ is 
				\begin{equation}
				\begin{split}
				ABC = \sum^d_{\nu=1}\sum^d_{\nu'=1} (ABC)_{\nu \nu'} \ket{\nu}\!\bra{\nu'} \ ,
				\end{split}
				\end{equation}	
after applying the bra-flipper operator to this expansion, we get
				\begin{equation}
				\label{eq:lket_ABC_a}
				\lket{ABC}  = \sum^d_{\nu=1}\sum^d_{\nu'=1} (ABC)_{\nu \nu'} \lket{\nu,\nu'}\ .
				\end{equation}			
But,
				\begin{equation}
				\begin{split}
				(ABC)_{\nu \nu'} & =
				\sum^d_{\nu''=1}\sum^d_{\nu'''=1} \matrixel{\nu}{A}{\nu''}\! \matrixel{\nu''}{B}{\nu'''}\!\matrixel{\nu'''}{C}{\nu'}\\
				& = \sum^d_{\nu''=1}\sum^d_{\nu'''=1} \lbraket{\nu,\nu''}{A}  \!\lbraket{\nu''',\nu'}{C}\!\lbraket{\nu'',\nu'''}{B}
				\end{split}
				\end{equation}		
which --- due to the identity in \eqref{eq:A_otimes_B^T} --- may be written as
				\begin{equation}
				\begin{split}
				(ABC)_{\nu \nu'} 
				& = \sum^d_{\nu''=1}\sum^d_{\nu'''=1} \lmatrixel{\nu,\nu'}{A \otimes C^T}{\nu'', \nu'''}\lbraket{\nu'',\nu'''}{B}\\
				& = \lmatrixel{\nu,\nu'}{A \otimes C^T}{B} \ .
				\end{split}
				\end{equation}
Substituting this into \eqref{eq:lket_ABC_a}, we finally obtain
				\begin{equation}
				\begin{split}
				\lket{ABC} & = \sum^d_{\nu=1}\sum^d_{\nu'=1}  \lket{\nu,\nu'}\lmatrixel{\nu,\nu'}{A \otimes C^T}{B} \\
				& = \left(A \otimes C^T\right)\lket{B} \ .
				\end{split}
				\end{equation}
The superket triple product identity is very important in Liouville space formalism and comes in handy, for example, when solving master equations. We shall see this application in the next section (Sec. \ref{sec:Applications}). Nonetheless, the following relations for the superket of the commutation $\left[ A,B \right]$ can be easily proved using \eqref{eq:triple_product}:
				\begin{equation}
				\label{eq:super-commutator_0}
				\lket{ \ \left[ A,B \right] \ } = \lshad A, \mathbb{I}_d \rshad \lket{B} = \lshad \mathbb{I}_d , B \rshad \lket{A} = \lshad A, B \rshad \lket{\mathbb{I}_d} \ ,
				\end{equation}	
where the ``super-commutator'' $\lshad X, Y\rshad $ is defined as
				\begin{equation}
				\label{eq:super-commutator}
				\lshad X, Y\rshad \equiv X \otimes Y^T - Y \otimes X^T \ .
				\end{equation}
We also note that, like the common commutator, Sec. \ref{subsec:pure_mixed}, $\lshad X, Y\rshad = - \lshad Y,X\rshad$. Analogously, we define the ``super-anticommutator'' $\lshad X , Y \rshad_+ $ as
				\begin{equation}
				\label{eq:super-anticommutator}
				\lshad X, Y\rshad_+ \equiv X \otimes Y^T + Y \otimes X^T \ .
				\end{equation}
As a result, $\lshad X, Y\rshad_+ = \lshad Y, X\rshad_+$ -- just like the common anti-commutator.
\subsection{Liouville space formalism and composite quantum systems}
\par The mathematical foundations for the Liouville space formalism laid above can be easily extend to any aggregate of quantum systems, each with a finite-dimensional state space. For simplicity, let us consider a bipartite quantum system $S_1+S_2$. Let $\mathcal{H}_{d_i}$ be the state space of system $S_i$, and of dimension $d_i$ ($i=1,2$). Also, let $\{\ket{\nu}\}$ and $\{\ket{\mu}\}$ be an orthonormal basis of $\mathcal{H}_{d_1}$ and $\mathcal{H}_{d_2}$, respectively. Then, the state space $\mathcal{H}_d$ of the composite system $S_1+S_2$ is given by the tensor product $\mathcal{H}_d = \mathcal{H}_{d_1} \otimes \mathcal{H}_{d_2}$, with $d=d_1 \cdot d_2$. The set $\{\ket{\nu} \otimes \ket{\mu}\}$ then constitutes an orthonormal basis for $\mathcal{H}_d$. 
\par Consider now the operator $\mathscr{Z} = A \otimes B$ acting on $\mathcal{H}_d$, where $A\in \mathcal{O}_{d_1}$ and $B \in \mathcal{O}_{d_2}$. We know $A=\sum^{d_1}_{\nu=1}\sum^{d_1}_{\nu'=1} A_{\nu\nu'} \ket{\nu}\!\bra{\nu'}$ and $B=\sum^{d_2}_{\mu=1}\sum^{d_2}_{\mu'=1} B_{\mu\mu'} \ket{\mu}\!\bra{\mu'}$, so
					\begin{equation}
					\begin{split}
					\mathscr{Z} & = \sum^{d_1}_{\nu=1}\sum^{d_1}_{\nu'=1} \sum^{d_2}_{\mu=1}\sum^{d_2}_{\mu'=1}   A_{\nu\nu'} B_{\mu\mu'}\ket{\nu}\!\bra{\nu'}\otimes \ket{\mu}\!\bra{\mu'}\\
					& = \sum^{d_1}_{\nu=1}\sum^{d_1}_{\nu'=1} \sum^{d_2}_{\mu=1}\sum^{d_2}_{\mu'=1}   A_{\nu\nu'} B_{\mu\mu'}\big(\ket{\nu}\otimes\ket{\mu}\big) \big(\bra{\nu'} \otimes\bra{\mu'}\big)
					\end{split}
					\end{equation}
where we have applied the mixed product rule of the Kronecker product, \eqref{eq:mixed_product_rule}. To transform $\mathscr{Z}$ into a superket, we simply apply to it the bra-flipper superoperator $\mho$, obtaining
					\begin{equation}
					\label{eq:Z_Liouville_1}
					\begin{split}
					\lket{\mathscr{Z}} =\mho \big[ \mathscr{Z}\big]& = \sum^{d_1}_{\nu=1}\sum^{d_1}_{\nu'=1} \sum^{d_2}_{\mu=1}\sum^{d_2}_{\mu'=1}   A_{\nu\nu'} B_{\mu\mu'} \ \mho \bigg[ \big(\ket{\nu}\otimes\ket{\mu}\big) \big(\bra{\nu'} \otimes\bra{\mu'}\big)\bigg]\\
					& = \sum^{d_1}_{\nu=1}\sum^{d_1}_{\nu'=1} \sum^{d_2}_{\mu=1}\sum^{d_2}_{\mu'=1}   A_{\nu\nu'} B_{\mu\mu'} \ \big(\ket{\nu}\otimes\ket{\mu}\big) \otimes \big(\ket{\nu'}^* \otimes\ket{\mu'}^*\big)\\
					& = \sum^{d_1}_{\nu=1}\sum^{d_1}_{\nu'=1} \sum^{d_2}_{\mu=1}\sum^{d_2}_{\mu'=1}   A_{\nu\nu'} B_{\mu\mu'} \ \lket{\nu \mu,\nu'\mu'}
					\end{split}
					\end{equation}
where $\lket{\nu \mu,\nu'\mu'}\equiv \ket{\nu}\otimes\ket{\mu} \otimes \ket{\nu'}^* \otimes\ket{\mu'}^*$. Because the set $\{\lket{\nu \mu,\nu'\mu'}\}$ is certainly an orthornomal superbasis of the Liouville space $\mathcal{L}_{d}$, we may also write
					\begin{equation}
					\label{eq:Z_Liouville_2}
					\lket{\mathscr{Z}} = \sum^{d_1}_{\nu=1}\sum^{d_1}_{\nu'=1} \sum^{d_2}_{\mu=1}\sum^{d_2}_{\mu'=1}   \mathscr{Z}_{\nu \mu, \nu' \mu'} \ \lket{\nu \mu,\nu'\mu'}
					\end{equation}
where
					\begin{equation}
					\mathscr{Z}_{\nu \mu, \nu' \mu'} = \lbraket{\nu \mu, \nu' \mu'}{\mathscr{Z}} = \lbraket{\nu \mu, \nu' \mu'}{A \otimes B} \ .
					\end{equation}
Comparing \eqref{eq:Z_Liouville_1} with \eqref{eq:Z_Liouville_2}, we conclude the following equations  ---  which are equivalent to each other  ---  must hold
					\begin{subequations}
					\label{eq:Z_elements}
					\begin{align}
					\mathscr{Z}_{\nu \mu, \nu' \mu'} & = A_{\nu\nu'} B_{\mu\mu'}\\
					\lbraket{\nu \mu, \nu' \mu'}{A \otimes B} & =  A_{\nu\nu'} B_{\mu\mu'} \\
					\lbraket{\nu \mu, \nu' \mu'}{A \otimes B} & =  \lbraket{\nu,\nu'}{A} \lbraket{\mu,\mu'}{B} \label{eq:Z_elements_c}
					\end{align}
					\end{subequations}	
where, in arriving at \eqref{eq:Z_elements_c}, we have made use of \eqref{eq:lmatrix_element}. These results can be easily extended to multipartite quantum systems.			 
\section{Applications}\label{sec:Applications}\label{sec:Applications}
In this section, we apply --- for the purpose of illustration --- the concepts and mathematical tools developed in the preceding section to some problems in quantum mechanics. In Sec. \ref{subsec:equations_motion_LS} we explore the derivation of equations of motion in the Schr\"odinger, Heisenberg and interaction pictures --- according to the Liouville space formalism.  We then discuss in Sec. \ref{subsec:solving_EMs} the Liouville space solution to a certain class of equations of motion. And in Sec. \ref{subsec:optics} we solve a quantum optical master equation for a two-level system using the Liouville space formalism.
\subsection{Equations of motion in Liouville space}\label{subsec:equations_motion_LS}
\subsubsection{Schr\"odinger picture in Liouville space.}
\par If we apply the bra-flipper superoperator on the Liouville-von Neumann equation,  \eqref{eq:LvN}, and employ the relation in \eqref{eq:super-commutator_0}, we obtain
					\begin{subequations}
					\begin{align}
					\frac{d}{dt} \lket{ \rho(t) } & = -\frac{i}{\hbar} \lket{ \ [H, \rho(t)] \ } \\
					\frac{d}{dt} \lket{\rho(t)} & = -\frac{i}{\hbar} \lshad H, \mathbb{I}_d \rshad \lket{\rho(t)} \equiv -\frac{i}{\hbar} \mathfrak{L} \lket{\rho(t)} \label{eq:LvN_in_LS_b}
					\end{align}
					\end{subequations}
where -- as mentioned in Sec. \ref{subsec:initial_consid} -- $\mathfrak{L} \in \mathcal{S}_d$ is called the Liouvillian, and it is the super-commutator, \eqref{eq:super-anticommutator},
					\begin{equation}
					\label{eq:Liouvillian}
					\mathfrak{L} = \lshad H, \mathbb{I}_d \rshad = H \otimes \mathbb{I}_d - \mathbb{I}_d \otimes H^T \ .
					\end{equation}
It is instructive to rederive \eqref{eq:LvN_in_LS_b} following a more laborious route considering an isolated system with a mixed state $\rho(t)$. Like \eqref{eq:rho}, we may write
\cite{art:Fano-1957,book:Blum-2012}
					\begin{equation}
					\label{eq:rho_t_isolated}
					\rho(t) = \sum^d_{k=1} P_k \ket{\psi_k(t)}\!\bra{\psi_k(t)}
					\end{equation}
Moreover, we know  ---  according to Schr\"odinger's equation, \eqref{eq:Schrodinger}  ---  that
\cite{art:Fano-1957,book:Blum-2012}
					\begin{equation}
					\label{eq:psi_n_t}
					\frac{d}{dt}\ket{\psi_k(t)} = -\frac{i}{\hbar} H \ket{\psi_k(t)} \ .
					\end{equation}
Now, upon the application of the bra-flipper superoperator, \eqref{eq:rho_t_isolated} becomes
					\begin{equation}
					\label{eq:lrho_t_isolated}
					\begin{split}
					\lket{\rho(t)} & = \sum^d_{k=1} P_k \ket{\psi_k(t)} \otimes \ket{\psi_k(t)}^* \ .
					\end{split}
					\end{equation}
Taking the time derivative, we get
					\begin{subequations}
					\label{eq:LvN_in_LS_2}
					\begin{align}
					\frac{d}{dt}\lket{\rho(t)} & = \sum^d_{k=1} P_k \left[ \left(\frac{d}{dt}\ket{\psi_k(t)}\right) \otimes \ket{\psi_k(t)}^* + \ket{\psi_k(t)} \otimes \left( \frac{d}{dt}\ket{\psi_k(t)}^*\right)\right]\\
					& = -\frac{i}{\hbar} \sum^d_{k=1} P_k \bigg[  H \ket{\psi_k(t)} \otimes \ket{\psi_k(t)}^* -\ket{\psi_k(t)} \otimes H^*\ket{\psi_k(t)}^*\bigg]\\
					& = -\frac{i}{\hbar} \sum^d_{k=1} P_k \bigg[  \big(H \otimes \mathbb{I}_d\big)\big( \ket{\psi_k(t)} \otimes \ket{\psi_k(t)}^*\big) -\big( \mathbb{I}_d \otimes H^*\big) \big(\ket{\psi_k(t)} \otimes \ket{\psi_k(t)}^*\big) \bigg]\\
					& = -\frac{i}{\hbar}  \bigg[  \big(H \otimes \mathbb{I}_d\big) -\big( \mathbb{I}_d \otimes H^*\big)  \bigg]\sum^d_{k=1} P_k \big( \ket{\psi_k(t)} \otimes \ket{\psi_k(t)}^*\big)\\
					& = -\frac{i}{\hbar}  \bigg[  H \otimes \mathbb{I}_d - \mathbb{I}_d \otimes H^T  \bigg] \lket{\rho(t)} = -\frac{i}{\hbar}  \mathfrak{L} \lket{\rho(t)} \label{eq:LvN_in_LS_2_e}
					\end{align}
					\end{subequations}
where in the last line we have exploited the fact that $H$ is Hermitian. We note that \eqref{eq:LvN_in_LS_2_e} agrees with \eqref{eq:LvN_in_LS_b} --- which confirms how logically consistent this approach is.
\par As a common practice in the literature
\cite{art:Fano-1959, art:Jeener-1982, book:Mukamel-1995, book:Ernst_Bodenhausen-1990}, we may want to express \eqref{eq:LvN_in_LS_b} in some orthonormal basis $\{\ket{\nu}\}$ of $\mathcal{H}_d$. This can easily be achieved as follows: surely, the set $\{\ket{\nu}\}$ generates the orthonormal basis $\{\lket{\nu,\nu'}\}$ for $\mathcal{L}_d$, and we may multiply \eqref{eq:LvN_in_LS_b} from the left by a generic superbra $\lbra{\nu,\nu'}$, obtaining
					\begin{subequations}
					\begin{align}
					\frac{d}{dt}\lbraket{\nu,\nu'}{\rho(t)} & = -\frac{i}{\hbar} \lmatrixel{\nu,\nu'}{\mathfrak{L}}{\rho(t)}\\
					\frac{d}{dt} \rho_{\nu\nu'}(t)& =  -\frac{i}{\hbar} \sum_{\nu'',\nu'''}\lmatrixel{\nu,\nu'}{\mathfrak{L}}{\nu'',\nu'''}\lbraket{\nu'',\nu'''}{\rho(t)} \\
					\frac{d}{dt} \rho_{\nu\nu'}(t) & = -\frac{i}{\hbar} \sum_{\nu'',\nu'''}\mathfrak{L}_{\nu\nu',\nu''\nu'''} \ \rho_{\nu''\nu'''}(t) \label{eq:LvN_matrixel_form_c}
					\end{align}
					\end{subequations}
where, naturally, $\mathfrak{L}_{\nu\nu',\nu''\nu'''} = \lmatrixel{\nu,\nu'}{\mathfrak{L}}{\nu'',\nu'''}$. We can easily find an explicit expression for the matrix elements  $\mathfrak{L}_{\nu\nu',\nu''\nu'''} $ using the definition in \eqref{eq:Liouvillian} for $\mathfrak{L}$ and the mixed product rule, \eqref{eq:mixed_product_rule}. As a matter of fact,
					\begin{subequations}
					\begin{align}
					\mathfrak{L}_{\nu\nu',\nu''\nu'''} & = \lmatrixel{\nu,\nu'}{\left( H \otimes \mathbb{I}_d - \mathbb{I}_d \otimes H^T \right)}{\nu'',\nu'''}\\
					& = \lmatrixel{\nu,\nu'}{H \otimes \mathbb{I}_d}{\nu'',\nu'''}- \lmatrixel{\nu,\nu'}{\mathbb{I}_d \otimes H^T}{\nu'',\nu'''} \label{eq:Liouvillian_matrixel}
					\end{align}
					\end{subequations}
Obviously, for the first term in \eqref{eq:Liouvillian_matrixel}, we have
					\begin{equation}
					\begin{split}
					\lmatrixel{\nu,\nu'}{H \otimes \mathbb{I}_d}{\nu'',\nu'''} & = \big(\bra{\nu} \otimes \bra{\nu'}^* \big) \big(H \otimes \mathbb{I}_d \big) \big(\ket{\nu''}\otimes \ket{\nu'''}^* \big)\\
					& = \big(\bra{\nu} \otimes \bra{\nu'}^* \big) \big(H\ket{\nu''} \otimes \mathbb{I}_d\ket{\nu'''}^* \big) \\
					& =\matrixel{\nu}{H}{\nu''} \otimes \matrixel{\nu'}{\mathbb{I}_d}{\nu'''}^*\\
					& = H_{\nu \nu''} \delta_{\nu'\nu'''} \ .
					\end{split}
					\end{equation}
On a similar note, the second term in \eqref{eq:Liouvillian_matrixel} gives
					\begin{subequations}
					\begin{align}	
					\lmatrixel{\nu,\nu'}{\mathbb{I}_d \otimes H^T}{\nu'',\nu'''} & = \lmatrixel{\nu}{\mathbb{I}_d}{\nu''} \otimes \lmatrixel{\nu'}{H}{\nu'''}^*\\
					& = \delta_{\nu \nu''} H^*_{\nu' \nu'''} \ .
					\end{align}
					\end{subequations}	
Thus, $\mathfrak{L}_{\nu\nu',\nu''\nu'''} = H_{\nu \nu''} \delta_{\nu'\nu'''} - \delta_{\nu \nu''} H^*_{\nu' \nu'''}$ and \eqref{eq:LvN_matrixel_form_c} becomes
					\begin{equation}
					\frac{d}{dt} \rho_{\nu\nu'}(t)  = -\frac{i}{\hbar} \sum_{\nu'',\nu'''}\left(  H_{\nu \nu''} \delta_{\nu'\nu'''} - \delta_{\nu \nu''} H^*_{\nu' \nu'''} \right) \ \rho_{\nu''\nu'''}(t) 
					\end{equation}
--- which is the expression one usually finds in the literature 
\cite{book:Mukamel-1995}.
\par Let us now solve \eqref{eq:LvN_in_LS_b} for $\lket{\rho(t)}$. Certainly, \eqref{eq:LvN_in_LS_b} is formally equivalent to the classic matrix differential equation in \eqref{eq:x_A_diff_eq} (or even the Schr\"odinger equation, \eqref{eq:Schrodinger}). Its solution is therefore
					\begin{equation}
					\label{eq:isolated_rho_SE}
					\lket{\rho(t)} = e^{-\frac{i}{\hbar} t \mathfrak{L}} \lket{\rho(0)} =  e^{-\frac{i}{\hbar} t \left[  H \otimes \mathbb{I}_d - \mathbb{I}_d \otimes H^T  \right]} \lket{\rho(0)} \ .
					\end{equation}
Because $H \otimes \mathbb{I}_d$ commutes with $\mathbb{I}_d \otimes H^T$, it follows that
					\begin{equation}
					\label{eq:rho_t_superket_Liouvillian}
					\lket{\rho(t)}  =  e^{-\frac{i}{\hbar} t \left[  H \otimes \mathbb{I}_d \right]} e^{\frac{i}{\hbar} t \left[\mathbb{I}_d \otimes H^T  \right]} \lket{\rho(0)} \ .
					\end{equation}
Applying the properties stated in \eqref{eq:function_Kronecker_prod} of the Kronecker product, we obtain
					\begin{equation}
					\begin{split}
					\label{eq:lket_rho_t_Schrodinger}
					\lket{\rho(t)} & =  \left( e^{-\frac{i}{\hbar} t H} \otimes \mathbb{I}_d \right) \left(\mathbb{I}_d \otimes e^{\frac{i}{\hbar} t  H^T }\right) \lket{\rho(0)}\\
					& = \left( e^{-\frac{i}{\hbar} t H} \otimes e^{\frac{i}{\hbar} t  H^T } \right)  \lket{\rho(0)} \ .
					\end{split}
					\end{equation}
This solution can be reversed to the square matrix form of $\rho(t)$ by resorting to the triple product identity in \eqref{eq:triple_product} and $\mho^{-1}$. Indeed, from \eqref{eq:triple_product}, \eqref{eq:lket_rho_t_Schrodinger} becomes
					\begin{equation}
					\lket{\rho(t)} = \lket{e^{-\frac{i}{\hbar} t H} \rho(0) e^{\frac{i}{\hbar} t  H}}\ .
					\end{equation}
If we now apply the inverse bra-flipper $\mho^{-1}$ on both sides, we get 
					\begin{equation}
					\label{eq:rho_t_O_d}
					\rho(t) = e^{-\frac{i}{\hbar} t H} \rho(0) e^{\frac{i}{\hbar} t  H} \ .
					\end{equation}
The solution in \eqref{eq:rho_t_O_d} is widely known and derived in every undergraduate textbook on quantum mechanics, but the purpose of deriving it here starting from \eqref{eq:LvN_in_LS_b} is to show how consistent the Liouville space formalism we developed earlier is with standard quantum mechanics in state space. 
\par To further appreciate the formal similarities between quantum mechanics in $\mathcal{H}_d$ and $\mathcal{L}_d$, we note that $ \left( e^{-\frac{i}{\hbar} t H} \otimes e^{\frac{i}{\hbar} t  H^T } \right)$ is the equivalent of evolution operator in $\mathcal{L}_d$. Hence, \eqref{eq:lket_rho_t_Schrodinger} may be written as 
					\begin{equation}
					\label{eq:evol_superop}
					\lket{\rho(t)} = \mathfrak{U}(t) \lket{\rho(0)} \ , \qquad  \ \mathfrak{U}(t) = e^{-\frac{i}{\hbar} t\mathfrak{L}} =  e^{-\frac{i}{\hbar} t H} \otimes e^{\frac{i}{\hbar} t  H^T } \ .
					\end{equation}
From its definition, we see that $\mathfrak{U}(0)=\mathbb{I}_d \otimes \mathbb{I}_d=\mathbb{I}_{d^2}$. Also, $\mathfrak{U}(t)$ is clearly unitary. Indeed, the Liouvillian $\mathfrak{L}$, \eqref{eq:Liouvillian}, is Hermitian, so
					\begin{equation}
					\mathfrak{U}(t) \mathfrak{U}^\dagger(t) =  \mathfrak{U}^\dagger(t)\mathfrak{U}(t) = \mathbb{I}_{d^2} \ .
					\end{equation}
Furthermore, after introducing \eqref{eq:evol_superop} into \eqref{eq:LvN_in_LS_b}, we observe that the evolution superoperator $\mathfrak{U}(t)$ satisfies the differential equation
					\begin{equation}
					\label{eq:equation_motion_U(t)_LS}
					\frac{d}{dt} \mathfrak{U}(t) = -\frac{i}{\hbar} \mathfrak{L} \ \mathfrak{U}(t)
					\end{equation}
--- as one would expect from the second equation in \eqref{eq:evol_superop}. Note that  \eqref{eq:equation_motion_U(t)_LS} has the same form as the equation of motion for the evolution operator in $\mathcal{H}_d$
\cite{book:Fayyazuddin-2013,book:Messiah-2014}.
\subsubsection{Heisenberg picture in Liouville space.} As it is well-known, in the Heisenberg picture, we let \emph{operators} other than the density matrix evolve in time
\cite{book:Fayyazuddin-2013, book:Messiah-2014}. If $A$ is a generic operator and element of $\mathcal{O}_d$, then it follows from \eqref{eq:evol_superop}  that the relations
				\begin{equation}
				\lbraket{\rho(t)}{A} = \lmatrixel{\rho(0)}{\mathfrak{U}^\dagger(t)}{A} = \lbraket{\rho(0)}{A_H(t)}
				\end{equation}		
hold, where $\lket{A_H(t)}$  ---  the Heisenberg representation of the superket $\lket{A}$  ---  is defined as
				\begin{equation}
				\label{eq:A_H_LS_EM}
				\lket{A_H(t)} := \mathfrak{U}^\dagger(t) \lket{A} \ .
				\end{equation}	
Taking the time derivative of \eqref{eq:A_H_LS_EM}, we obtain
				\begin{equation}
				\label{eq:Heisenberg_eq_motion}
				\frac{d}{dt} \lket{A_H(t)} = \frac{i}{\hbar} \mathfrak{L} \lket{A_H(t)} 
				+ \mathfrak{U}^\dagger(t) \ \frac{\partial}{\partial t}\lket{A}
				\end{equation}		
where the last term becomes zero for time-independent $A$. Equation \eqref{eq:Heisenberg_eq_motion}	is the operator $A$'s equation of motion in the Heisenberg picture in Liouville space.
\subsubsection{Interaction picture in Liouville space.} In the interaction picture, the time evolution of the system is shared between the density matrix and other operators 
\cite{book:Fayyazuddin-2013, book:Messiah-2014}. The interaction picture is very useful when dealing with perturbations. Consider, for example, the equation of motion
				\begin{equation}
				\label{eq:Interaction_eq_motion}
				\frac{d}{dt}\lket{\rho(t)} = -\frac{i}{\hbar} \mathfrak{L}(t)\lket{\rho(t)}
				\end{equation}
where the Liouvillian $\mathfrak{L}(t)$ is now time-dependent and given by the sum
				\begin{equation}
				\mathfrak{L}(t) = \mathfrak{L}_o + \mathfrak{L}'(t)
				\end{equation}	
where $\mathfrak{L}_o$ originates from a known time-independent Hamiltonian $H_o$, \emph{i.e.} 
$\mathfrak{L}_o = \lshad H_o , \mathbb{I}_d \rshad $, and $\mathfrak{L}'(t)$ is a perturbation respect to $\mathfrak{L}_o$. For Hermitian $H_o$, $\mathfrak{L}_o$ remains Hermitian. As usual, we may write $\lket{\rho(t)} = \mathfrak{U}(t)\lket{\rho(0)}$, which together with \eqref{eq:Interaction_eq_motion} leads to the equation of motion
				\begin{equation}
				\label{eq:diff_superpropa_Int}
				\frac{d}{dt}\mathfrak{U}(t) = -\frac{i}{\hbar} \big[ \mathfrak{L}_o + \mathfrak{L}'(t) \big] \mathfrak{U}(t) \ .
				\end{equation}
The superpropagator $\mathfrak{U}(t)$ here is different from that in \eqref{eq:evol_superop}. We may decompose $\mathfrak{U}(t)$ according to the product
				\begin{equation}
				\label{eq:superprogatator_Int}
				\mathfrak{U}(t) = \mathfrak{U}_o(t)\mathfrak{U}_I(t)
				\end{equation}
where we set $\mathfrak{U}_o(t)=e^{-\frac{i}{\hbar}t \mathfrak{L}_o}$. Given that $\mathfrak{U}(0)=\mathbb{I}_{d^2}$ and $\mathfrak{U}_o(0)=\mathbb{I}_{d^2}$, we find that $\mathfrak{U}_I(0)=\mathbb{I}_{d^2}$. Substituting \eqref{eq:superprogatator_Int} into \eqref{eq:diff_superpropa_Int}, we get
				\begin{equation}
				\begin{split}
				\mathfrak{U}_o(t) \frac{d}{dt}\mathfrak{U}_I(t) = -\frac{i}{\hbar} \mathfrak{L}'(t)\mathfrak{U}_o(t) \mathfrak{U}_I(t)
				\end{split}
				\end{equation}
from which we obtain the final result
				\begin{equation}
				\label{eq:diff_U_I_t}
				\frac{d}{dt}\mathfrak{U}_I(t) = -\frac{i}{\hbar} \mathfrak{L}'_I(t) \mathfrak{U}_I(t) \ , \qquad \ \text{where }\ \mathfrak{L}'_I(t) \equiv \mathfrak{U}^\dagger_o(t) \mathfrak{L}'(t)\mathfrak{U}_o(t) \ .
				\end{equation}
An iterative integration of the differential equation in \eqref{eq:diff_U_I_t} for $\mathfrak{U}_I(t)$ shows that
				\begin{equation}
				\label{eq:U_I_expansion_interaction_pic}
				\begin{split}
				\mathfrak{U}_I(t) & = \mathbb{I}_{d^2} - \frac{i}{\hbar} \int^t_0 dt_1 \ \mathfrak{L}'_I(t_1)  + \left( -\frac{i}{\hbar}\right)^2 \int^t_0 dt_1 \int^{t_1}_0 dt_2 \ \mathfrak{L}'_I(t_1)  \mathfrak{L}'_I(t_2) \\
				& \qquad + \left( -\frac{i}{\hbar}\right)^3 \int^t_0 dt_1 \int^{t_1}_0 dt_2 \int^{t_2}_0 dt_3 \ \mathfrak{L}'_I(t_1)  \mathfrak{L}'_I(t_2) \mathfrak{L}'_I(t_3) + \ldots \\
				& \equiv \mathcal{T} \exp\left[ -\frac{i}{\hbar} \int^t_0 dt' \  \mathfrak{L}'_I(t') \right]
				\end{split}
				\end{equation}
where $\mathcal{T}$ denotes the time-ordering operator
\cite{book:Mukamel-1995,book:Fayyazuddin-2013}. Note that even though the above equations formally still hold when $\mathfrak{L}'(t)$ is not a perturbation with respect to $\mathfrak{L}_o$, the expansion in \eqref{eq:U_I_expansion_interaction_pic} diverges if $\mathfrak{L}'(t)$ is not sufficiently small with respect to $\mathfrak{L}_o$.
\par With $\mathfrak{U}(t)$ as defined in \eqref{eq:superprogatator_Int}, it turns out that
				\begin{equation}
				\lket{\rho(t)} = \mathfrak{U}(t)\lket{\rho(0)}= \mathfrak{U}_o(t)\lket{\rho_I(t)}
				\end{equation}
where
				\begin{equation}
				\label{eq:rho_I_t}
				\lket{\rho_I(t)} \equiv \mathfrak{U}_I(t)\lket{\rho(0)}= \mathfrak{U}^\dagger_o(t) \lket{\rho(t)} \ .
				\end{equation}
$\lket{\rho_I(t)}$ is the superket of the density matrix in the interaction picture, and its equation of motion is easily found to be
				\begin{equation}
				\frac{d}{dt} \lket{\rho_I(t)} = - \frac{i}{\hbar} \mathfrak{L}'_I(t)\lket{\rho_I(t)}
				\end{equation}
in accordance with \eqref{eq:diff_U_I_t}.
\par In regards to the time evolution of other operators, let $A$ be a generic operator belonging to $\mathcal{O}_d$. Then, the inner product $\lbraket{\rho(t)}{A}$ yields 
				\begin{equation}
				\lbraket{\rho(t)}{A} = \lmatrixel{\rho_I(t)}{\mathfrak{U}^\dagger_o(t)}{A} = \lbraket{\rho_I(t)}{A_I(t)}
				\end{equation}
where we have employed \eqref{eq:rho_I_t}, and
				\begin{equation}
				\label{eq:A_I_t}
				\lket{A_I(t)} \equiv \mathfrak{U}^\dagger_o(t)\lket{A} \ .
				\end{equation}
$\lket{A_I(t)}$ is said to be the interaction picture representation of the superket $\lket{A}$. Its corresponding equation of motion is easily derived from \eqref{eq:A_I_t} to be
				\begin{equation}
				\label{eq:Interaction_picture_eq_mot_A}
				\frac{d}{dt}\lket{A_I(t)} = \frac{i}{\hbar} \mathfrak{L}_{o}(t)\lket{A_I(t)} +  \mathfrak{U}^\dagger_o(t)\ \frac{\partial}{\partial t}\lket{A} 
				\end{equation}
for Hermitian $\mathfrak{L}_o$. Equation \eqref{eq:Interaction_picture_eq_mot_A} is the Liouville space equation of motion of $A$ in the interaction picture. All these results bear very close similarities, formally speaking, to their equivalent counterparts in standard quantum mechanics
\cite{book:Fayyazuddin-2013, book:Messiah-2014}.
\subsection{Solving master equations in Liouville space and the connection to non-Hermitian quantum mechanics.} \label{subsec:solving_EMs}
We have so far dealt only with isolated quantum systems  ---  where the Liouvillian $\mathfrak{L}\in \mathcal{S}_d$, \eqref{eq:Liouvillian}, is Hermitian. This makes it possible to apply the quantum mechanical spectral decomposition theorem
\cite{book:Jordan-2006, book:Moretti-2018, book:Dennery-1996}, \eqref{eq:B_spectral_decomp_0}, which is an important result frequently applied in what is commonly referred to as \emph{Hermitian} quantum mechanics
\cite{art:Bender-1998, art:Bender-2007, book:Moiseyev-2011}  ---  \emph{i.e.} the kind of quantum mechanics where observables are postulated to be represented by Hermitian operators. However, not all elements of $\mathcal{S}_d$ are Hermitian or skew-Hermitian (see below) and it is commonplace to encounter non-Hermitian (or non-skew-Hermitian) superoperators when working in Liouville space. For these superoperators, the spectral decomposition theorem, \eqref{eq:B_spectral_decomp_0}, cannot be applied.
\par To explore these problems, we consider in this subsection the solution to equations of motion of the type
				\begin{equation}
				\label{eq:Liouvillian_time_independent}
				\frac{d}{dt} \lket{\rho(t)} = \mathfrak{L}\lket{\rho(t)}
				\end{equation}
where the generator of the dynamics $\mathfrak{L}$ is time-independent. It is important to note that, in general, all equations of motions for $\rho(t)$ which are linear in the latter can be reduced to the form in \eqref{eq:Liouvillian_time_independent} by means of the superket triple product identity in \eqref{eq:triple_product}. However, in the general case, $\mathfrak{L}$ may be time-dependent. The solution to \eqref{eq:Liouvillian_time_independent} is evidently,
				\begin{equation}
				\label{eq:general_rho_EM}
				\lket{\rho(t)} = e^{t\mathfrak{L}}\lket{\rho(0)} \ .
				\end{equation}
We may want to proceed further by expanding $e^{t\mathfrak{L}}\lket{\rho(0)}$ in the `eigenvectors' of $\mathfrak{L}$, just as we commonly do in traditional quantum mechanics in state space. Caution is needed here because, as a matrix, $\mathfrak{L} \in \mathcal{S}_d$ may be diagonalizable (\emph{i.e.} has $d^2$ independent \emph{genuine} eigenvectors
\cite{book:Lax-2007} (more on this latter in Sec. \ref{sec:generalized_basis}); another set of criteria is given in \cite{art:Abate-1997}) or non-diagonalizable. As we shall shortly see, for isolated quantum systems, $\mathfrak{L}$ in \eqref{eq:Liouvillian_time_independent} is skew-Hermitian (\emph{i.e.} $\mathfrak{L}^\dagger = - \mathfrak{L}$) and therefore diagonalizable. On the other hand, for open quantum systems $\mathfrak{L}$ is neither Hermitian nor skew-Hermitian, so could be diagonalizable or not. For isolated systems, $e^{t\mathfrak{L}}$ can be expanded in an orthonormal basis of $\mathcal{S}_d$ which diagonalizes the former. But for open quantum systems, the analogous expansion of $e^{t\mathfrak{L}}$ will require either a \emph{biorthonormal basis} (Sec. \ref{sec:biorthonormal}) or a \emph{generalized basis} (Sec. \ref{sec:generalized_basis}) of $\mathcal{S}_d$, depending on the nature of $\mathfrak{L}$. It goes without saying that orthonormal and biorthonormal bases are special instances of generalized bases. 
\subsubsection{Isolated quantum systems. Orthonormal basis expansion of $e^{t\mathfrak{L}}$.}\label{sec:orthonormal}
\par If we go back to \eqref{eq:LvN_in_LS_b} and set 
				\begin{equation}
				\label{eq:L_isolated}
				\mathfrak{L}\equiv -\frac{i}{\hbar}\lshad H, \mathbb{I}_d \rshad  = -\frac{i}{\hbar}\left(H \otimes \mathbb{I}_d - \mathbb{I}_d \otimes H^T \right) \ ,
				\end{equation}
equation \eqref{eq:isolated_rho_SE} then becomes $\lket{\rho(t)} = e^{t \mathfrak{L}}\lket{\rho(0)}$, which is equivalent to the expression in \eqref{eq:general_rho_EM}. As remarked earlier, $\lshad H, \mathbb{I}_d \rshad$ is Hermitian (for Hermitian Hamiltonian $H$), so $\mathfrak{L}$ in \eqref{eq:L_isolated} is skew-Hermitian. Thus, according to the spectral decomposition theorem, \eqref{eq:B_spectral_decomp_0}, we can find an orthonormal basis of $\mathcal{S}_d$ which diagonalizes $\lshad H, \mathbb{I}_d \rshad$, and by extension $\mathfrak{L}$, \eqref{eq:L_isolated}. The first thing we need to observe is that $\mathfrak{L}$ is proportional to the sum of two Hermitian superoperators $H \otimes \mathbb{I}_d$ and $\mathbb{I}_d \otimes H^T$. Secondly, since they commute, \emph{i.e.} $\left[H \otimes \mathbb{I}_d,\mathbb{I}_d \otimes H^T \right]=0$, the two superoperators must share the same eigen-supervectors. 
\par Now, since $H$ is Hermitian, let $H=\sum_\nu \epsilon_\nu \ket{\nu}\!\bra{\nu}$ be its spectral decomposition, where $\epsilon_\nu \equiv \matrixel{\nu}{H}{\nu}$ and $\{\epsilon_\nu\}$ are the eigenvalues of $H$. If we take $H \otimes \mathbb{I}_d$, we see that
				\begin{equation}
				\label{eq:H_otimes_I_0}
				H \otimes \mathbb{I}_d = \sum_\nu \epsilon_\nu \ket{\nu}\!\bra{\nu}\otimes \mathbb{I}_d = \sum_\nu \epsilon_\nu \ket{\nu}\!\bra{\nu}\otimes \sum_{\nu'} \big(\ket{\nu'}\!\bra{\nu'}\big)^T 
				 = \sum_\nu \epsilon_\nu \ket{\nu}\!\bra{\nu}\otimes \sum_{\nu'} \big(\ket{\nu'}\!\bra{\nu'}\big)^*
 				\end{equation}
where we have made use of the completeness of the orthonormal basis $\{\ket{\nu}\}$, \eqref{eq:identity_I}, and the fact that $\mathbb{I}_d$ is invariant under transpose. After applying the mixed product rule of the Kronecker product  ---  \eqref{eq:mixed_product_rule}  ---  on \eqref{eq:H_otimes_I_0}, we get
				\begin{subequations}
				\begin{align}
				H \otimes \mathbb{I}_d  &= \sum_\nu \sum_{\nu'} \epsilon_\nu \ \big(\ket{\nu} \otimes  \ket{\nu'}^*\big) \big( \bra{\nu} \otimes \bra{\nu'}^*\big) \\
				& = \sum_\nu \sum_{\nu'} \epsilon_\nu \ \lket{\nu,\nu'} \!\lbra{\nu,\nu'} \ . \label{eq:H_otimes_I_1_b}
				\end{align}
				\end{subequations}
Comparing \eqref{eq:H_otimes_I_1_b} with \eqref{eq:B_spectral_decomp}, we clearly see that \eqref{eq:H_otimes_I_1_b} is the spectral decomposition of the superoperator $\big(H \otimes \mathbb{I}_d \big)$, and $\epsilon_\nu = \left(H \otimes \mathbb{I}_d \right)_{\nu\nu',\nu \nu'}= \lmatrixel{\nu,\nu'}{H \otimes \mathbb{I}_d}{\nu, \nu'}$. Likewise, for the superoperator $\big(\mathbb{I}_d \otimes H^T \big)$, it is easy to prove that
				\begin{equation}
				\mathbb{I}_d \otimes H^T = \sum_\nu \sum_{\nu'} \epsilon_{\nu'} \ \lket{\nu,\nu'}\! \lbra{\nu,\nu'} \label{eq:I_otimes_H_0} \ .
				\end{equation}
Thus, putting these last two equations together, it follows from \eqref{eq:L_isolated} that
				\begin{equation}
				\label{eq:L_isolated_diagonal}
				\mathfrak{L} =  -\frac{i}{\hbar}\sum_\nu \sum_{\nu'} \big(\epsilon_\nu - \epsilon_{\nu'}\big) \ \lket{\nu,\nu'}\! \lbra{\nu,\nu'}
				\end{equation}
which is the spectral decomposition of $\mathfrak{L}$. So, the eigenvalues of $\mathfrak{L}$ are simply proportional to the energy gap $(\epsilon_\nu -\epsilon_{\nu'})$ between the eigenstates of $H$. We also observe that 
\begin{enumerate}
\item for those eigen-superkets $\lket{\nu, \nu'}$ where $\nu=\nu'$, their corresponding eigenvalue is identically zero. Hence, $\mathfrak{L}$ is a \emph{singular} matrix (that is, \emph{not} invertible).
\item for fixed $\nu$ and $\nu'$, $\mathfrak{L}_{\nu\nu',\nu\nu'}=-\mathfrak{L}_{\nu'\nu,\nu'\nu}$. This implies that $\mathfrak{L}$ is a zero-trace superoperator (an observation which can alternatively be proved from the definition of $\mathfrak{L}$ in \eqref{eq:L_isolated} by taking the trace).
\end{enumerate}
From \eqref{eq:L_isolated_diagonal}, it is derived that
				\begin{equation}
				\label{eq:L_isolated_diagonalized_skewed_H}
				\mathfrak{L} \lket{\nu,\nu'} = -\frac{i}{\hbar} \big(\epsilon_\nu - \epsilon_{\nu'}\big) \ \lket{\nu,\nu'}
				\end{equation}
from which follows that, for any positive integer $n$,
				\begin{equation}
				\label{eq:L^n_nu_n'}
				\mathfrak{L}^n \lket{\nu,\nu'} = \left[-\frac{i}{\hbar}\big(\epsilon_\nu - \epsilon_{\nu'}\big)\right]^n  \lket{\nu,\nu'} \ .
				\end{equation}	

\par The fact that the eigen-superkets $\{\lket{\nu,\nu'}\}$ of $\mathfrak{L}$ form a complete orthonormal basis for the Liouville space $\mathcal{L}_d$  ---  leading therefore to the completeness relation in \eqref{eq:I_d^2_a}  ---  can be used to greatly simplify derivations and calculations. Back to $\lket{\rho(t)}$, \eqref{eq:general_rho_EM}, for example, we can expand $e^{t\mathfrak{L}}$ in the basis $\{\lket{\nu,\nu'}\}$ as follows:
				\begin{equation}
				\label{eq:L_exp_isolated_0}
				\lket{\rho(t)}  = e^{t\mathfrak{L}}\lket{\rho(0)} =e^{t\mathfrak{L}} \ \mathbb{I}_{d^2} \lket{\rho(0)} 
				 = \sum_{\nu, \nu'} e^{t\mathfrak{L}} \lket{\nu,\nu'}\lbraket{\nu,\nu'}{\rho(0)} 
				\end{equation}
But, from \eqref{eq:exp(tmahfrakL)} and \eqref{eq:L^n_nu_n'}, we have that
				\begin{equation}
				\begin{split}
				e^{t\mathfrak{L}} \lket{\nu,\nu'} & = \sum^\infty_{n=0} \frac{t^n }{n!}\  \mathfrak{L}^n\lket{\nu,\nu'} = \sum^\infty_{n=0} \frac{t^n }{n!}\  \left[-\frac{i}{\hbar}\big(\epsilon_\nu - \epsilon_{\nu'}\big)\right]^n  \lket{\nu,\nu'}\\
				& = e^{-i\frac{t}{\hbar}(\epsilon_\nu-\epsilon_{\nu'})} \lket{\nu,\nu'}\ .
				\end{split}
				\end{equation}
Thus, \eqref{eq:L_exp_isolated_0} may finally be written as
				\begin{subequations}
				\begin{align}
				\lket{\rho(t)} & = \sum_{\nu, \nu'} e^{-i\frac{t}{\hbar}(\epsilon_\nu-\epsilon_{\nu'})} \lket{\nu,\nu'}\lbraket{\nu,\nu'}{\rho(0)}\label{eq:L_exp_isolated}\\
				& = \sum_\nu \rho_{\nu\nu} \lket{\nu,\nu} + \sum_{\nu \neq \nu'} e^{-i\frac{t}{\hbar}(\epsilon_\nu-\epsilon_{\nu'})} \rho_{\nu\nu'}\lket{\nu,\nu'}\label{eq:rho_diagonalized_isolated}
				\end{align}
				\end{subequations}
where $\rho_{\nu \nu'} = \lbraket{\nu,\nu'}{\rho(0)}$. It is clear from \eqref{eq:rho_diagonalized_isolated} that the populations $\lbraket{\nu, \nu}{\rho(t)}$ remain invariant during the evolution. Namely, $\lbraket{\nu, \nu}{\rho(t)} = \lbraket{\nu, \nu}{\rho(0)}=\rho_{\nu\nu}$. Meanwhile, each initial coherence simply gains a phase factor for $t>0$; \emph{i.e.}  $\lbraket{\nu, \nu'}{\rho(t)} = e^{-i\frac{t}{\hbar}(\epsilon_\nu-\epsilon_{\nu'})} \lbraket{\nu, \nu'}{\rho(0)}$, $\nu \neq \nu'$. These observations are in agreement with standard quantum mechanics of isolated systems and deducible from \eqref{eq:rho_t_O_d}. 
\subsubsection{Open quantum systems. Biorthonormal basis expansion of $e^{t\mathfrak{L}}$.}\label{sec:biorthonormal}
\par In standard formulation of quantum mechanics (commonly referred to as \emph{Hermitian} quantum mechanics), it is postulated that all physical observables are represented by Hermitian operators.  In \emph{non-Hermitian} quantum mechanics
\cite{art:Bender-1998, art:Bender-2007, book:Moiseyev-2011, art:Brody-2016}, the formal definition of an operator representing a physical observable posits on entirely different criteria and are more relaxed with respect to the Hermiticity requirement in Hermitian quantum mechanics. In the so-called $\mathcal{PT}$-symmetric formulation
\cite{art:Bender-1998, art:Bender-2007, book:Moiseyev-2011, art:Brody-2016}, for example, an operator qualifies as an observable if it is simultaneously invariant with respect to both parity ($\mathcal{P}$) and time-reversal ($\mathcal{T}$) operations. The concept of \emph{biorthonormal basis}
\cite{art:Brody-2014,art:Gardas-2016} plays an important role in non-Hermitian quantum mechanics. And it should come as no surprise if we encounter it here in Liouville space formalism dealing with open quantum systems because non-Hermitian quantum mechanics handles exceptionally well resonance phenomena
\cite{book:Moiseyev-2011}, which  ---  conceptually speaking  ---  are a consequence of a quantum system being open.
\par In the study of open quantum systems, for example, the Gorini-Kossakowski-Sudarshan-Lindblad (GKSL, or simply Lindblad) master equation plays a crucial role and has found diverse applications: from tackling fundamental problems like the quantum-to-classical transition to the development of quantum technologies
\cite{art:Manzano-2020,  book:Breuer-2007,  book:Nielsen-2011, book:Tarasov-2008, book:Schlosshauer-2007, book:Zeh-2003}. For $\rho(t)$ defined on a finite-dimensional state space of dimension $d$, the GKSL equation reads
\cite{art:Manzano-2020, misc:Lidar-2019, art:Pearle-2012, book:Breuer-2007}
			\begin{equation}
			\label{eq:Lindblad}
			\frac{d}{dt}\rho(t) = -\frac{i}{\hbar} \left[H , \rho(t) \right] + \sum^{d^2-1}_{k=1} \gamma_k \left( A_k \rho(t) A^\dagger_k - \frac{1}{2} A^\dagger_k A_k \rho(t) - \frac{1}{2} \rho(t) A^\dagger_k A_k \right)
			\end{equation}
where the first and second terms represent the unitary and the dissipative parts of the system's evolution, respectively. The constants $\{\gamma_k\}$ are positive and represent relaxation rate constants (if the $A_k$ are dimensionless). Naturally, $H$ and $\{A_k\}$ are operators; so, even though the GSKL equation is linear in $\rho(t)$, it does not easily lend itself to a solution. However, we can circumvent the problem by transitioning to the Liouville space. Namely, we first apply the bra-flipper operator on both sides of \eqref{eq:Lindblad} and then make use of the superket triple product identity in \eqref{eq:triple_product} to factorize out $\rho(t)$ as $\lket{\rho(t)}$ on the r.h.s. of \eqref{eq:Lindblad}. This leads to a differential equation of the same form as that in \eqref{eq:Liouvillian_time_independent}, and the solution in \eqref{eq:general_rho_EM} still holds, with
			\begin{equation}
			\label{eq:L_semigroup}
			\mathfrak{L} = -\frac{i}{\hbar} \lshad H, \mathbb{I}_d \rshad + \sum^{d^2-1}_{k=1} \gamma_k \left(  A_k \otimes A^*_k - \frac{1}{2} \ \lshad A^\dagger_k A_k , \mathbb{I}_d \rshad_+ \right)
			\end{equation}
where we have made use of the definitions in \eqref{eq:super-commutator} and \eqref{eq:super-anticommutator} (see also Appendix A of Ref. \cite{art:Albert-2014}). Unlike the generator $\mathfrak{L}$ for an isolated system, the generator $\mathfrak{L}$ in \eqref{eq:L_semigroup} is not skew-Hermitian (nor is it Hermitian).
This means $\mathfrak{L}$ and $\mathfrak{L}^\dagger$, in general, do not share the same eigen-superkets as one would expect for a Hermitian (or skew-Hermitian) superoperator. 
\par Suppose the generator $\mathfrak{L} \in \mathcal{S}_d$ in \eqref{eq:L_semigroup} is diagonalizable. This will mean that there exists an invertible matrix $\mathfrak{A} \in \mathcal{S}_d$
\cite{book:Lax-2007, book:Strang-2006, book:Smirnov-1961} such that
			\begin{equation}
			\label{eq:L_S-1_S_decomp}
			\mathfrak{A}^{-1} \mathfrak{L}\mathfrak{A} = \mathfrak{D}
			\end{equation}
where $\mathfrak{D}$ is a $(d^2 \times d^2)$ diagonal matrix. In matrix theory, $\mathfrak{A}$ and $\mathfrak{D}$ are the eigenvector and eigenvalue matrices, respectively, of $\mathfrak{L}$. Let $\{\lambda_1, \lambda_2, \ldots, \lambda_{d^2} \}$ be the eigenvalues of $\mathfrak{L}$. There could be repeated eigenvalues. Then,
			\begin{equation}
			\label{eq:D_decomp}
			\mathfrak{D} = \sum^{d^2}_{k=1} \lambda_k \lket{k}\!\lbra{k}
			\end{equation}
where $\lket{k}$ is the $k-$th element of $\mathcal{L}_d$'s canonical basis. That is, $\lket{k}$ is the $(d^2 \times 1)$ column vector whose entries are all equal to zero except for that at the $k-$th row, where the entry has value $1$, \emph{i.e.}
			\begin{equation}
			\label{eq:canonical_k}
			\lket{k} =\begin{pmatrix}
			\underbrace{ \begin{matrix} 0 & \ldots & 0 \end{matrix} }_{k-1} & 1 & \underbrace{ \begin{matrix}0 & \ldots & 0 \end{matrix}}_{d^2-k}  
			\end{pmatrix}^T \ .
			\end{equation}
Surely, 
			\begin{equation}
			\label{eq:k_ortho_complete}
			\lbraket{k}{k'}=\delta_{k,k'} \qquad \text{and} \qquad \sum^{d^2}_{k=1} \lket{k}\!\lbra{k} = \mathbb{I}_{d^2} \ .
			\end{equation}
Substituting \eqref{eq:D_decomp} into \eqref{eq:L_S-1_S_decomp}, we derive that
			\begin{equation}
			\label{eq:L_in_biorthogornal_form}
			\begin{split}
			 \mathfrak{L} & = \sum^{d^2}_{k=1} \lambda_k \ \mathfrak{A}\lket{k}\!\lbra{k}\mathfrak{A}^{-1} 
			  = \sum^{d^2}_{k=1} \lambda_k \ \lket{\zeta_k}\!\lbra{\xi_k}
			 \end{split}
			\end{equation}
where
			\begin{equation}
			\label{eq:zeta_xi_A_A_inv_def}
			\lket{\zeta_k} \equiv \mathfrak{A}\lket{k} \ \qquad \ \lbra{\xi_k} \equiv \lbra{k}\mathfrak{A}^{-1} \ .
			\end{equation}
Note that, in general, $\lket{\zeta_k} \neq \lbra{\xi_k}^\dagger$. We also observe that
			\begin{equation}
			\label{eq:biorthogornal_orthogonality}
			\lbraket{\xi_k}{\zeta_{k'}} = \lmatrixel{k}{\mathfrak{A}^{-1}\mathfrak{A}}{k'}=\lbraket{k}{k'}=\delta_{k,k'} \ .
			\end{equation}
Moreover, multiplying from the left the completeness relation $\mathbb{I}_{d^2}=\sum^{d^2}_{k=1} \lket{k}\!\lbra{k}$ by $\mathfrak{A}$, and then from the right by $\mathfrak{A}^{-1}$, we end up with the following completeness relation
			\begin{equation}
			\label{eq:biorthogornal_completeness_relation}
			\mathbb{I}_{d^2}=\sum^{d^2}_{k=1} \lket{\zeta_k}\!\lbra{\xi_k} \ .
			\end{equation}
We hence conclude from \eqref{eq:biorthogornal_completeness_relation} that the superoperators $\{\lket{\zeta_k}\!\lbra{\xi_k}\}$ also form a complete basis for $\mathcal{S}_d$. In addition, the fact that $\lbra{\xi_k}$ is orthogornal to $\lket{\zeta_{k'}}$, \eqref{eq:biorthogornal_orthogonality}, even though the two sets $\{\lbra{\xi_k}\}$ and $\{\lket{\zeta_{k}}\}$ are not conjugate transpose of each other, implies that the superoperators $\{\lket{\zeta_k}\!\lbra{\xi_k}\}$ constitute what we call a \emph{biorthonormal basis}
\cite{art:Brody-2014, art:Gardas-2016} for $\mathcal{S}_d$. In particular, $\{\lket{\zeta_{k}}\}$ are the \emph{right} eigenvectors (or \emph{right} eigen-supervectors) of $\mathfrak{L}$, while $\{\lbra{\xi_k}\}$ are the corresponding \emph{left} eigen-supervectors. To see why this is so, let us go back to \eqref{eq:L_in_biorthogornal_form} where we saw that $\mathfrak{L}= \sum^{d^2}_{k=1} \lambda_k \ \lket{\zeta_k}\!\lbra{\xi_k}$. If we multiply this relation from the left by $\lbra{\xi_{k'}}$, we get
			\begin{equation}
			\label{eq:left_eigen-supervector_L}
			\begin{split}
			\lbra{\xi_{k'}}\mathfrak{L} & = \sum^{d^2}_{k=1} \lambda_k \ \lbraket{\xi_{k'}}{\zeta_k}\lbra{\xi_k} = \sum^{d^2}_{k=1} \lambda_k \ \delta_{k,k'}\lbra{\xi_k}\\
			& = \lambda_{k'} \lbra{\xi_{k'}} \ .
			\end{split}
			\end{equation}
Similarly, multiplying \eqref{eq:L_in_biorthogornal_form} from the right by $\lket{\zeta_{k'}}$ yields
			\begin{equation}
			\label{eq:right_eigen-supervector_L}
			\mathfrak{L}\lket{\zeta_{k'}} = \lambda_{k'}\lket{\zeta_{k'}} \ .
			\end{equation}
From \eqref{eq:left_eigen-supervector_L} and \eqref{eq:right_eigen-supervector_L}, we readily derive that, for an arbitrary positive integer $n$,
			\begin{subequations}
			\begin{align}
			\lbra{\xi_{k}}\mathfrak{L}^n & =   \lambda^n_{k} \lbra{\xi_{k}}\\
			\mathfrak{L}^n\lket{\zeta_{k}} & = \lambda^n_{k}\lket{\zeta_{k}} \label{eq:biorthonormal_L^n_zeta}
			\end{align}
			\end{subequations}
\par Finally, if we now go back to \eqref{eq:general_rho_EM}, and introduce the biorthonormal completeness relation in \eqref{eq:biorthogornal_completeness_relation}, the result is
			\begin{equation}
			\lket{\rho(t)} = e^{t\mathfrak{L}}\lket{\rho(0)} 
			= e^{t\mathfrak{L}} \mathbb{I}_{d^2} \lket{\rho(0)} 
			= \sum^{d^2}_{k=1} e^{t\mathfrak{L}} \lket{\zeta_k}\!\lbraket{\xi_k}{\rho(0)}
			\end{equation}
-- which, from \eqref{eq:exp(tmahfrakL)} and \eqref{eq:biorthonormal_L^n_zeta}, becomes
			\begin{equation}
			\label{eq:biorthonormal_expansion_e^Lt}
			\lket{\rho(t)} = \sum^{d^2}_{k=1} e^{t \lambda_k} \lket{\zeta_k}\lbraket{\xi_k}{\rho(0)} \ .
			\end{equation}		
We observe from \eqref{eq:biorthonormal_expansion_e^Lt} that we have expressed $\lket{\rho(t)}$ as a linear combination of the right eigen-supervectors $\{\lket{\zeta_k}\}$ of $\mathfrak{L}$. And in the same expansion, the left eigen-supervectors $\{\lbra{\xi_k}\}$, the eigenvalues $\{\lambda_k\}$ and $\lket{\rho(0)}$ determine the coefficients in the expansion. The eigenvalues $\{\lambda_k\}$ are, generally speaking, complex. Moreover, the stability condition requires that $\Re[\lambda_k] \leq 0, \ \forall k$. The corresponding equilibrium superket $\lket{\rho(\infty)}$ of \eqref{eq:biorthonormal_expansion_e^Lt} is obtained by taking the limit $\lim_{t\to +\infty} \lket{\rho(t)}$. We note that -- if the stability condition $\Re[\lambda_k] \leq 0$ is satisfied --, then in taking the limit $t\to +\infty$ of \eqref{eq:biorthonormal_expansion_e^Lt}, only the right eigen-supervectors $\lket{\zeta_k}$ with eigenvalue $\Re[\lambda_k]=0$ survive. This means that $\lket{\rho(\infty)}$ is a linear combination of those $\{\lket{\zeta_k}\}$ with $\Re[\lambda_k]=0$
\cite{art:Albert-2014, art:Manzano-2020, art:Thingna-2016}. 

\subsubsection{Open quantum systems. Generalized basis expansion of $e^{t\mathfrak{L}}$.}\label{sec:generalized_basis}
\par For non-diagonalizable $\mathfrak{L}$, the spectral theorem
\cite{book:Lax-2007} of linear algebra asserts that it is possible to expand $\mathfrak{L}$ in its \emph{generalized} eigen-supervectors. Let us see briefly what we mean by \emph{generalized} eigenvectors. 
\par For \emph{ordinary} or \emph{genuine} (right) eigenvectors of the superoperator $\mathfrak{L}$ we mean the non-null superkets satisfying the relation
\cite{book:Dennery-1996, book:Lax-2007, book:Strang-2006, book:Smirnov-1961}
				\begin{equation}
				\label{eq:eigenvalue_prob_ord}
				\mathfrak{L} \lket{\zeta_\alpha} = \lambda_\alpha\lket{\zeta_\alpha}
				\end{equation}
(as we saw in \eqref{eq:L_isolated_diagonalized_skewed_H} and \eqref{eq:right_eigen-supervector_L}) -- which may also be written as
				\begin{equation}
				\left(\mathfrak{L} - \lambda_\alpha \mathbb{I}_{d^2} \right)\lket{\zeta_\alpha} = 0 
				\end{equation}
where $\lambda_\alpha$, a complex scalar, is the eigenvalue of $\mathfrak{L}$ associated with the eigenvector $\lket{\zeta_\alpha}$. Say $\mathfrak{L}$ has $L$ distinct eigenvalues, which we indicate as $\{\lambda_\alpha\}, \ \alpha=1,2,\ldots,L$. Let $p_\alpha$ be the multiplicity of the eigenvalue $\lambda_\alpha$. Naturally, $\sum_\alpha p_\alpha = d^2$. The positive integer $p_\alpha$ is also commonly referred to as the \emph{algebraic multiplicity} of $\lambda_\alpha$
\cite{book:Strang-2006}. Let $g_\alpha$ indicate the number of independent \emph{genuine} eigenvectors associated with the eigenvalue $\lambda_\alpha$. The positive integer $g_\alpha$ is called the \emph{geometric multiplicity} of $\lambda_\alpha$
\cite{book:Strang-2006}. In general, $g_\alpha \leq p_\alpha$. This also means, $\sum_\alpha g_\alpha \leq d^2$. We thus have that $\mathfrak{L}$ is diagonalizable when $g_\alpha = p_\alpha$ for all $\alpha$  ---  or, in other terms, when $\sum_\alpha g_\alpha = d^2$. On the other hand, it only takes one eigenvalue $\lambda_{\alpha'}$ with $g_{\alpha'} < p_{\alpha'}$ to make  $\mathfrak{L}$ non-diagonalizable.
\par If a given eigenvalue $\lambda_\alpha$ has $g_\alpha < p_\alpha$, with $p_\alpha \geq 2$,  the spectral theorem
\cite{book:Dennery-1996, book:Lax-2007, book:Strang-2006, book:Smirnov-1961} of linear algebra asserts that it is possible to find $p_\alpha$ independent eigenvectors, $\{\lket{\zeta_\alpha(m)}\}, \ m=1,2,\ldots,p_\alpha$, all with eigenvalue $\lambda_\alpha$. The vectors $\{\lket{\zeta_\alpha(m)}\}$ are called \emph{generalized} eigenvectors and are defined as follows
				\begin{subequations}
				\begin{align}
				(\mathfrak{L}-\lambda_\alpha \mathbb{I}_{d^2}) \lket{\zeta_\alpha(1)} & = 0\\
				(\mathfrak{L}-\lambda_\alpha \mathbb{I}_{d^2}) \lket{\zeta_\alpha(m)} & = \lket{\zeta_\alpha(m-1)} \ \qquad \ m=2,\ldots, p_\alpha \label{eq:L_zeta_alpha_m_def}
				\end{align}
				\end{subequations}
from which follows that
				\begin{equation}
				\label{eq:generalized_eigen_prob}
				(\mathfrak{L}-\lambda_\alpha \mathbb{I}_{d^2})^m \lket{\zeta_\alpha(m)} = 0 \ \qquad \ m= 1, 2, \ldots, p_\alpha \ .
				\end{equation}
The vector $\lket{\zeta_\alpha(m)}$ satisfying \eqref{eq:generalized_eigen_prob} is said to be the generalized (right) eigenvector of $\mathfrak{L}$ of rank $m$
\cite{book:Dennery-1996, book:Lax-2007, book:Puri-2001}. An important implication of \eqref{eq:generalized_eigen_prob} is that
				\begin{equation}
				\label{eq:generalized_eigen_prob_2}
				(\mathfrak{L}-\lambda_\alpha \mathbb{I}_{d^2})^n \lket{\zeta_\alpha(m)} = 0 \ \qquad \ \text{for }n \geq m \ .
				\end{equation}				
\par We may collect the vectors $\{\lket{\zeta_\alpha(m)}\}$ into a single matrix $\mathfrak{B}$ according to the arrangement 
				\begin{equation}
				\label{eq:B_Jordan}
				\mathfrak{B} = \begin{pmatrix}
				 &  & &  &  &  & \\
				  &  & &  &  &  & \\
				\lket{\zeta_1(1)} & \ldots & \lket{\zeta_1(p_1)} & \ldots & \lket{\zeta_L(1)} & \ldots & \lket{\zeta_L(p_L)}\\
				 &  & &  &  &  & \\
				  &  & &  &  &  & 
				\end{pmatrix} \ .
				\end{equation}
That is, the vector $\lket{\zeta_\alpha(m)}$ occupies the $n(\alpha,m)-$th column of $\mathfrak{B}$, where
				\begin{equation}
				\label{eq:n_alpha_m}
				n(\alpha,m) = \mathscr{F}(\alpha-1) + m \ ,  \qquad \ \mathscr{F}(\alpha')=\begin{cases}0 \ , & \text{if } \alpha'=0 \\ \sum^{\alpha'}_{\alpha=1} p_\alpha \ , & \text{otherwise}\end{cases} \ .
				\end{equation}
The superoperator $\mathfrak{B}$ is invertible and the similarity transformation
				\begin{equation}
				\mathfrak{B}^{-1} \mathfrak{L} \mathfrak{B} = \mathfrak{J}
				\end{equation}
is called the \emph{Jordan canonical form}
\cite{book:Dennery-1996, book:Lax-2007, book:Strang-2006, book:Smirnov-1961, book:Puri-2001} of $\mathfrak{L}$. For non-diagonalizable $\mathfrak{L}$, the Jordan form $\mathfrak{J}$ is \emph{almost} diagonal -- meaning, $\mathfrak{J}$ is composed of $L$ blocks of matrices
				\begin{equation}
				\mathfrak{J} = \begin{pmatrix}
				\mathfrak{J}_1 &                &                     &        &  \\
				               & \ddots &       &   	              &        &  \\
				               &                & \mathfrak{J}_\alpha &        &   \\
				               &                & 					  & \ddots &   \\
				               &                &                     &        & \mathfrak{J}_L
				\end{pmatrix}
				\end{equation}
where the block matrix $\mathfrak{J}_\alpha$ is a triangular square matrix of dimension $p_\alpha$, whose diagonal has the fixed value $\lambda_\alpha$ and its immediate upper diagonal has the fixed value of $1$, while all other entries are zero. That is,
				\begin{equation}
				\mathfrak{J}_\alpha = \begin{pmatrix}
				\lambda_\alpha &   1   &     &      &    \\
							   & \lambda_\alpha &  1  &      &    \\
							    &                & \ddots &  \ddots  &       \\
								&		         &         & \lambda_\alpha &  1 \\
								&				&		         &         & \lambda_\alpha   
				\end{pmatrix} \ .
				\end{equation}
\par With the concept of generalized eigenvectors in our possession, we can now proceed to show how we can conveniently expand $e^{t\mathfrak{L}}\lket{\rho(0)}$ in terms of the generalized eigen-superkets $\{\lket{\zeta_\alpha(m)}\}$ of $\mathfrak{L}$. To begin with, let $\{\lket{k}\} \ (k=1,\ldots, d^2)$ be once again the canonical basis for $\mathcal{L}_d$, \eqref{eq:canonical_k}. It is easy to see that the column vector $\mathfrak{B}\lket{k}$ is non other but the $k-$th column of $\mathfrak{B}$. Since there is a one-to-one correspondence between the set of integers $\{n(\alpha,m)\}$, \eqref{eq:n_alpha_m}, and $\{k\}$, it follows that for fixed $\alpha$ and $m$, if
				\begin{equation}
				\mathfrak{B}\lket{k} = \lket{\zeta_\alpha(m)} \qquad \text{then } \qquad k=n(\alpha,m)
				\end{equation}
--- in virtue of \eqref{eq:B_Jordan} --- ,  and we may therefore write
				\begin{equation}
				\label{eq:B_zeta_alpha}
				\mathfrak{B}\lket{n(\alpha,m)} = \lket{\zeta_\alpha(m)} \ .
				\end{equation}
In other words, the integers $\{n(\alpha,m)\}$ are simply a two-indexed representation of the integers $\{k\}$. Thus, from \eqref{eq:k_ortho_complete}, we see that the following chain of completeness relations hold
				\begin{equation}
				\label{eq:n_alpha_m_completeness_relation}
				\mathbb{I}_{d^2} = \sum^{d^2}_{k=1} \lket{k}\!\lbra{k} = \sum^L_{\alpha=1} \sum^{p_\alpha}_{m=1} \lket{n(\alpha,m)}\!\lbra{n(\alpha,m)}
				\end{equation}
and 
				\begin{equation}
				\label{eq:n_alpha_m_orthogonality}
				\lbraket{n(\alpha',m')}{n(\alpha,m)}=\delta_{\alpha',\alpha}\delta_{m',m} \ .
				\end{equation}
If we multiply \eqref{eq:n_alpha_m_completeness_relation} from the left by $\mathfrak{B}$ and from the right by $\mathfrak{B}^{-1}$, we get
				\begin{equation}
				\label{eq:n_alpha_m_completeness_relation_2}
				\mathbb{I}_{d^2} = \sum^L_{\alpha=1} \sum^{p_\alpha}_{m=1} \lket{\zeta_\alpha(m)}\!\lbra{\xi_\alpha(m)}
				\end{equation}
where
				\begin{equation}
				\label{eq:B_xi_alpha}
				\lbra{\xi_\alpha(m)} = \lbra{n(\alpha,m)} \mathfrak{B}^{-1} \ .
				\end{equation}
Note that $\lbra{n(\alpha,m)} \mathfrak{B}^{-1}$ is the $n(\alpha,m)-$th row vector of $\mathfrak{B}^{-1}$. $\lbra{\xi_\alpha(m)}$ is the left generalized eigenvector of $\mathfrak{L}$ of rank $m$ and with eigenvalue $\lambda_\alpha$. From   \eqref{eq:B_zeta_alpha}, \eqref{eq:B_xi_alpha} and \eqref{eq:n_alpha_m_orthogonality}, we readily conclude that
				\begin{equation}
				\lbraket{\xi_\alpha(m)}{\zeta_{\alpha'}(m')} = \delta_{\alpha',\alpha}\delta_{m',m}
				\end{equation}
which is the orthogonality relation between these generalized left and right eigenvectors.
\par With the aid of the completeness relation in \eqref{eq:n_alpha_m_completeness_relation_2}, we have
				\begin{equation}
				\begin{split}
				\lket{\rho(t)} & = e^{t\mathfrak{L}}\lket{\rho(0)} = \sum^L_{\alpha=1} \sum^{p_\alpha}_{m=1} e^{t\mathfrak{L}}\lket{\zeta_\alpha(m)}\lbraket{\xi_\alpha(m)}{\rho(0)}\\
				& = \sum^L_{\alpha=1} e^{\lambda_\alpha t}\sum^{p_\alpha}_{m=1} e^{t(\mathfrak{L}-\lambda_\alpha \mathbb{I}_{d^2})}\lket{\zeta_\alpha(m)}\lbraket{\xi_\alpha(m)}{\rho(0)}\\
				& = \sum^L_{\alpha=1} e^{\lambda_\alpha t}\sum^{p_\alpha}_{m=1} \sum^\infty_{n=0} \frac{t^n}{n!}\left( \mathfrak{L}-\lambda_\alpha \mathbb{I}_{d^2} \right)^n \lket{\zeta_\alpha(m)}\lbraket{\xi_\alpha(m)}{\rho(0)}
				\end{split}
				\end{equation}
which leads to the final result (see also \cite{book:Puri-2001} for a similar result in $\mathcal{H}_d$)
				\begin{equation}
				\label{eq:generalized_eigenvectors_e^Lt}
				\lket{\rho(t)} = \sum^L_{\alpha=1} e^{\lambda_\alpha t}\sum^{p_\alpha}_{m=1} \sum^{m-1}_{n=0} \frac{t^n}{n!} \lket{\zeta_\alpha(m-n)}\lbraket{\xi_\alpha(m)}{\rho(0)}
				\end{equation}
where we have made use of   \eqref{eq:L_zeta_alpha_m_def} and \eqref{eq:generalized_eigen_prob_2}. Here too, the stability of $\lket{\rho(t)}$ demands that $\Re[\lambda_\alpha]\leq 0, \ \forall \lambda_\alpha$. Naturally, the generalized right eigen-superkets $\lket{\zeta_\alpha(m)}$ with $\Re[\lambda_\alpha]=0$ define the equilibrium state $\lket{\rho(\infty)}$ of the dynamics. More importantly, the stability of the state requires that each eigenvalue $\lambda_\alpha$ with $\Re[\lambda_\alpha]=0$ must necessarily have a geometric multiplicity of $1$, so that $\lket{\rho(\infty)}= \lim_{t \to + \infty} \lket{\rho(t)}$ is a well-defined state.
\par We conclude the discussion by noting that if $\mathfrak{L}$ is diagonalizable but not skew-Hermitian (nor Hermitian), \eqref{eq:generalized_eigenvectors_e^Lt} reduces to \eqref{eq:biorthonormal_expansion_e^Lt}; while for skew-Hermitian $\mathfrak{L}$ (\emph{i.e.} isolated systems), \eqref{eq:generalized_eigenvectors_e^Lt} simply reduces to \eqref{eq:L_exp_isolated}. In fact, \eqref{eq:generalized_eigenvectors_e^Lt} is a generalization of \eqref{eq:L_exp_isolated} and \eqref{eq:biorthonormal_expansion_e^Lt}. These  observations then lead us to the following proposition:
\begin{description}\label{prop:1}
\item \textbf{P1}: \textit{for an open quantum system with $\lket{\rho(t)}=e^{t\mathfrak{L}}\lket{\rho(0)}$, if the stability conditions are satisfied, the corresponding equilibrium superket $\lket{\rho(\infty)}$ is a linear combination of all the generalized right eigen-superkets $\lket{\alpha}$ of $\mathfrak{L}$ whose eigenvalues $\lambda_{\alpha}$ are such that $\Re[\lambda_{\alpha}]=0$.
} 
\end{description}
\subsection{Quantum optical master equation: Two-level system interacting with a quantized electrmagnetic field.}\label{subsec:optics}
In Sec. \ref{subsec:solving_EMs}, we made general considerations on how to solve a certain class of master equations using the Liouville space formalism without applying the techniques discussed to any concrete example. We wish to fill the gap here, for pedagogical reasons. 
\par Now, suppose we want to describe the time-evolution of a two-level system (TLS) interacting with a radiation field (or light, for short). Excluding a completely classical description, we are left with two main alternatives here: we can either resort to a semiclassical description where we treat the TLS as a quantum entity, while we treat the radiation as a classical entity. Or, in alternative, we treat both as quantum entities. The last method is the primary \emph{modus operandi} of \emph{quantum optics}
\cite{book:Grynberg-2010}. We show in this section how the Liouville space formalism can be employed to effectively describe the time-evolution of a quantum TLS interacting with a quantized radiation field. We start with a short introduction to the problem in Sec. \ref{subsub:TLS_background}. In Sec. \ref{subsub:TLS_ equation_mot_rho}, the equation of motion to be solved is presented. Sec.s \ref{subsub:TLS_dissipator_term_only}-\ref{subsub:TLS_Lamb-shift_Hamil} are devoted to solving the master equation and writing the solution in different forms using the Liouville space formalism. 
\subsubsection{Some background.}\label{subsub:TLS_background}
\par Quantum optics
\cite{book:Grynberg-2010}, which became a vibrant established field of research after the invention of the laser, deals with phenomena which arise out of light-matter interactions whose correct theoretical description requires both the radiation and matter to be fully quantized. One well-known example is the process of spontaneous emission of light by an excited atom
\cite{book:Grynberg-2010, book:Craig-1998, art:Bradshaw-2020}. Research in quantum optics today also plays a central role in our quest to develop quantum technologies (for example, the production of entangled photons for quantum communication
\cite{art:Yuan-2010} and the realization of so-called optical lattice clocks \cite{art:Ludlow-2015} for the measurement of time with unprecedented precision). 
\par Back to our TLS-radiation interaction, say we choose to have the radiation also quantized
\cite{book:Grynberg-2010, book:Craig-1998, art:Torre-2005}. (This will be necessary for the correct description of one-atom lasers
\cite{art:McKeever-2003}, for example.) Let $\mathcal{H}_S$ and $\mathcal{H}_R$ be the state space of the TLS and the quantized radiation field, respectively. $\mathcal{H}_{S}$ is then a two dimensional Hilbert space and we may indicate its basis as $\{\ket{g},\ket{e}\}$, where $\ket{g}$ is the ground state and $\ket{e}$ is the excited state. In terms of matrix representation, we may have
			\begin{equation}
			\ket{e} = \begin{pmatrix}
			1 \\
			0
			\end{pmatrix} \qquad
			\ket{g} = \begin{pmatrix}
			0 \\
			1
			\end{pmatrix} \ .
			\end{equation}
Note that the state space of the bipartite TLS-plus-radiation is $\mathcal{H}_{S+R}=\mathcal{H}_S \otimes \mathcal{H}_R$. We consider the TLS-plus-radiation as a closed quantum system with Hamiltonian $H_{S+R}$,
			\begin{equation}
			H_{S+R} = H_{S}\otimes \mathbb{I}_R + \mathbb{I}_S \otimes H_R + H_I
			\end{equation}
-- where $\mathbb{I}_S$ and $\mathbb{I}_R$ are the identity operators on $\mathcal{H}_S$ and $\mathcal{H}_R$, respectively (in particular, for the TLS, $\mathbb{I}_S=\mathbb{I}_2$); $H_S$ is the Hamiltonian of the free TLS, $H_R$ is the Hamiltonian of the free radiation field, and $H_I$ is the TLS-radiation interaction Hamiltonian. Let $\rho_{S+R}(t)$ be the density matrix of the TLS-plus-radiation system. Then, since the bipartite system is assumed closed, the Liouville-von Neumann equation, \eqref{eq:LvN}, applies, and in the Schr\"odinger picture we may write:
			\begin{equation}
			\frac{d}{dt}\rho_{S+R}(t) = -\frac{i}{\hbar} \left[ H_{S+R}, \rho_{S+R}(t) \right] \ .
			\end{equation}
\par In our discussion, we will make use of the fact that any operator $X$ on a two-dimensional state space can be expressed as a linear combination of the identity operator $\mathbb{I}_2$ and the Pauli matrices $\sigma_1$, $\sigma_2$, $\sigma_3$ as follows:
			\begin{equation}
			\label{eq:X_qubit_Pauli_comb_0}
			X = \frac{1}{2} \big[c_o \mathbb{I}_2 + c_1 \sigma_1 + c_2 \sigma_2 + c_3 \sigma_3 \big]
			\end{equation}
-- where 
			\begin{equation}
			c_o = \Tr[X] \quad c_1 = \Tr[X\sigma_1] \quad c_2 = \Tr[X \sigma_2] \quad c_3 = \Tr[X \sigma_3]
			\end{equation}
and
			\begin{equation}
			\sigma_1 = 
			\begin{pmatrix}
			0 & 1 \\
			1 & 0
			\end{pmatrix} \qquad
			\sigma_2 = 
			\begin{pmatrix}
			0 & -i \\
			i & 0
			\end{pmatrix} \qquad
			\sigma_3= 
			\begin{pmatrix}
			1 & 0 \\
			0 & -1
			\end{pmatrix}	 \ .		
			\end{equation}
We will also find it helpful to know the relation between Pauli operators and the basis $\{\ket{e},\ket{g}\}$:
			\begin{subequations}
			\begin{align}
			\sigma_1 & = \ket{g}\!\bra{e} + \ket{e}\!\bra{g}\\
			\sigma_2 & = i \ket{g}\!\bra{e} - i \ket{e}\!\bra{g} \\
			\sigma_3 & = \ket{e}\!\bra{e} - \ket{g}\!\bra{g} \label{eq:sigma_3_in_e_g} \ .
			\end{align}
			\end{subequations}						 
For simplicity, suppose the Hamiltonian $H_S$ of the free TLS is diagonal in the basis $\{\ket{e},\ket{g}\}$, \emph{i.e.}
			\begin{equation}
			H_S \ket{g} = E_g \ket{g} \qquad H_S \ket{e} = E_e \ket{e} \ .
			\end{equation}
That is, the ground state $\ket{g}$ corresponds to the energy $E_g$, where we may write  $E_g\equiv\epsilon_o - \frac{1}{2}\hbar \omega_o$, while the excited state $\ket{e}$ corresponds to the energy $E_e\equiv \epsilon_o + \frac{1}{2}\hbar \omega_o$ ($\epsilon_o$ and $\omega_o$ are constants, characteristic of the TLS; moreover, $\omega_o > 0$). Then, by the spectral decomposition theorem, \eqref{eq:spectral_decomposition_thm}, we know $H_S= E_g \ket{g}\!\bra{g} + E_e \ket{e}\!\bra{e}$, from which follows that (using \eqref{eq:sigma_3_in_e_g} and the closure relation, \eqref{eq:identity_I}):
			\begin{equation}
			H_S = \epsilon_o \mathbb{I}_2+ \frac{1}{2}\hbar \omega_o \sigma_3
			\end{equation}
-- which we note is a linear combination of $\mathbb{I}_2$ and the Pauli matrices, \eqref{eq:X_qubit_Pauli_comb_0}. We thus see that $\hbar \omega_o$ is the energy separation between the ground state $\ket{g}$ and the excited state $\ket{e}$. Without loss of generality, we set $\epsilon_o=0$ in the following. This is just shifting the energy levels of the TLS by a fixed constant, and it does not change the ultimate physics of the system.
\par Let us further assume an electric dipole approximation of $H_I$
\cite{book:Craig-1998} (\emph{i.e.} we are assuming that the wavelength of the radiation is longer with respect to the dimensions of the TLS, such that the latter only sees an homogeneous field) so that
			\begin{equation}
			H_I = - \mathbf{D} \cdot \mathbf{E}
			\end{equation}
where $\mathbf{D}$ is the electric dipole vector operator of the TLS in the state space $\mathcal{H}_{S+R}$, namely
			\begin{equation}
			\mathbf{D} = \left(D_x \otimes \mathbb{I}_R\right) \ \mathbf{e}_x + \left(D_y \otimes \mathbb{I}_R\right) \ \mathbf{e}_y + \left(D_z \otimes \mathbb{I}_R\right) \ \mathbf{e}_z
			\end{equation}
-- ($D_x,D_y,D_z$ are the components of the free TLS' electric dipole vector operator in the state space $\mathcal{H}_S$ along the axes $\mathbf{e}_x,\mathbf{e}_y, \mathbf{e}_z$, respectively). Analogously, $\mathbf{E}$ is the electric field vector operator of the radiation field in the state space $\mathcal{H}_{S+R}$,
			\begin{equation}
			\mathbf{E} = \left(\mathbb{I}_2 \otimes E_x\right) \ \mathbf{e}_x + \left(\mathbb{I}_2 \otimes E_y\right) \ \mathbf{e}_y + \left(\mathbb{I}_2 \otimes E_z\right) \ \mathbf{e}_z 
			\end{equation}
(similarly, $E_x,E_y, E_z$ are the electric field components of the radiation field along the axes $\mathbf{e}_x,\mathbf{e}_y, \mathbf{e}_z$, respectively, in $\mathcal{H}_R$). Since the radiation field is quantized, the components $E_x,E_y, E_z$ and the Hamiltonian $H_R$ are expressed in terms of photon creation and annihilation operators. For exact expressions, see 
\cite{book:Breuer-2007, book:Craig-1998, art:Torre-2005, book:Carmichael-1993}. We do not report them here because it is beyond the scope of the paper. 

\subsubsection{The equation of motion for $\rho_S(t)$.}\label{subsub:TLS_ equation_mot_rho}
\par What interests us is that, after a chain of algebraic manipulations, introduction of some approximations and assumptions, and tracing out of the radiation field's degrees of freedom, one arrives at the following equation of motion for the TLS' density matrix $\rho_S(t) \in \mathcal{H}_S$ (in the Schr\"odinger picture)
\cite{book:Breuer-2007, book:Carmichael-1993}:
			\begin{equation}
			\label{eq:TLS_reduced}
			\frac{d}{dt} \rho_S(t) = -\frac{i}{\hbar}\left[H_{LS}+H_S, \rho_S(t) \right] + \mathcal{D}\left[ \rho_S(t)\right]
			\end{equation}
where $H_{LS}$ is the so-called Lamb shift Hamiltonian. It describes a modification of the free TLS' Hamiltonian due to its interaction with the radiation field. Moreover, it always commutes with $H_S$
\cite{book:Breuer-2007}. The second term in \eqref{eq:TLS_reduced} describes the TLS' relaxation due to its  interaction with the radiation field, and it is given by the expression
\cite{book:Breuer-2007, book:Carmichael-1993}
			\begin{multline}
			\label{eq:TLS_dissipator}
			\mathcal{D}\left[ \rho_S(t)\right] \equiv  \gamma_o (1+\mathscr{N}(\omega_o)) \left[\sigma_- \rho_S(t) \sigma_+ - \frac{1}{2}\sigma_+ \sigma_- \rho_S(t) - \frac{1}{2} \rho_S(t)\sigma_+ \sigma_-  \right]\\
		+ \gamma_o \mathscr{N}(\omega_o) \left[\sigma_+ \rho_S(t) \sigma_- - \frac{1}{2}\sigma_- \sigma_+ \rho_S(t) - \frac{1}{2} \rho_S(t)\sigma_- \sigma_+ \right]
			\end{multline}
where \cite{book:Breuer-2007,book:Craig-1998, book:Carmichael-1993}\footnote{The expression for $\gamma_o$ reported here is different from that in Ref. [8] because the former was derived using the equation for the quantized electric field in Ref. [51, Eq. (2.8.12)]. See also Ref.s [52] and [57].} 
			\begin{subequations}
			\begin{align}
			\gamma_o & = \frac{\omega_o^3 \lvert \mathbf{d}\rvert^2}{3 \pi \varepsilon_o \hbar c^3} \qquad \mathbf{d} \equiv \matrixel{g}{D_x \mathbf{e}_x + D_y  \mathbf{e}_y + D_z \mathbf{e}_z}{e}\\
			\mathscr{N}(\omega_o) & = \frac{1}{e^{\beta \hbar \omega_o}-1} \label{eq:Planck_dist}
			\end{align}
			\end{subequations}
where $c$ is the speed of light and $\varepsilon_o$ is the vacuum permittivity. Here, $\mathscr{N}(\omega_o)$ is the Planck distribution centered on the frequency $\omega_o$ (\emph{i.e.} the average number of photons in the radiation field in a mode of frequency $\omega_o$). The constant $\gamma_o$ is the rate of spontaneous emission of photons by the TLS. In addition, the rates of (thermally induced) emission and absorption of photons by the TLS coincide, and they are given by the product $\gamma_o \mathscr{N}(\omega_o)$. We thus see that $\gamma_o (1+\mathscr{N}(\omega_o))$ is the rate of total emission of photons and the first term in \eqref{eq:TLS_dissipator} describes the  TLS' emission process, while the second term describes its absorption process.  Finally, the operators $\sigma_\pm$ are the usual Pauli ladder operators defined as
			\begin{equation}
			\sigma_\pm = \frac{1}{2} \left[\sigma_1 \pm i \sigma_2 \right]
			\end{equation}
and their corresponding matrix representations are
			\begin{equation}
			\label{eq:sigma_pm}
			\sigma_+ = 
			\begin{pmatrix}
			0 & 1 \\
			0 & 0
			\end{pmatrix}= \ket{e}\!\bra{g} \qquad
			\sigma_- =
			\begin{pmatrix}
			0 & 0 \\
			1 & 0
			\end{pmatrix} = \ket{g}\!\bra{e} \ .
			\end{equation}
Equations of motions like \eqref{eq:TLS_reduced} describing the interaction of an $n-$level system interacting with a quantized radiation field are commonly referred to as \emph{quantum optical master equations}. \Eref{eq:TLS_reduced} has the properties of what is called \emph{quantum Markovian master equation}: these are master equations where -- among other properties which are beyond our scope here -- the time-evolution of the system is independent of its past history. For more on quantum Markovian master equations (or Lindblad master equations), see for example \cite{art:Pearle-2012, book:Breuer-2007,  book:Alicki-2007}. 

\par There are a number of ways of solving \eqref{eq:TLS_reduced} for $\rho_S(t)$. One could, for example, express $\rho_S(t)$ in terms of $\mathbb{I}_2$  and the Pauli matrices, as mentioned in \eqref{eq:X_qubit_Pauli_comb_0}, obtaining
				\begin{equation}
				\label{eq:TLS_rho_in_Pauli}
				\rho_S(t) = \frac{1}{2} \big[\mathbb{I}_2 + \big<\sigma_1(t)\big> \sigma_1 + \big<\sigma_2(t)\big> \sigma_2 + \big<\sigma_3(t)\big> \sigma_3 \big]
				\end{equation}
or, alternatively,
				\begin{equation}
				\label{eq:TLS_rho_in_Pauli_+-}
				\rho_S(t) = \frac{1}{2} \big[\mathbb{I}_2 + 2\big<\sigma_-(t)\big> \sigma_+ + 2\big<\sigma_+(t)\big> \sigma_- + \big<\sigma_3(t)\big> \sigma_3 \big]
				\end{equation}
-- where
				\begin{equation}
				\big<\sigma_\mu(t)\big> \equiv \Tr \big[ \rho_S(t)\ \sigma_\mu^\dagger \big] \qquad \mu = {1,2,3,\pm}
				\end{equation}
-- and solve \eqref{eq:TLS_reduced} for the scalars $\big<\sigma_\mu(t)\big>$ . This is, however, not easy to scale-up for an arbitrary $n-$dimensional system. Our aim is to solve \eqref{eq:TLS_reduced} for $\rho_S(t)$ using the Liouville space formalism. We will see that besides yielding the same results as other methods, it has the advantage of making it easy to work out the \emph{Kraus operator sum representation} of the solution (more on this latter in Sec. \ref{subsub:TLS_Kraus_OSR}). The method can be easily adapted for an arbitrary $n-$dimensional system. The computational cost can, however, increase rapidly as $n$ increases, since it involves matrix diagonalization. 
\subsubsection{Rewriting the master equation in Liouville space formalism.} 
\label{subsub:ME_in_LS}
We begin by applying the bra-flipper operator to both sides of \eqref{eq:TLS_reduced}:
			\begin{equation}
			\label{eq:TLS_reduced_LS}
			\begin{split}
			\frac{d}{dt} \mho\big[ \rho_S(t)\big] & = -\frac{i}{\hbar}\mho\bigg[\left[H_{LS}+H_S, \rho_S(t) \right]\bigg] + \mho\bigg[\mathcal{D}\left[ \rho_S(t)\right]\bigg] \ .
			\end{split}
			\end{equation}
Certainly, $\mho\big[ \rho_S(t)\big]=\lket{\rho_S(t)}$. We now consider the two superkets on the r.h.s. of \eqref{eq:TLS_reduced_LS}, separately. For the first superket, we have
			\begin{equation}
			\label{eq:TLS_preL_a}
			\begin{split}
			-\frac{i}{\hbar}\mho\bigg[\left[H_{LS}+H_S, \rho_S(t) \right]\bigg] & = -\frac{i}{\hbar}\lket{\ \left[H_{LS}+H_S, \rho_S(t) \right]\ } = \mathfrak{L}_a \lket{\rho_S(t)}
			\end{split}
			\end{equation}
with 
			\begin{equation}
			\label{eq:TLS_L_a}
			\mathfrak{L}_a \equiv -\frac{i}{\hbar} \lshad H_{LS}+H_S, \mathbb{I}_2  \rshad
			\end{equation}
-- where in obtaining the final result in \eqref{eq:TLS_preL_a}, we have applied the superket triple product identity, \eqref{eq:triple_product} -- or better, \eqref{eq:super-commutator_0}. From the definition of the super-commutator, \eqref{eq:super-commutator}, we know 
			\begin{equation}
			\label{eq:TLS_L_a_1}
			\mathfrak{L}_a = -\frac{i}{\hbar} \lshad H_{LS}+H_S, \mathbb{I}_2  \rshad = -\frac{i}{\hbar} \bigg[\big( H_{LS}+H_S \big)\otimes \mathbb{I}_2 - \mathbb{I}_2 \otimes \big( H_{LS}+H_S \big)^T\bigg] \ .
			\end{equation}
As mentioned above, $H_{LS}$ is known to commute with $H_S$ 
\cite{book:Breuer-2007}; so, without loss of generality, we may simply choose $H_{LS}$ to be of the form
			\begin{equation}
			H_{LS} = -\frac{1}{2}\hbar \Omega(\omega_o)\sigma_3
			\end{equation}
where $\Omega(\omega_o)$ is some real scalar function of $\omega_o$. Then,
			\begin{equation}
			H_{S} + H_{LS} = \frac{\hbar}{2} \Delta(\omega_o)\ \sigma_3 \qquad \Delta(\omega_o) \equiv  \left[\omega_o - \Omega(\omega_o)\right]
			\end{equation}
and the superoperator $\mathfrak{L}_a$, \eqref{eq:TLS_L_a_1}, becomes 
			\begin{equation}
			\label{eq:L_a_in_sigma_3_I_2}
			\mathfrak{L}_a = -i \frac{\Delta(\omega_o)}{2} \lshad \sigma_3 , \mathbb{I}_2 \rshad = -i  \frac{\Delta(\omega_o)}{2} \big[\sigma_3 \otimes \mathbb{I}_2 - \mathbb{I}_2 \otimes \sigma_3 \big] \ .
			\end{equation}	
(Note that $\sigma_3^T = \sigma_3$.) The matrix representation of $\mathfrak{L}_a$ is then
			\begin{equation}
			\label{eq:TLS_L_a_matrix_rep}
			\mathfrak{L}_a = 
			\begin{pmatrix}
			0     &       0     &    0     &    0     \\
			0     &   -i \Delta(\omega_o) &   0    &   0   \\
			0     &        0     &  i \Delta(\omega_o) &   0      \\
			0     &       0     &    0     &    0     \\
			\end{pmatrix} \ .
			\end{equation}					
\par For the dissipator term in \eqref{eq:TLS_reduced_LS}, $\mho\bigg[\mathcal{D}\left[ \rho_S(t)\right]\bigg]$, it follows from \eqref{eq:TLS_dissipator}  and the linear property of $\mho$, \eqref{eq:mho_linear}, that
			\begin{multline}
			\label{eq:TLS_precond_dissi}
			\mho\bigg[\mathcal{D}\left[ \rho_S(t)\right]\bigg]  = \Gamma_1 \left(\mho\big[\sigma_- \rho_S(t) \sigma_+ \big]- \frac{1}{2} \mho\big[\sigma_+ \sigma_- \rho_S(t)\big] - \frac{1}{2} \mho\big[\rho_S(t)\sigma_+ \sigma_-\big]  \right)\\
		+ \Gamma_2 \left(\mho\big[\sigma_+ \rho_S(t) \sigma_-\big] - \frac{1}{2}\mho\big[\sigma_- \sigma_+ \rho_S(t)\big] - \frac{1}{2} \mho\big[\rho_S(t)\sigma_- \sigma_+ \big] \right) 
			\end{multline}
which, upon applying the superket triple product identity, \eqref{eq:triple_product} -- with the scope of factoring out $\rho_S(t)$ as the superket $\lket{\rho_S(t)}$ on the r.h.s. of \eqref{eq:TLS_precond_dissi} --, becomes
			\begin{equation}
			\label{eq:TLS_L_b_intro_dissip}
			\mho\bigg[\mathcal{D}\left[ \rho_S(t)\right]\bigg]  = \mathfrak{L}_b \lket{\rho_S(t)}
			\end{equation}
where
			\begin{equation}
			\label{eq:L_TLS_dissipator}
			\mathfrak{L}_b \equiv  \Gamma_1 \left(\sigma_- \otimes \sigma^T_+ - \frac{1}{2} \lshad \sigma_+ \sigma_- ,  \mathbb{I}_2 \rshad_+  \right) + \Gamma_2 \left(\sigma_+ \otimes \sigma^T_- - \frac{1}{2}\lshad \sigma_- \sigma_+ , \mathbb{I}_2\rshad_+ \right) 
			\end{equation}
(recall, for example, that $\lshad \sigma_- \sigma_+ , \mathbb{I}_2\rshad_+ = \big(\sigma_- \sigma_+ \big)\otimes \mathbb{I}_2 + \mathbb{I}_2 \otimes \big( \sigma_- \sigma_+ \big)^T$, \eqref{eq:super-anticommutator}), and
			\begin{equation}
			\label{eq:Gamma_1_Gamma_2}
			\Gamma_1 \equiv \gamma_o (1+\mathscr{N}(\omega_o)) \ , \qquad 
			\Gamma_2 \equiv \gamma_o \mathscr{N}(\omega_o) \ , \qquad 
			\Gamma \equiv \Gamma_1 + \Gamma_2 = \gamma_o (2\mathscr{N}(\omega_o)+1) \ .
			\end{equation}
The matrix representation of $\mathfrak{L}_b$ is easily found by carrying out the Kronecker products and sums, and recalling the matrix representation of the operators $\sigma_\pm$ given in \eqref{eq:sigma_pm}:
			\begin{equation}
			\label{eq:TLS_L_b_matrix_rep}
			\mathfrak{L}_b =
			\begin{pmatrix}
			- \Gamma_1 &   0                &    0               &   \Gamma_2 \\
			0          & -\frac{1}{2}\Gamma &    0			     &  0          \\
			0          & 0                  & -\frac{1}{2}\Gamma &  0          \\
			\Gamma_1   & 0                  &    0               &  - \Gamma_2          
			\end{pmatrix} \ .
			\end{equation}
With \eqref{eq:TLS_preL_a} and \eqref{eq:TLS_L_b_intro_dissip}, we see that \eqref{eq:TLS_reduced_LS} may be rewritten as
			\begin{equation}
			\label{eq:TLS_qmme_LS}
			\frac{d}{dt}\lket{\rho_S(t)} = \mathfrak{L}\lket{\rho_S(t)} \ , \qquad \mathfrak{L} \equiv \mathfrak{L}_a + \mathfrak{L}_b 
			\end{equation}
which is of the same form as the type of master equations discussed in the previous section, \eqref{eq:Liouvillian_time_independent}. Here too, we note that $\mathfrak{L}$ in \eqref{eq:TLS_qmme_LS} is time-independent, so the solution to the differential equation is still given by \eqref{eq:general_rho_EM}, \emph{i.e.}
			\begin{equation}
			\label{eq:rho_TLS_LS_1}
			\lket{\rho_S(t)} = e^{t\mathfrak{L}} \lket{\rho_S(0)} = e^{t(\mathfrak{L}_a + \mathfrak{L}_b)} \lket{\rho_S(0)} \ .
			\end{equation}
Interestingly, the superoperators $\mathfrak{L}_a$ and $\mathfrak{L}_b$ commute: $\mathfrak{L}_a \mathfrak{L}_b = \mathfrak{L}_b \mathfrak{L}_a$. This can be easily verified using their matrix representations in \eqref{eq:TLS_L_a_matrix_rep} and \eqref{eq:TLS_L_b_matrix_rep}. With this observation in hand, it thus follows from \eqref{eq:rho_TLS_LS_1} that
			\begin{equation}
			\label{eq:rho_TLS_LS_2}
			\lket{\rho_S(t)} =  e^{t\mathfrak{L}_a} e^{t\mathfrak{L}_b} \lket{\rho_S(0)} =   e^{t\mathfrak{L}_b} e^{t\mathfrak{L}_a} \lket{\rho_S(0)} \ .
			\end{equation}
Note that $e^{t\mathfrak{L}_a} e^{t\mathfrak{L}_b}$ is a composition of two superoperators; so, for example, $e^{t\mathfrak{L}_a} e^{t\mathfrak{L}_b} \lket{\rho_S(0)}$ must be interpreted as $e^{t\mathfrak{L}_a} \left[e^{t\mathfrak{L}_b} \lket{\rho_S(0)}\right] $ -- where it must be observed that $e^{t\mathfrak{L}_b} \lket{\rho_S(0)}$ is a superket of $\mathcal{L}_d (=\mathcal{L}_2)$. 
\par As we saw in the previous section, Sec. \ref{subsec:solving_EMs}, we may expand $e^{t \mathfrak{L}} \lket{\rho_S(0)}$ in terms of the (generalized left and right) eigen-superkets of $\mathfrak{L}$. In the following, we carry out this expansion considering $\lket{\rho_S(t)} =  e^{t\mathfrak{L}_a} \left[ e^{t\mathfrak{L}_b} \lket{\rho_S(0)}\right]$. We first concentrate on $e^{t\mathfrak{L}_b} \lket{\rho_S(0)}$ -- Sec. \ref{subsub:TLS_dissipator_term_only}--, and we then apply the superoperator $e^{t\mathfrak{L}_a}$ on the resulting expansion in Sec. \ref{subsub:TLS_Lamb-shift_Hamil}.
\subsubsection{Considering only the dissipator superoperator, $\mathfrak{L}_b$.}\label{subsub:TLS_dissipator_term_only} 
The Liouvillian $\mathfrak{L}_a$ is not of particular interest because, as we shall soon see, it only accounts for the unitary part of the dynamics -- which is uncharacteristic of an open quantum system like the TLS under consideration. The real dynamics of the open quantum system is encapsulated in the Liouvillian $\mathfrak{L}_b$.  If we put $\mathfrak{L}_a=0$, then \eqref{eq:rho_TLS_LS_2} becomes
			\begin{equation}
			\label{eq:TLS_master_equation_with_L_a}
			\lket{\rho_S(t)} =    e^{t\mathfrak{L}_b} \lket{\rho_S(0)} \ .
			\end{equation}			 
To expand $e^{t\mathfrak{L}_b}$ as we discussed in Sec. \ref{subsec:solving_EMs}, we need to determine whether $\mathfrak{L}_b$ is diagonalizable or not.
From its matrix representation in \eqref{eq:TLS_L_b_matrix_rep}, we note that $\mathfrak{L}_b$ is neither skew-Hermitian nor Hermitian), so its ``diagonalizability" is not guaranteed and we cannot be in the case of Sec. \ref{sec:orthonormal}. If it happens to be diagonalizable, then we are in the case of Sec. \ref{sec:biorthonormal}, else we are in the case of Sec. \ref{sec:generalized_basis}.
\par To find the eigenvalues of $\mathfrak{L}_b$, we solve $\det[\mathfrak{L}-\lambda  \mathbb{I}_2]=0$ for $\lambda$, which yields
			\begin{equation}
			\lambda \cdot (\Gamma +2\lambda)^2 \cdot \left( \Gamma  +\lambda  \right) =0
			\end{equation}
-- from which we deduce that the eigenvalues of $\mathfrak{L}_b$ are 
			\begin{equation}
			\lambda_1 = 0 \qquad \lambda_2 = \lambda_3 = -\frac{\Gamma}{2} \qquad \lambda_4 = -\Gamma \ .
			\end{equation}
Note that $\gamma_o >0$ and $\mathscr{N}(\omega_o) >0$, so all the eigenvalues of $\mathfrak{L}_b$ are such that $\Re[\lambda_i] \leq 0, \quad i=1,2,3,4$ -- as we would expect from the stability conditions for the dynamics. In particular, the right eigen-superket corresponding to $\lambda_1=0$ will be proportional to the equilibrium state of the TLS. 
\par The superoperator $\mathfrak{L}_b$ is a $4\times 4$ matrix, so it will be diagonalizable if it has four independent (right) eigenvectors. After some algebra, one finds that $\mathfrak{L}_b$ has, indeed, four independent (right) eigenvectors -- so we are in the case discussed in Sec. \ref{sec:biorthonormal} -- and the similarity transformation which diagonalizes $\mathfrak{L}_b$ is -- in analogy to \eqref{eq:L_S-1_S_decomp} -- ,
				\begin{equation}
				\label{eq:TLS_diagonalized_L}
				\mathfrak{A}^{-1} \mathfrak{L}_b \mathfrak{A} = \sum^4_{k=1} \lambda_k \lket{k}\!\lbra{k} 
				\end{equation}
where
				\begin{equation}
				\label{eq:TLS_A_A-1}
				\mathfrak{A} = 
			\begin{pmatrix}
			\frac{\Gamma_2}{\Gamma_1} &   0                &    0               &   -1 \\
			0                        &   1                &    0			    &  0          \\
			0                        &   0                &    1                &  0          \\
			1   					 & 0                  &    0               &  1          
			\end{pmatrix}
\qquad
				\mathfrak{A}^{-1} = 
			\begin{pmatrix}
			\frac{\Gamma_1}{\Gamma} &   0                &    0               &   \frac{\Gamma_1}{\Gamma} \\
			0                        &   1                &    0			    &  0          \\
			0                        &   0                &    1                &  0          \\
			-\frac{\Gamma_1}{\Gamma}   					 & 0                  &    0               &  \frac{\Gamma_2}{\Gamma}          
			\end{pmatrix}
				\end{equation}
and $\{\lket{k}\} \ (k=1,2,3,4)$	is the canonical basis in four-dimensional space:
			\begin{equation}
			\label{eq:TLS_canonical_kets}
			\lket{1} = 
			\begin{pmatrix}
			1 \\
			0 \\
			0 \\
			0
			\end{pmatrix} \quad
			\lket{2} = 
			\begin{pmatrix}
			0 \\
			1 \\
			0 \\
			0
			\end{pmatrix} \quad
			\lket{3} = 
			\begin{pmatrix}
			0 \\
			0 \\
			1 \\
			0
			\end{pmatrix} \quad
			\lket{4} = 
			\begin{pmatrix}
			0 \\
			0 \\
			0 \\
			1
			\end{pmatrix} \ .
			\end{equation}	
In terms of the basis $\{\ket{e},\ket{g}\}$, we see that 
			\begin{subequations}
			\label{eq:TLS_lkets(k)_in_e_g}
			\begin{align}
			\lket{1} & = \ket{e}\otimes \ket{e}^* = \lket{e,e}\\
			\lket{2} & = \ket{e} \otimes \ket{g}^* = \lket{e,g}\\
			\lket{3} & = \ket{g} \otimes \ket{e}^* = \lket{g,e}\\
			\lket{4} & = \ket{g}\otimes \ket{g}^* = \lket{g,g}\ .
			\end{align}
			\end{subequations}
And from \eqref{eq:TLS_diagonalized_L}, we derive that
			\begin{equation}
			\mathfrak{L}_b = \sum^4_{k=1} \lambda_k \ \mathfrak{A}\lket{k}\!\lbra{k} \mathfrak{A}^{-1} = \sum^4_{k=1} \lambda_k \lket{\zeta_k}\!\lbra{\xi_k} 
			\end{equation}		
where  -- like \eqref{eq:zeta_xi_A_A_inv_def} -- $\lket{\zeta_k}= \mathfrak{A}\lket{k}$ and $\lbra{\xi_k} = \lbra{k} \mathfrak{A}^{-1}$. Concretely,
			\begin{equation}
			\label{eq:TLS_column_vectors}
			\lket{\zeta_1}=
			\begin{pmatrix}
			\frac{\Gamma_2}{\Gamma_1}\\
			0 \\
			0 \\
			1
			\end{pmatrix} \quad
			\lket{\zeta_2} = 
			\begin{pmatrix}
			0 \\
			1 \\
			0 \\
			0
			\end{pmatrix} \quad
			\lket{\zeta_3} =
			\begin{pmatrix}
			0 \\
			0 \\
			1 \\
			0
			\end{pmatrix} \quad
			\lket{\zeta_4} =
			\begin{pmatrix}
			-1\\
			0 \\
			0 \\
			1
			\end{pmatrix}						 
			\end{equation}	
while
			\begin{multline}
			\label{eq:TLS_row_vectors}
			\lbra{\xi_1} = 
			\begin{pmatrix}
			\frac{\Gamma_1}{\Gamma} & 0 & 0 & \frac{\Gamma_1}{\Gamma}
			\end{pmatrix} \quad
			\lbra{\xi_2} = 
			\begin{pmatrix}
			0 & 1 & 0 & 0
			\end{pmatrix} \quad
			\lbra{\xi_3} = 
			\begin{pmatrix}
			0 & 0 & 1 & 0
			\end{pmatrix} \\ 
			\lbra{\xi_4} =
			\begin{pmatrix}
			- \frac{\Gamma_1}{\Gamma} & 0 & 0 & \frac{\Gamma_2}{\Gamma} 
			\end{pmatrix} \ .
			\end{multline}			
From these expressions, we note, indeed, that, in general, the right (\emph{i.e.} $\lket{\zeta_k}$) and left (\emph{i.e.} $\lbra{\xi_k}$) eigen-superkets of $\mathfrak{L}_b$ are not related through a simple operation of conjugate transpose: \emph{i.e.} $\lket{\zeta_k} \neq \lbra{\xi_k}^\dagger$. One can also verify that $\lbraket{\xi_k}{\zeta_{k'}}=\delta_{k,k'}$ which is the condition of biorthornomality, \eqref{eq:biorthogornal_orthogonality}. In fact, $\{\lket{\zeta_k}\!\lbra{\xi_k}\}$ constitute a biorthonormal basis for the Liouville space $\mathcal{L}_2$. And it can be verified that the closure relation $\sum^4_{k=1} \lket{\zeta_k}\!\lbra{\xi_k} = \mathbb{I}_4$ is satisfied (see \eqref{eq:biorthogornal_completeness_relation}).
\par With the help of the closure relation, we may write (in analogy to \eqref{eq:biorthonormal_expansion_e^Lt})
			\begin{equation}
			\label{eq:TLS_rho_solved_LS_lambda_k}
			\begin{split}
			\lket{\rho_S(t)} & = e^{t\mathfrak{L}_b}\lket{\rho_S(0)} = \sum^4_{k=1} e^{t\mathfrak{L}_b}\lket{\zeta_k}\lbraket{\xi_k}{\rho_S(0)}\\
			& = \sum^4_{k=1} e^{t\lambda_k}\lket{\zeta_k}\lbraket{\xi_k}{\rho_S(0)} \ .
			\end{split}
			\end{equation}
Putting in the values of $\lambda_k$, we have
			\begin{multline}
			\label{eq:TLS_rho_solved_LS}
			\lket{\rho_S(t)} = \lket{\zeta_1}\lbraket{\xi_1}{\rho_S(0)} + e^{-\frac{\Gamma}{2}t}\lket{\zeta_2}\lbraket{\xi_2}{\rho_S(0)} + e^{-\frac{\Gamma}{2}t}\lket{\zeta_3}\lbraket{\xi_3}{\rho_S(0)} \\
			+ e^{-\Gamma t}\lket{\zeta_4}\lbraket{\xi_4}{\rho_S(0)}
			\end{multline}
-- from which it is evident that the corresponding equilibrium superket is $\lket{\rho_S(\infty)} = \lket{\zeta_1}\lbraket{\xi_1}{\rho_S(0)}$.
\par Now, if we want to get the square matrix $\rho_S(t)$, we just apply the inverse bra-flipper operator $\mho^{-1}$ to both sides of \eqref{eq:TLS_rho_solved_LS}:
			\begin{multline}
			\label{eq:TLS_rho_solved_HS_0}
			\mho^{-1}[\lket{\rho_S(t)}] = \mho^{-1}\big[\lket{\zeta_1}\big]\lbraket{\xi_1}{\rho_S(0)} + e^{-\frac{\Gamma}{2}t}\ \mho^{-1}\big[\lket{\zeta_2}\big]\lbraket{\xi_2}{\rho_S(0)}\\
			 + e^{-\frac{\Gamma}{2}t}\ \mho^{-1}\big[\lket{\zeta_3}\big]\lbraket{\xi_3}{\rho_S(0)} 
			+ e^{-\Gamma t}\ \mho^{-1}\big[\lket{\zeta_4}\big]\lbraket{\xi_4}{\rho_S(0)} \ .
			\end{multline}			
Naturally, 	$\mho^{-1}[\lket{\rho_S(t)}]=\mho^{-1}\big[ \mho[\rho_S(t)]\big]=\rho_S(t)$. Let us now determine the square matrices $\mho^{-1}[\lket{\zeta_k}]$ in terms of the qubit operators $\mathbb{I}_2, \sigma_\pm, \sigma_3$. We begin with $\lket{\zeta_1}$. From \eqref{eq:TLS_column_vectors} and \eqref{eq:TLS_canonical_kets}, we note that
			\begin{equation}
			\lket{\zeta_1} = 
			\frac{\Gamma_2}{\Gamma_1} \begin{pmatrix}
			1 \\
			0\\
			0\\
			0
			\end{pmatrix} 
			+ 
			\begin{pmatrix}
			0 \\
			0\\
			0\\
			1
			\end{pmatrix} = \frac{\Gamma_2}{\Gamma_1} \lket{1} + \lket{4} \ .
			\end{equation}
Using now \eqref{eq:TLS_lkets(k)_in_e_g}, it follows that
			\begin{equation}
			\label{eq:TLS_mho_zeta_1_0}
			\lket{\zeta_1} =  \frac{\Gamma_2}{\Gamma_1} \left[\ket{e}\otimes \ket{e}^*\right] + \left[\ket{g}\otimes \ket{g}^*\right] \ .
			\end{equation}
Thus,
			\begin{equation}
			\label{eq:TLS_mho_zeta_1}
			\begin{split}
			\mho^{-1}\left[\lket{\zeta_1} \right] & = \frac{\Gamma_2}{\Gamma_1} \ \mho^{-1}\left[\ket{e}\otimes \ket{e}^*\right] + \mho^{-1}\left[\ket{g}\otimes \ket{g}^*\right]\\
			& = \frac{\Gamma_2}{\Gamma_1} \ket{e}\!\bra{e} + \ket{g}\!\bra{g}
			\end{split}
			\end{equation}
where we have employed \eqref{eq:def_mho_inverse}.
One can verify that $\ket{e}\!\bra{e} = \frac{1}{2} \left[\mathbb{I}_2 + \sigma_3 \right]$ and $\ket{g}\!\bra{g} = \frac{1}{2} \left[\mathbb{I}_2 - \sigma_3 \right]$. Introducing these expressions into \eqref{eq:TLS_mho_zeta_1}, we finally obtain
			\begin{equation}
			\label{eq:TLS_mho_inv_zeta_1}
			\mho^{-1}\left[\lket{\zeta_1} \right]  =   \frac{\Gamma}{2\Gamma_1} \mathbb{I}_2 -  \frac{\Gamma_1-\Gamma_2}{2\Gamma_1}\sigma_3 \ .
			\end{equation}
Analogously, we deduce from \eqref{eq:TLS_column_vectors} and \eqref{eq:TLS_canonical_kets} that
			\begin{subequations}
			\begin{align}
			\lket{\zeta_2} & = \ket{e} \otimes \ket{g}^* \label{eq:TLS_mho_zeta_2}\\
			\lket{\zeta_3} & = \ket{g} \otimes \ket{e}^* \label{eq:TLS_mho_zeta_3}\\
			\lket{\zeta_4} & = -\ket{e} \otimes \ket{e}^*+\ket{g} \otimes \ket{g}^*\label{eq:TLS_mho_zeta_4}
			\end{align}
			\end{subequations}
from which follows that
			\begin{subequations}
			\begin{align}
			\mho^{-1}\left[\lket{\zeta_2}\right] &  = \ket{e}\!\bra{g} = \sigma_+ \label{eq:TLS_mho_inv_zeta_2}\\
			\mho^{-1}\left[\lket{\zeta_3}\right] & = \ket{g}\!\bra{e} = \sigma_- \label{eq:TLS_mho_inv_zeta_3}\\
			\mho^{-1}\left[ \lket{\zeta_4}\right] & = -\ket{e} \! \bra{e}+\ket{g} \!\bra{g}= -\sigma_3 \label{eq:TLS_mho_inv_zeta_4}
			\end{align}
			\end{subequations}
-- where we have made use of \eqref{eq:sigma_pm} and \eqref{eq:sigma_3_in_e_g}. It then follows from \eqref{eq:TLS_mho_inv_zeta_1}, \eqref{eq:TLS_mho_inv_zeta_2}, \eqref{eq:TLS_mho_inv_zeta_3}, \eqref{eq:TLS_mho_inv_zeta_4} that in terms of the operators $\mathbb{I}_2, \sigma_\pm, \sigma_3$, \eqref{eq:TLS_rho_solved_HS_0} may be written as	
			\begin{multline}
			\label{eq:TLS_rho_solved_HS_3}
			\rho_S(t) =\left[\frac{\Gamma}{2\Gamma_1}\lbraket{\xi_1}{\rho_S(0)}\right]\mathbb{I}_2  
			+ \left[e^{-\frac{\Gamma}{2}t}\lbraket{\xi_2}{\rho_S(0)}\right]\sigma_+
			 + \left[e^{-\frac{\Gamma}{2}t}\lbraket{\xi_3}{\rho_S(0)}\right] \sigma_-\\
			-\left[\frac{\Gamma_1-\Gamma_2}{2\Gamma_1}\lbraket{\xi_1}{\rho_S(0)}+ e^{-\Gamma t}\lbraket{\xi_4}{\rho_S(0)}\right]\sigma_3
			\end{multline}	
At this point, we may also want to express the scalars 	$\lbraket{\xi_k}{\rho_S(0)}$ in terms of known parameters. We begin with the superbras $\lbra{\xi_k}$. From \eqref{eq:TLS_row_vectors}, we see that
			\begin{subequations}
			\begin{align}
			\lbra{\xi_1} & = \frac{\Gamma_1}{\Gamma} \begin{pmatrix} 1 & 0 & 0 & 0 \end{pmatrix} + \frac{\Gamma_1}{\Gamma} \begin{pmatrix} 0 & 0 & 0 & 1 \end{pmatrix} 
			 = \frac{\Gamma_1}{\Gamma} \bra{e}\otimes \bra{e}^* + \frac{\Gamma_1}{\Gamma} \bra{g}\otimes \bra{g}^* \nonumber \\
			& = \frac{\Gamma_1}{\Gamma} \lbra{e,e} + \frac{\Gamma_1}{\Gamma} \lbra{g,g}\label{eq:xi_1_in_row}\\
			\lbra{\xi_2} & = \begin{pmatrix} 0 & 1 & 0 & 0 \end{pmatrix} = \bra{e} \otimes \bra{g}^* = \lbra{e,g} \label{eq:xi_2_in_row}\\
			\lbra{\xi_3} & = \begin{pmatrix} 0 & 0 & 1 & 0 \end{pmatrix} = \bra{g} \otimes \bra{e}^* = \lbra{g,e}\label{eq:xi_3_in_row}\\
			\lbra{\xi_4}  & =
			\begin{pmatrix}
			- \frac{\Gamma_1}{\Gamma} & 0 & 0 & \frac{\Gamma_2}{\Gamma} 
			\end{pmatrix} = - \frac{\Gamma_1}{\Gamma}\bra{e}\otimes \bra{e}^* + \frac{\Gamma_2}{\Gamma}\bra{g}\otimes \bra{g}^*\nonumber \\
			 & = -\frac{\Gamma_1}{\Gamma} \lbra{e,e} + \frac{\Gamma_2}{\Gamma} \lbra{g,g} \label{eq:xi_4_in_row} \ . 
			\end{align}
			\end{subequations}
Now, let $P_{j,j'}(0)\equiv \matrixel{j}{\rho_S(0)}{j'} \ , \ j,j'\in \{e,g\}$ -- then, using the property in \eqref{eq:lmatrix_element}, it follows from \eqref{eq:xi_1_in_row} and \eqref{eq:xi_4_in_row} that
			\begin{equation}
			\begin{split}
			\lbraket{\xi_1}{\rho_S(0)} & =  \frac{\Gamma_1}{\Gamma} \lbraket{e,e}{\rho_S(0)} + \frac{\Gamma_1}{\Gamma} \lbraket{g,g}{\rho_S(0)}  
			 = \frac{\Gamma_1}{\Gamma} \big[ P_{e,e}(0) + P_{g,g}(0)\big] \\
			 & = \frac{\Gamma_1}{\Gamma} \ .
			\end{split}
			\end{equation}
			\begin{equation}
			\begin{split}
			\lbraket{\xi_4}{\rho_S(0)} & = -\frac{\Gamma_1}{\Gamma} \lbraket{e,e}{\rho_S(0)} + \frac{\Gamma_2}{\Gamma} \lbraket{g,g}{\rho_S(0)}
			 = -\frac{\Gamma_1}{\Gamma} P_{e,e}(0) + \frac{\Gamma_2}{\Gamma} P_{g,g}(0)		\\
			& = P_{g,g}(0) - \frac{\Gamma_1}{\Gamma}	
			\end{split}
			\end{equation}
-- respectively. Here, we have used the fact that $P_{e,e}(0) + P_{g,g}(0) = \Tr[\rho_S(0)]=1$. Proceeding, we also have from \eqref{eq:xi_2_in_row} and \eqref{eq:xi_3_in_row} that
			\begin{equation}
			\lbraket{\xi_2}{\rho_S(0)}  = \lbraket{e,g}{\rho_S(0)} = P_{e,g}(0)
			\end{equation}
			\begin{equation}
			\lbraket{\xi_3}{\rho_S(0)}  = \lbraket{g,e}{\rho_S(0)} = P_{g,e}(0) \ .
			\end{equation}				
With these new expressions for the scalars $\lbraket{\xi_k}{\rho_S(0)}$, we may rewrite \eqref{eq:TLS_rho_solved_HS_3} as 
			\begin{equation}
			\label{eq:TLS_traditional_rho_solution}
			\rho_S(t) =\frac{1}{2} \big[\mathbb{I}_2 + 2 \left<\sigma_-(t)\right>\sigma_+ + 2 \left<\sigma_+(t)\right>\sigma_- + \left<\sigma_3(t)\right> \sigma_3 \big]
			\end{equation}
with
			\begin{subequations}
			\begin{align}
			\left<\sigma_-(t)\right> & = e^{-\frac{\Gamma}{2}t}P_{e,g}(0)\\
			\left<\sigma_+(t)\right> & = e^{-\frac{\Gamma}{2}t}P_{g,e}(0) \\
			 \left<\sigma_3(t)\right>  & = -\left[\frac{\Gamma_1-\Gamma_2}{\Gamma}+2 e^{-\Gamma t}\left( P_{g,g}(0) - \frac{\Gamma_1}{\Gamma} \right)\right] \ .
			\end{align}
			\end{subequations}
We also remark that the ratios $\frac{\Gamma_1}{\Gamma}$ and $\frac{\Gamma_2}{\Gamma}$ are the equilibrium populations of the ground state ($\ket{g}$) and excited state ($\ket{e}$) of the TLS, respectively. Indeed, using \eqref{eq:Gamma_1_Gamma_2} and \eqref{eq:Planck_dist}, it follows that
			\begin{equation}
			\frac{\Gamma_1}{\Gamma} = \frac{e^{\beta \hbar \omega_o/2}}{\Tr[e^{-\beta H_S}]} \ , \qquad \frac{\Gamma_2}{\Gamma} = \frac{e^{-\beta \hbar \omega_o/2}}{\Tr[e^{-\beta H_S}]} \ .
			\end{equation}
\subsubsection{Kraus operator sum representation.}\label{subsub:TLS_Kraus_OSR}
\par It is known that the solution to a quantum Markovian master equation like the quantum optical master equation under consideration, \eqref{eq:TLS_reduced}, may be written in the form (for $\rho_S \in \mathcal{H}_d$
)\cite{misc:Lidar-2019, book:Breuer-2007,book:Alicki-2007,book:Hayashi-2006, book:Watrous-2018, book:Kraus-1983}:
			\begin{equation}
			\label{eq:general_Kraus_OR}
			\rho_S(t) = \sum^{D-1}_{\alpha=0} K_\alpha(t) \rho_S(0) K^\dagger_\alpha(t) \ , \qquad D \leq d^2
			\end{equation}
where -- like a complete set of quantum measurement operators, \eqref{eq:completeness_M_m} -- the operators $K_\alpha(t)$ (called \emph{Kraus operators}) satisfy the completeness relation 
			\begin{equation}
			\label{eq:Kraus_completeness_rel}
			\sum^{D-1}_{\alpha=0} K^\dagger_\alpha(t) K_\alpha(t) = \mathbb{I}_d \ , \qquad D \leq d^2 \ .
			\end{equation}
\Eref{eq:general_Kraus_OR} is referred to as the \emph{Kraus operator sum representation} (or simply \emph{Kraus representation}). The similarities between \eqref{eq:general_Kraus_OR} and \eqref{eq:non-selective_measurement} must be noted. In fact, the Kraus representation allows us to interpret the evolution of the density matrix $\rho_S$ as a non-selective measurement, where the quantum measurement operators are the Kraus operators. It, indeed, reinforces the connection between measurement and interaction
\cite{book:Kraus-1983}. However, deriving the Kraus operators from the underlying quantum Markovian master equation is not, in general, an easy problem. Nonetheless, some procedures have been proposed in the literature
\cite{art:Nakazato-2006, art:Anderson-2007}. One relatively easy way to construct the Kraus operators is through what is normally called the Choi-state construction method 
\cite{book:Hayashi-2006, book:Watrous-2018} -- which is a basic tool in quantum information theory but beyond the scope of this paper. We show here how one can still derive the Kraus operators using the Liouville space formalism without knowing much about quantum information theory.
\par First of all, we need to know how the Kraus representation, \eqref{eq:general_Kraus_OR}, plays out in Liouville space. This can easily be done by applying the bra-flipper operator $\mho$ to both sides of \eqref{eq:general_Kraus_OR}: 
			\begin{subequations}
			\begin{align}
			\mho\big[\rho_S(t) \big] & = \sum^{D-1}_{\alpha=0} \mho \big[K_\alpha(t) \rho_S(0) K^\dagger_\alpha(t) \big] \\
			\lket{\rho_S(t)} & = \sum^{D-1}_{\alpha=0} \lket{K_\alpha(t) \rho_S(0) K^\dagger_\alpha(t) }
			\end{align}
			\end{subequations}
and then applying the superket triple product identity, \eqref{eq:triple_product}, on the r.h.s.:
			\begin{equation}
			\lket{\rho_S(t)}  = \sum^{D-1}_{\alpha=0} \big[K_\alpha(t) \otimes K^*_\alpha(t) ] \lket{\rho_S(0)} \label{eq:Kraus_LS}\ .
			\end{equation}
Comparing \eqref{eq:Kraus_LS} with \eqref{eq:TLS_master_equation_with_L_a}, we see that the relation 
			\begin{equation}
			\label{eq:e^L_b_Kraus_general}
			e^{t\mathfrak{L}_b}= \sum^{D-1}_{\alpha=0} K_\alpha(t) \otimes K^*_\alpha(t) 
			\end{equation}
must hold for a set of operators $\{K_\alpha(t)\}$ of $\mathcal{O}_d$. Therefore, if we are able to express the superoperator $e^{t\mathfrak{L}_b}$ in the form $\sum^{D-1}_{\alpha=0} \big[K_\alpha(t) \otimes K^*_\alpha(t) ]$, then the resulting operators $K_\alpha$ must be the desired Kraus operators. Since in our case the operators $K_\alpha$ operate on a two-dimensional state space, we know we can express each one of them as a linear combination of the operators $\{\mathbb{I}_2, \sigma_1, \sigma_2,\sigma_3\}$, \eqref{eq:X_qubit_Pauli_comb_0} -- or, in alternative $\{\mathbb{I}_2, \sigma_+, \sigma_-,\sigma_3\}$ (see, for example, \eqref{eq:TLS_rho_in_Pauli_+-}).  And that is what we aim to achieve below.
\par We start with the expression for $\lket{\rho_S(t)}$ in \eqref{eq:TLS_rho_solved_LS}. We see that we may rewrite it as:
			\begin{multline}
			\label{eq:TLS_rho_solved_Kraus}
			\lket{\rho_S(t)} = \bigg[\lket{\zeta_1}\!\lbra{\xi_1} + e^{-\frac{\Gamma}{2}t}\lket{\zeta_2}\!\lbra{\xi_2} + e^{-\frac{\Gamma}{2}t}\lket{\zeta_3}\!\lbra{\xi_3} 
			+ e^{-\Gamma t}\lket{\zeta_4}\!\lbra{\xi_4}\bigg]\lket{\rho_S(0)} \ .
			\end{multline}
Comparing \eqref{eq:TLS_rho_solved_Kraus} with \eqref{eq:TLS_rho_solved_LS_lambda_k}, it is clear that
			\begin{equation}
			\label{eq:e^L_b_home_coming}
			e^{t \mathfrak{L}_b} = \lket{\zeta_1}\!\lbra{\xi_1} + e^{-\frac{\Gamma}{2}t}\lket{\zeta_2}\!\lbra{\xi_2} + e^{-\frac{\Gamma}{2}t}\lket{\zeta_3}\!\lbra{\xi_3} 
			+ e^{-\Gamma t}\lket{\zeta_4}\!\lbra{\xi_4} \ . 
			\end{equation}
We begin by writing each superoperator $\lket{\zeta_k}\!\lbra{\xi_k}$  in \eqref{eq:e^L_b_home_coming} as a sum of tensor products of the form $\sum_i c_{ij} A_i \otimes A^*_j$, where the $\{c_{ij}\}$ are scalars. In doing so, we have to rely heavily on the mixed product rule of the Kronecker product, \eqref{eq:mixed_product_rule}. If we take $\lket{\zeta_1}\!\lbra{\xi_1}$, for example, it follows from \eqref{eq:TLS_mho_zeta_1_0} and \eqref{eq:xi_1_in_row} that
			\begin{equation}
			\label{eq:Zeta_1_xi_1}
			\begin{split}
			\lket{\zeta_1}\!\lbra{\xi_1}  =&  \left(\frac{\Gamma_2}{\Gamma_1} \ket{e}\otimes \ket{e}^*\big] + \big[\ket{g}\otimes \ket{g}^*\right) \left( \frac{\Gamma_1}{\Gamma} \bra{e}\otimes \bra{e}^* + \frac{\Gamma_1}{\Gamma} \bra{g}\otimes \bra{g}^* \right)\\
			 = & \frac{\Gamma_2}{\Gamma}\ket{e}\!\bra{e}\otimes (\ket{e}\!\bra{e})^* 
			+ \frac{\Gamma_2}{\Gamma}\ket{e}\!\bra{g}\otimes (\ket{e}\!\bra{g})^*
			+ \frac{\Gamma_1}{\Gamma}\ket{g}\!\bra{e}\otimes (\ket{g}\!\bra{e})^*\\
			& + \frac{\Gamma_1}{\Gamma}\ket{g}\!\bra{g}\otimes (\ket{g}\!\bra{g})^*
			\end{split}
			\end{equation}
For $\lket{\zeta_2}\!\lbra{\xi_2}$, in accordance with \eqref{eq:TLS_mho_zeta_2} and \eqref{eq:xi_2_in_row}, we have
			\begin{equation}
			\label{eq:Zeta_2_xi_2}
			\begin{split}
			\lket{\zeta_2}\!\lbra{\xi_2} &  = \left(\ket{e} \otimes \ket{g}^* \right) \left(\bra{e} \otimes \bra{g}^* \right)\\
			& = \ket{e}\!\bra{e}\otimes (\ket{g}\!\bra{g})^*
			\end{split}
			\end{equation}
Similarly, for $\lket{\zeta_3}\!\lbra{\xi_3}$, we get
			\begin{equation}
			\label{eq:Zeta_3_xi_3}
			\begin{split}
			\lket{\zeta_3}\!\lbra{\xi_3} &  = \left(\ket{g} \otimes \ket{e}^* \right) \left(\bra{g} \otimes \bra{e}^* \right)\\
			& = \ket{g}\!\bra{g}\otimes (\ket{e}\!\bra{e})^*
			\end{split}			
			\end{equation}
after employing \eqref{eq:TLS_mho_zeta_3} and \eqref{eq:xi_3_in_row}. Finally, for  $\lket{\zeta_4}\!\lbra{\xi_4}$ it follows from \eqref{eq:TLS_mho_zeta_4} and \eqref{eq:xi_4_in_row} that
			\begin{equation}
			\label{eq:Zeta_4_xi_4}
			\begin{split}
			\lket{\zeta_4}\!\lbra{\xi_4}   =& \bigg( -\ket{e} \otimes \ket{e}^*+\ket{g} \otimes \ket{g}^* \bigg) \bigg(-\frac{\Gamma_1}{\Gamma} \bra{e}\otimes\bra{e}^* + \frac{\Gamma_2}{\Gamma} \bra{g}\otimes\bra{g}^*  \bigg)\\
			 =& \frac{\Gamma_1}{\Gamma}\ket{e}\!\bra{e}\otimes (\ket{e}\!\bra{e})^* 
			- \frac{\Gamma_2}{\Gamma}\ket{e}\!\bra{g}\otimes (\ket{e}\!\bra{g})^*
			- \frac{\Gamma_1}{\Gamma}\ket{g}\!\bra{e}\otimes (\ket{g}\!\bra{e})^*\\
			& + \frac{\Gamma_2}{\Gamma}\ket{g}\!\bra{g}\otimes (\ket{g}\!\bra{g})^* \ .
			\end{split}			
			\end{equation}
Substituting the results obtained in \eqref{eq:Zeta_1_xi_1}-\eqref{eq:Zeta_4_xi_4} into \eqref{eq:e^L_b_home_coming}, we get, after some algebra,
			\begin{equation}
			\label{eq:TLS_Kraus_precursor_2}
			\begin{split}
			e^{t \mathfrak{L}_b}  
			 = & \left(\frac{\Gamma_2}{\Gamma}+ e^{-\Gamma t}\frac{\Gamma_1}{\Gamma}\right)\ket{e}\!\bra{e}\otimes (\ket{e}\!\bra{e})^*  
			+ \left(\frac{\Gamma_1}{\Gamma} + e^{-\Gamma t}\frac{\Gamma_2}{\Gamma} \right)\ket{g}\!\bra{g}\otimes (\ket{g}\!\bra{g})^*\\
			& + e^{-\frac{\Gamma}{2}t}\ket{e}\!\bra{e}\otimes (\ket{g}\!\bra{g})^* + e^{-\frac{\Gamma}{2}t}\ket{g}\!\bra{g}\otimes (\ket{e}\!\bra{e})^*\\ 
			& + \frac{\Gamma_2}{\Gamma}(1-e^{-\Gamma t}) \ \sigma_+\otimes \sigma_+^*
			+ \frac{\Gamma_1}{\Gamma}(1-e^{-\Gamma t}) \ \sigma_- \otimes \sigma_-^* \ .
			\end{split}
			\end{equation}
where we have made use of \eqref{eq:sigma_pm}. Note that, whereas the first four terms in \eqref{eq:TLS_Kraus_precursor_2} are not expressed as Kronecker products of operators which are linear combinations of $\{\mathbb{I}_2, \sigma_\pm, \sigma_3\}$, the last two terms are. As a matter of fact, each of the last two terms represents Kronecker products of the form $K_\alpha(t) \otimes K^*_\alpha(t)$, where $K_\alpha(t)$ is a Kraus operator. If we take $\frac{\Gamma_2}{\Gamma}(1-e^{-\Gamma t}) \ \sigma_+\otimes \sigma_+^*$, for example, we see that we may rewrite it as
			\begin{equation}
			\frac{\Gamma_2}{\Gamma}(1-e^{-\Gamma t}) \ \sigma_+\otimes \sigma_+^* = \bigg(\sqrt{\frac{\Gamma_2}{\Gamma}(1-e^{-\Gamma t})} \ \sigma_+ \bigg) \otimes \bigg(\sqrt{\frac{\Gamma_2}{\Gamma}(1-e^{-\Gamma t})} \ \sigma_+ \bigg)^*
			\end{equation}
Note that such a decomposition is possible because the coefficients multiplying $\sigma_\pm \otimes \sigma_\pm^*$ in \eqref{eq:TLS_Kraus_precursor_2} are always positive for $t \geq 0$. 
\par To derive the Kraus operators related to the sum of the first four terms of \eqref{eq:TLS_Kraus_precursor_2}, it is important to observe that the factors $\frac{\Gamma_1}{\Gamma}$ and $\frac{\Gamma_2}{\Gamma}$ in the coefficients multiplying the various superoperators are crucial, as they inform us on which physical process (emission or absorption) is the superoperator related to. Recall that $\Gamma_1$ and $\Gamma_2$ are the rates of total emission and absorption of photons, respectively, by the TLS (see \eqref{eq:TLS_dissipator} and \eqref{eq:Gamma_1_Gamma_2}). It is therefore easy to realize that \emph{all} superoperators in \eqref{eq:TLS_Kraus_precursor_2} proportional to $\frac{\Gamma_1}{\Gamma}$ are related to the total emission process of photons by the TLS. Similarly, those proportional to $\frac{\Gamma_2}{\Gamma}$ are related to the absorption process. If we take the term with $\sigma_+ \otimes \sigma^*_+$, for example, we see that it is proportional to $\frac{\Gamma_2}{\Gamma}$ so it must be related to the absorption process. In fact, $\sigma_+$ transitions the state of the TLS from the ground state $\ket{g}$ to the excited state $\ket{e}$, \emph{i.e.} $\sigma_+ \ket{g} = \ket{e}$. 
\par The sum of the first four terms in \eqref{eq:TLS_Kraus_precursor_2} may therefore be decomposed into two sums: one proportional to $\frac{\Gamma_1}{\Gamma}$, and the other proportional to $\frac{\Gamma_2}{\Gamma}$. In order to achieve this, we need to recall that $\left(\frac{\Gamma_1}{\Gamma} + \frac{\Gamma_2}{\Gamma}\right)=1$ -- (see \eqref{eq:Gamma_1_Gamma_2}) -- so the third and fourth terms in \eqref{eq:TLS_Kraus_precursor_2} may be multiplied by $\left(\frac{\Gamma_1}{\Gamma} + \frac{\Gamma_2}{\Gamma}\right)$ in order to make each term be either proportional to $\Gamma_1$ or $\Gamma_2$. With this understanding, we may therefore rewrite the sum of the first four terms in \eqref{eq:TLS_Kraus_precursor_2} as
			\begin{equation}
			\label{eq:first_4_terms_mathfrak_B_j}
			\begin{split}
			 \bigg(\frac{\Gamma_2}{\Gamma}& + e^{-\Gamma t}\frac{\Gamma_1}{\Gamma}\bigg) \ket{e}\!\bra{e}\otimes (\ket{e}\!\bra{e})^*  
			+ \left(\frac{\Gamma_1}{\Gamma} + e^{-\Gamma t}\frac{\Gamma_2}{\Gamma} \right)\ket{g}\!\bra{g}\otimes (\ket{g}\!\bra{g})^*\\
			 & + e^{-\frac{\Gamma}{2}t} \left(\frac{\Gamma_2}{\Gamma}+ \frac{\Gamma_1}{\Gamma}\right) \ket{e}\!\bra{e}\otimes (\ket{g}\!\bra{g})^* + e^{-\frac{\Gamma}{2}t}\left(\frac{\Gamma_2}{\Gamma}+ \frac{\Gamma_1}{\Gamma}\right) \ket{g}\!\bra{g}\otimes (\ket{e}\!\bra{e})^*\\
			 & = \frac{\Gamma_1}{\Gamma} \mathfrak{B}_1(t) + \frac{\Gamma_2}{\Gamma} \mathfrak{B}_2(t)
			\end{split}
			\end{equation}
where $\mathfrak{B}_1(t), \mathfrak{B}_2(t) \in \mathcal{S}_2$ and
			\begin{multline}
			\label{eq:mathfrak_B_1}
			\mathfrak{B}_1(t)= \bigg[  e^{-\Gamma t}\ket{e}\!\bra{e}\otimes (\ket{e}\!\bra{e})^* + \ket{g}\!\bra{g}\otimes (\ket{g}\!\bra{g})^*+  e^{-\frac{\Gamma}{2}t}\ket{e}\!\bra{e}\otimes (\ket{g}\!\bra{g})^* \\
			+  e^{-\frac{\Gamma}{2}t}\ket{g}\!\bra{g}\otimes (\ket{e}\!\bra{e})^*\bigg]
			\end{multline}	
			\begin{multline}
			\label{eq:mathfrak_B_2}
			\mathfrak{B}_2(t)= \bigg[  \ket{e}\!\bra{e}\otimes (\ket{e}\!\bra{e})^* + e^{-\Gamma t} \ket{g}\!\bra{g}\otimes (\ket{g}\!\bra{g})^*+  e^{-\frac{\Gamma}{2}t}\ket{e}\!\bra{e}\otimes (\ket{g}\!\bra{g})^* \\
			+  e^{-\frac{\Gamma}{2}t}\ket{g}\!\bra{g}\otimes (\ket{e}\!\bra{e})^*\bigg]
			\end{multline}				
It is clear at this point that realizing the decomposition $\mathfrak{B}_i (t) = \sum_i c_i A_i \otimes A_i^*$ ($i =1,2$ and $c_i >0$) will enable us to derive the Kraus operators. To this end, note that all the terms in \eqref{eq:mathfrak_B_1} and \eqref{eq:mathfrak_B_2} are Kronecker products involving only two operators: $\ket{e}\!\bra{e}$ and $\ket{g}\!\bra{g}$. But,-- considering that our ultimate goal is to express the Kraus operators as expansions of $\{\mathbb{I}, \sigma_+,\sigma_-,\sigma_3\}$ -- we note that the operators $\ket{e}\!\bra{e}$ and $\ket{g}\!\bra{g}$, are to be found only in the definitions of $\mathbb{I}_2(=\ket{e}\!\bra{e} + \ket{g}\!\bra{	g})$ and $\sigma_3(=\ket{e}\!\bra{e} - \ket{g}\!\bra{	g})$, \eqref{eq:sigma_3_in_e_g}. We may thus express $\mathfrak{B}_j(t)$ as $A_j(t) \otimes A_j(t)^*$, where $A_j(t) \in \mathcal{O}_d$ is a linear combination of $\mathbb{I}_2$ and $\sigma_3$. That is, we may write
			\begin{subequations}
			\label{eq:mathfrak_B_j_A_j}
			\begin{align}
			\mathfrak{B}_j(t) = A_j(t) \otimes A_j(t)^* = \bigg(a_j(t) \mathbb{I}_2 + b_j(t) \sigma_3 \bigg) \otimes \bigg(a_j(t) \mathbb{I}_2 + b_j(t) \sigma_3  \bigg)^* \qquad j\in \{ 1,2\}
			\end{align}
			\end{subequations}
where $a_j(t), b_j(t)$ are complex scalars. Expressing the operators $\mathbb{I}_2$ and $\sigma_3$ in terms of $\ket{e}\!\bra{e}$ and $\ket{g}\!\bra{g}$, we find that
			\begin{equation}
			\label{eq:A_j_otimes_A_j}
			\begin{split}
			A_j \otimes A^*_j & = \left(a_j \mathbb{I}_2 + b_j \sigma_3 \right)\otimes \left(a_j \mathbb{I}_2 + b_j \sigma_3 \right)^*\\
			& = \bigg( \vert a_j \vert^2 + a_j b^*_j + a^*_j b_j + \vert b_j \vert^2 \bigg)\ket{e}\!\bra{e}\otimes (\ket{e}\!\bra{e})^* \\
			& + \bigg( \vert a_j \vert^2 - a_j b^*_j - a^*_j b_j + \vert b_j \vert^2 \bigg)\ket{g}\!\bra{g}\otimes (\ket{g}\!\bra{g})^*\\
			& + \bigg( \vert a_j \vert^2 - a_j b^*_j + a^*_j b_j - \vert b_j \vert^2 \bigg)\ket{e}\!\bra{e}\otimes (\ket{g}\!\bra{g})^*\\
			& + \bigg( \vert a_j \vert^2 + a_j b^*_j - a^*_j b_j - \vert b_j \vert^2 \bigg)\ket{g}\!\bra{g}\otimes (\ket{g}\!\bra{g})^*
			\end{split}
			\end{equation}
where, for convenience, we have dropped all indicators of time-dependence. Comparing \eqref{eq:A_j_otimes_A_j} with \eqref{eq:mathfrak_B_1} and \eqref{eq:mathfrak_B_2}, and equating coefficients of the same superoperator, leads to the following systems of equations for the coefficients $a_j,b_j$:
			\begin{equation}
			\begin{cases}
			\vert a_1 \vert^2 + a_1 b^*_1 + a^*_1 b_1 + \vert b_1 \vert^2 = e^{-\Gamma t} \\
			\vert a_1 \vert^2 - a_1 b^*_1 - a^*_1 b_1 + \vert b_1 \vert^2 = 1 \\
			\vert a_1 \vert^2 - a_1 b^*_1 + a^*_1 b_1 - \vert b_1 \vert^2 = e^{-\frac{\Gamma}{2}t}\\
			\vert a_1 \vert^2 + a_1 b^*_1 - a^*_1 b_1 - \vert b_1 \vert^2 = e^{-\frac{\Gamma}{2}t}
			\end{cases} \qquad
			\begin{cases}
			\vert a_2 \vert^2 + a_2 b^*_2 + a^*_2 b_2 + \vert b_2 \vert^2 = 1 \\
			\vert a_2 \vert^2 - a_2 b^*_2 - a^*_2 b_2 + \vert b_2 \vert^2 = e^{-\Gamma t} \\
			\vert a_2 \vert^2 - a_2 b^*_2 + a^*_2 b_2 - \vert b_2 \vert^2 = e^{-\frac{\Gamma}{2}t}\\
			\vert a_2 \vert^2 + a_2 b^*_2 - a^*_2 b_2 - \vert b_2 \vert^2 = e^{-\frac{\Gamma}{2}t}
			\end{cases} \ .
			\end{equation}
The coefficients $a_j,b_j$ are easily found to be 
			\begin{equation}
			a_1 = a_2= \frac{1+e^{-\frac{\Gamma}{2}t}}{2} \qquad b_1 = -b_2 = - \frac{1-e^{-\frac{\Gamma}{2}t}}{2}
			\end{equation}
(Without loss of generality, we have reported above only the real solutions for $a_j,b_j$.)  Thus,
			\begin{subequations}
			\label{eq:finals_A_j(t)}
			\begin{align}
			A_1(t) & = \frac{1+e^{-\frac{\Gamma}{2}t}}{2} \mathbb{I}_2 - \frac{1-e^{-\frac{\Gamma}{2}t}}{2}\sigma_3		\\	
			A_2(t) & = \frac{1+e^{-\frac{\Gamma}{2}t}}{2} \mathbb{I}_2 + \frac{1-e^{-\frac{\Gamma}{2}t}}{2}\sigma_3 \ .
			\end{align}
			\end{subequations}		
From \eqref{eq:mathfrak_B_j_A_j}, it follows then that \eqref{eq:first_4_terms_mathfrak_B_j} may be rewritten as
			\begin{equation}
			\label{eq:first_4_terms_mathfrak_B_j_2}
			\begin{split}
			 \bigg(\frac{\Gamma_2}{\Gamma}& + e^{-\Gamma t}\frac{\Gamma_1}{\Gamma}\bigg) \ket{e}\!\bra{e}\otimes (\ket{e}\!\bra{e})^*  
			+ \left(\frac{\Gamma_1}{\Gamma} + e^{-\Gamma t}\frac{\Gamma_2}{\Gamma} \right)\ket{g}\!\bra{g}\otimes (\ket{g}\!\bra{g})^*\\
			 & + e^{-\frac{\Gamma}{2}t} \left(\frac{\Gamma_2}{\Gamma}+ \frac{\Gamma_1}{\Gamma}\right) \ket{e}\!\bra{e}\otimes (\ket{g}\!\bra{g})^* + e^{-\frac{\Gamma}{2}t}\left(\frac{\Gamma_2}{\Gamma}+ \frac{\Gamma_1}{\Gamma}\right) \ket{g}\!\bra{g}\otimes (\ket{e}\!\bra{e})^*\\
			 & = \left(\sqrt{\frac{\Gamma_1}{\Gamma}} A_1(t)\right) \otimes \left(\sqrt{\frac{\Gamma_1}{\Gamma}} A_1(t)\right)^* + \left(\sqrt{\frac{\Gamma_2}{\Gamma}} A_2(t)\right) \otimes \left(\sqrt{\frac{\Gamma_2}{\Gamma}} A_2(t)\right)^*
			\end{split}
			\end{equation}
with $A_j(t)$ given in \eqref{eq:finals_A_j(t)}. The last equation in \eqref{eq:first_4_terms_mathfrak_B_j_2} gives the Kraus representation of the first four terms in \eqref{eq:TLS_Kraus_precursor_2}. Substituting \eqref{eq:first_4_terms_mathfrak_B_j_2} into \eqref{eq:TLS_Kraus_precursor_2} yields the final result
			\begin{equation}
			\label{eq:TLS_e^Lt_Kraus}
			e^{t \mathfrak{L}_b} = \sum^3_{\alpha=0} K_\alpha(t) \otimes K^*_\alpha(t) \ .
			\end{equation}	
with
			\begin{subequations}
			\begin{align}
			K_0(t) & \equiv \sqrt{\frac{\Gamma_1}{\Gamma}} \left( \frac{1+e^{-\frac{\Gamma}{2}t}}{2}  \mathbb{I}_2 - \frac{1-e^{-\frac{\Gamma}{2}t}}{2}\sigma_3 \right)\\
			K_1(t) & \equiv \sqrt{\frac{\Gamma_2}{\Gamma}} \left( \frac{1+e^{-\frac{\Gamma}{2}t}}{2}  \mathbb{I}_2 + \frac{1-e^{-\frac{\Gamma}{2}t}}{2}\sigma_3 \right)\\
			K_2(t) & \equiv \left[\frac{\Gamma_1}{\Gamma}(1-e^{-\Gamma t}) \right]^{1/2} \sigma_-	\\	
			K_3(t) & \equiv 	\left[\frac{\Gamma_2}{\Gamma}(1-e^{-\Gamma t}) \right]^{1/2} \sigma_+ \ .
			\end{align}
			\end{subequations}
Putting together \eqref{eq:TLS_master_equation_with_L_a} and \eqref{eq:TLS_e^Lt_Kraus}, we see that
			\begin{equation}
			\label{eq:TLS_e^Lt_Kraus_2}
			\lket{\rho_S(t)} = \left[\sum^3_{\alpha=0} K_\alpha(t) \otimes K^*_\alpha(t)\right] \lket{\rho_S(0)} = \sum^3_{\alpha=0} \lket{K_\alpha(t) \rho_S(0) K^\dagger_\alpha(t)} 
			\end{equation}	
where, in the last step, we have applied the superket triple product identity, \eqref{eq:triple_product}. 
If we now apply the inverse bra-flipper on both sides of \eqref{eq:TLS_e^Lt_Kraus_2}, we find that
			\begin{subequations}
			\begin{align}
			\mho^{-1}\lket{\rho_S(t)} & = \sum^3_{\alpha=0} \mho^{-1}\lket{K_\alpha(t) \rho_S(0) K^\dagger_\alpha(t)}\\
			\rho_S(t) & = \sum^3_{\alpha=0} K_\alpha(t) \rho_S(0) K^\dagger_\alpha(t) \label{eq:TLS_L_b_Kraus_form}
			\end{align}
			\end{subequations}	
-- which is the desired Kraus operator sum representation, \eqref{eq:general_Kraus_OR}. In addition, it can easily be verified that
			\begin{equation}
			\sum^3_{\alpha=0} K^\dagger_\alpha(t) K_\alpha(t) = \mathbb{I}_2 \ 
			\end{equation}
-- as it is expected of Kraus operators in quantum Markovian master equations, \eqref{eq:Kraus_completeness_rel}.
\subsubsection{Considering also the unitary Liouvillian, $\mathfrak{L}_a$.}\label{subsub:TLS_Lamb-shift_Hamil}
\par We now consider the full generator $\mathfrak{L}$, \eqref{eq:TLS_qmme_LS}. As remarked earlier, $\mathfrak{L}_a$ and $\mathfrak{L}_b$ commute. This means they share the same left and right eigenvectors. This also implies that $\mathfrak{L}$ has the same left and right eigenvectors of $\mathfrak{L}_b$. In fact, the similarity trasformation which diagonalizes $\mathfrak{L}$ is
				\begin{equation}
				\label{eq:TLS_similarity_L}
				\mathfrak{A}^{-1} \mathfrak{L}\mathfrak{A} = \sum^d_{k=1} \lambda_k \lket{k}\!\lbra{k}
				\end{equation}
where the matrix representations of $\mathfrak{A}$ and $\mathfrak{A}^{-1}$ are still given by \eqref{eq:TLS_A_A-1}, except that the eigenvalues $\{\lambda_k\}$ of $\mathfrak{L}$ are now
			\begin{equation}
			\label{eq:TLS_eigenvalues_of_L}
			\lambda_1 = 0 \qquad 
			\lambda_2 =-\frac{\Gamma}{2} - i \Delta(\omega_o) \qquad 
			\lambda_3 = -\frac{\Gamma}{2} + i \Delta(\omega_o)\qquad
			\lambda_4 = -\Gamma \ .
			\end{equation}	
and can be easily verified by summing \eqref{eq:TLS_L_a_matrix_rep} and \eqref{eq:TLS_L_b_matrix_rep}. The expansion of $e^{t \mathfrak{L}}\lket{\rho_S(0)}$, \eqref{eq:rho_TLS_LS_1}, in the left and right eigenvectors of $\mathfrak{L}$ readily follows from \eqref{eq:TLS_rho_solved_LS_lambda_k} and \eqref{eq:TLS_eigenvalues_of_L}:
			\begin{multline}
			\label{eq:TLS_rho_solved_LS_complete}
			\lket{\rho_S(t)} = \lket{\zeta_1}\lbraket{\xi_1}{\rho_S(0)} 
			+ e^{-\frac{\Gamma}{2}t- i \Delta(\omega_o)t}\lket{\zeta_2}\lbraket{\xi_2}{\rho_S(0)} \\
			+ e^{-\frac{\Gamma}{2}t+ i \Delta(\omega_o)t}\lket{\zeta_3}\lbraket{\xi_3}{\rho_S(0)} 
			+ e^{-\Gamma t}\lket{\zeta_4}\lbraket{\xi_4}{\rho_S(0)} \ .
			\end{multline}
Applying the inverse bra-flipper $\mho^{-1}$ to both sides of \eqref{eq:TLS_rho_solved_LS_complete}, we may write the resulting square matrix $\rho_S(t)$ as in \eqref{eq:TLS_traditional_rho_solution}, but this time,
			\begin{equation}
			\left<\sigma_+(t)\right>  = e^{-\frac{\Gamma}{2}t - i\Delta(\omega_o)t }P_{g,e}(0) \ , \qquad \left<\sigma_-(t)\right>  = e^{-\frac{\Gamma}{2}t + i \Delta(\omega_o)t}P_{e,g}(0)\ .
			\end{equation}
\par The Kraus representation of the complete solution can also be easily obtained.  Indeed, from \eqref{eq:rho_TLS_LS_1}
			\begin{equation}
			\label{eq:TLS_complete_L_Kraus_precursor_0}
			\begin{split}
			\lket{\rho_S(t)} & = e^{t (\mathfrak{L}_a + \mathfrak{L}_b)} \lket{\rho_S(0)} \\
			& = e^{t \mathfrak{L}_a} \left[e^{t \mathfrak{L}_b} \lket{\rho_S(0)}\right]\\
			& = \left[ e^{-i\frac{\Delta(\omega_o)t}{2} \sigma_3 } \otimes e^{i\frac{\Delta(\omega_o)t}{2} \sigma_3 }\right]\left[e^{t \mathfrak{L}_b} \lket{\rho_S(0)}\right]
			\end{split}
			\end{equation}
where we have made use of the definition of $\mathfrak{L}_a$, \eqref{eq:L_a_in_sigma_3_I_2}, and the Kronecker product property we saw in \eqref{eq:function_Kronecker_prod} (see also the derivation of \eqref{eq:lket_rho_t_Schrodinger}). But from \eqref{eq:TLS_e^Lt_Kraus} and \eqref{eq:TLS_e^Lt_Kraus_2}, we know $e^{t \mathfrak{L}_b} \lket{\rho_S(0)} = \sum^3_{\alpha=0} \lket{K_\alpha(t) \rho_S(0) K^\dagger_\alpha(t)}$, so \eqref{eq:TLS_complete_L_Kraus_precursor_0} may be rewritten as
			\begin{equation}
			\begin{split}
			\lket{\rho_S(t)} & = \left[ e^{-i\frac{\Delta(\omega_o)t}{2} \sigma_3 } \otimes e^{i\frac{\Delta(\omega_o)t}{2} \sigma_3 }\right]\left[\sum^3_{\alpha=0} \lket{K_\alpha(t) \rho_S(0) K^\dagger_\alpha(t)}\right]\\
			& = \sum^3_{\alpha=0}\left[ e^{-i\frac{\Delta(\omega_o)t}{2} \sigma_3 } \otimes e^{i\frac{\Delta(\omega_o)t}{2} \sigma_3 }\right] \lket{K_\alpha(t) \rho_S(0) K^\dagger_\alpha(t)}\\
			& = \sum^3_{\alpha=0}\lket{ e^{-i\frac{\Delta(\omega_o)t}{2} \sigma_3 } K_\alpha(t) \rho_S(0) K^\dagger_\alpha(t) e^{i\frac{\Delta(\omega_o)t}{2} \sigma_3 } }
			\end{split}
			\end{equation}
where we have applied the superket triple product identity, \eqref{eq:triple_product}.
Applying the inverse bra-flipper operator $\mho^{-1}$ finally yields
			\begin{equation}
			\begin{split}
			\mho^{-1}\big[ \lket{\rho_S(t)} \big] & = \sum^3_{\alpha=0} \ \mho^{-1} \bigg[ \lket{ e^{-i\frac{\Delta(\omega_o)t}{2} \sigma_3 } K_\alpha(t) \rho_S(0) K^\dagger_\alpha(t) e^{i\frac{\Delta(\omega_o)t}{2} \sigma_3 } } \bigg]\\
			\rho_S(t) & = \sum^3_{\alpha=0}  \mathscr{K}_\alpha(t) \rho_S(0) \mathscr{K}^\dagger_\alpha(t)
			\end{split}
			\end{equation}
where
			\begin{equation}
			\mathscr{K}_\alpha(t) \equiv e^{-i\frac{\Delta(\omega_o)t}{2} \sigma_3 } K_\alpha(t) \ .
			\end{equation}	
Here too, the completeness relation expected of Kraus operators is easily seen to be satisfied:
			\begin{equation}
			\sum^3_{\alpha=0} \mathscr{K}^\dagger_\alpha(t) \mathscr{K}_\alpha(t) = \mathbb{I}_2 \ .
			\end{equation}						
\section{Concluding Remarks}
We have illustrated how a finite-dimensional Liouville space $\mathcal{L}_d$ can be built from a finite-dimensional state space $\mathcal{H}_d$, and how one can effectively do quantum mechanics in $\mathcal{L}_d$ using essentially the same mathematical tools and concepts (most notably, the Dirac bra-ket notation) students learn in their first course in non-relativistic quantum mechanics. These mathematical tools are only augmented by the use of the Kronecker product (Sec. \ref{sec:Kronecker-product}) and the introduction of the bra-flipper operator $\mho$ and its inverse $\mho^{-1}$ (Sec.s \ref{sec:bra-flipper} and \ref{sec:inverse_bra-flipper}, respectively). In particular, $\mho$ allows us to transform elements of $\mathcal{O}_d$ (the Hilbert space of linear operators on $\mathcal{H}_d$) -- which are square matrices of dimension $d$ -- into elements of $\mathcal{L}_d$ (which are $(d^2\times 1)$ column vectors); $\mho^{-1}$, on the contrary, does the inverse. We have also stressed that, like $\mathcal{H}_d$, the Liouville space $\mathcal{L}_d$ is also a Hilbert space. Whereas the vectors in $\mathcal{H}_d$ may be said to be pure states, the quantum state vectors in $\mathcal{L}_d$ may be either pure or mixed. In this view, $\mathcal{L}_d$ is an enriched extension of $\mathcal{H}_d$. Both Hilbert spaces also share the same form of scalar product, namely, the extended Hilbert-Schmidt inner product, \eqref{eq:inner_prod_Hilbert-Schmidt}. The same type of inner product also applies to the Hilbert spaces $\mathcal{O}_d$ and $\mathcal{S}_d$ (\tref{tab:Hilbert_spaces}). 
\par From a conceptually practical perspective, an important argument can be further made for the Liouville space formalism --- and it relates to spectroscopy. In ordinary quantum mechanics in $\mathcal{H}_d$, the eigenvalues of observables  ---  like the Hamiltonian or magnetization vector  ---  , which are absolute quantities (at least, up to a constant), are the natural occurrences. Meanwhile, what we experimentally measure in spectroscopic experiments like nuclear magnetic resonance (NMR) are quantities related to the differences between these eigenvalues
\cite{misc:Gyamfi-2019}. We need not look further than the resonance conditions of such experiments to see this is the case. A formalism in quantum mechanics where these differences naturally appear as proportional to the eigenvalues of a corresponding operator would be much more practical. And the Liouville space formalism satisfies this need. (If we look at the Liouvillian $\mathfrak{L}$ in \eqref{eq:L_isolated} for an isolated quantum system, for example, it is clear from its diagonal form in \eqref{eq:L_isolated_diagonal} that its eigenvalues are proportional to the energy differences $(\epsilon_\nu -\epsilon_{\nu'})$.) This makes the Liouville space formalism particularly suited for theoretical studies in diverse spectroscopies
\cite{inbook:Fano-1964, book:Mukamel-1995, book:Ernst_Bodenhausen-1990}. The usefulness of the formalism is further accentuated by the superket triple product identity \eqref{eq:triple_product}  ---  which enables one to, at least, formally solve any kind of master equation linear in the density matrix $\rho(t)$ (or any other operator which is the subject of the differential equation). 
\par Furthermore, it is worth noting that there is not a higher Hilbert space than the Liouville space $\mathcal{L}_d$ where one can still do quantum mechanics as in $\mathcal{H}_d$. To see why, recall that $\mathcal{L}_d$ is nothing but the Hilbert space of the column `vectorized' operators on $\mathcal{H}_d$ (\tref{tab:Hilbert_spaces}). Technically speaking, one can column `vectorize' the superoperators on $\mathcal{L}_d$ (that is, the elements of $\mathcal{S}_d$) and create another Hilbert space $\mathcal{T}_d$ whose dimension will be $d^4$. The $(d^2 \times d^2)$ matrices of $\mathcal{S}_d$ will now be $(d^4 \times 1)$ column vectors in $\mathcal{T}_d$. The operators on $\mathcal{T}_d$, which will be $(d^4 \times d^4)$ matrices will also constitute a Hilbert space whose elements can be column `vectorized' to get another Hilbert space $\mathcal{U}_d$ of dimension $d^{16}$, and so on. The point is, mathematically speaking, we can build an infinite sequence of finite-dimensional Hilbert spaces from $\mathcal{H}_d$.  However, we cannot do quantum mechanics beyond $\mathcal{L}_d$ because the density matrix $\rho(t)$, as far as current quantum physics goes, is the most general way of describing a quantum state, and in $\mathcal{L}_d$ it becomes a column vector --- which is the most rudimentary element a finite-dimensional linear space can have. A square matrix can be reduced to a column vector, but a column vector cannot be reduced any further into a simpler entity without losing any information. The transition from $\mathcal{H}_d$ to $\mathcal{L}_d$, in regards to quantum mechanics,  was made possible simply because we can reduce the matrix $\rho(t)$ to the column vector $\lket{\rho(t)}$ without losing information on the quantum state. In trying to represent $\lket{\rho(t)}$ in higher Hilbert spaces than $\mathcal{L}_d$, it generally becomes impossible not to lose some information on the quantum state (mainly populations and coherences).
\par Finally, we remark that the formalism discussed above can be extended to any kind of Liouville space $\mathcal{L}$ (separable or continuous)
\cite{book:Mukamel-1995, inbook:Petrosky_Prigogine-1996}, following the general theory of Hilbert spaces expounded, for example, in \cite{book:Jordan-2006, book:Moretti-2018, book:Conway-2007, book:Riesz-1955}.
\ack
The author would like to express his gratitude to Prof. Vittorio Giovannetti for the many interesting discussions. He is also grateful to Mr. Matteo Bruschi and Dr. Andrea Piserchia for proofreading the first draft of the manuscript and sharing their thoughts. He would also like to thank the anonymous referees for their helpful suggestions.

\appendix
\section*{Appendix}
\setcounter{section}{1}
\subsection{Vector spaces, Hilbert spaces  ---  a brief overview.}\label{appendix:A1}
The definitions and propositions listed below can be found in books like
\cite{book:Jordan-2006, book:Conway-2007, book:Riesz-1955, book:Tarasov-2008, book:Dennery-1996, book:Lax-2007, book:Strang-2006, book:Smirnov-1961}. 
\begin{itemize}
\item A \emph{vector space} or \emph{linear space} $\mathscr{X}$ over a field $\mathbb{F}$ is a set of elements (called \emph{vectors}) endowed with two operations: 
	\begin{enumerate}
	\item \emph{addition}: if $x$ and $y$ are both elements of $\mathscr{X}$ then $x+y$ is also an element of $\mathscr{X}$
	\item \emph{scalar multiplication}: if $c \in \mathbb{F}$ and $x \in \mathscr{X}$, then $c x \in \mathscr{X}$
	\end{enumerate}
such that for any vectors $x,y,z \in \mathscr{X}$ and scalars $c,c' \in \mathbb{F}$
	\begin{enumerate}
	\item $x+y=y+x$
	\item $x + (y+z)=(x+y)+z$
	\item there exists a unique vector $0$ such that $x+0=x$
	\item $c(x+y) = cx + cy$
	\item $(c+c')x=cx+c'x$
	\item $c(c'x)=(cc')x$
	\end{enumerate}
$\mathscr{X}$ is said to be a \emph{complex} vector space if $\mathbb{F}=\mathbb{C}$ (the set of complex numbers). If $\mathbb{F}=\mathbb{R}$ (the set of real numbers), then $\mathscr{X}$ is said to be \emph{real}.
\item Let $S$ be a subset of $\mathscr{X}$. $S$ is said to be a \emph{linear manifold} if it is a linear space itself.
\item A set of vectors $\{x_1,x_2, \ldots , x_l\}$ is said to be \emph{linearly independent} if the only solution to the equation $\sum^l_{i=1} c_i x_i=0$ in the variables $c_1, c_2, \ldots,c_l$ is that $c_1 = c_2 = \cdots = c_l=0$. Otherwise, the set is said to be \emph{linearly dependent}.  
\item A set of vectors $\{x_1,x_2, \ldots , x_l\}$ is said to \emph{span} a linear space $\mathscr{X}$ defined over the field $\mathbb{F}$ if every element $x$ of $\mathscr{X}$ can be expressed as a linear combination of $\{x_1,x_2, \ldots , x_l\}$: \emph{i.e.} $x=c_1 x_1 + c_2 x_2 + \ldots + c_l x_l$, where $c_1,c_2, \ldots,c_l \in \mathbb{F}$.
\item A set of vectors $\{x_1,x_2, \ldots , x_l\}$ is said to be the \emph{basis} of a linear space $\mathscr{X}$ if: a) it is linearly independent, and b) spans $\mathscr{X}$.
\item A linear space $\mathscr{X}$ is $n-$dimensional if and only if its basis consists of $n$ vectors.
\item An \emph{inner product} or \emph{scalar product} for a linear space $\mathscr{X}$ is a map $F$ which assigns to each pair of vectors $x,y \in \mathscr{X}$ a scalar, denoted symbolically $\big< x, y \big>$. An inner product for $\mathscr{X}$ must satisfy the following properties for vectors $x,y,z \in \mathscr{X}$ and scalar $c$:
	\begin{enumerate}
	\item $\big< x, y + z\big> = \big< x,y\big> + \big< x, z \big>$
	\item $\big< x, cy \big> = c \big< x,y \big>$
	\item $\big< x, y \big> = \big< y, x\big>^*$
	\item $\big< x, x \big> \geq 0$, the equality holding only if $x=0$
	\end{enumerate}
\item For a given inner product $F$ on linear space $\mathscr{X}$, the \emph{norm} or \emph{length} (according to $F$) of the vector $x \in \mathscr{X}$ is the non-negative real number $\Vert x \Vert := \sqrt{\big< x, x \big>}$.  
\item A \emph{normed space} is a linear space with a norm.
\item Two vectors $x,y \in \mathscr{X}$ are said to be \emph{orthogonal} to each other if $\big< x, y \big> =0$.
\item Two vectors $x,y \in \mathscr{X}$ are said to be \emph{orthonormal} if: a) they are orthogonal, and b) $\Vert x \Vert = \Vert y \Vert = 1$.
\item We say a \emph{sequence of vectors} $\{w_n\} \ \left( w_n=\sum^n_{k=1} c_k x_k\right)$ converges to a limit vector $w$ if $\Vert w_n - w \Vert \to 0$ as $n \to \infty$.
\item A \emph{Cauchy sequence} of vectors is a sequence of vectors $\{w_n\}$ such that $\Vert w_m - w_n \Vert \to 0$ as $m,n \to \infty$.
\item A linear space $\mathscr{X}$ is said to be \emph{complete} if every Cauchy sequence of vectors converges to a limit vector which also belongs to $\mathscr{X}$.
\item Every finite-dimensional complex linear space is complete. (See \cite{book:Jordan-2006} for proof.)
\item A linear space $\mathscr{X}$ is called a \emph{Hilbert space} if $\mathscr{X}$ is: a) normed and b) complete.
\item A \emph{separable} Hilbert space is one whose orthonormal basis consists of a countable (finite or infinite) number of vectors.
\end{itemize}
\subsection{Injective, surjective and bijective maps}\label{appendix:inj_sur_bij}
\par Let $\mathcal{X}$ and $\mathcal{Y}$ be two sets. Consider the map $f:\mathcal{X} \to \mathcal{Y}$. That is, $f$ is a rule which associates to each element $x$ of $\mathcal{X}$ an element $y$ of $\mathcal{Y}$. This assigned element $y$ is usually indicated as $f(x)$ -- usually called the \emph{image} of $x$ under the map $f$.
\par The map $f:\mathcal{X} \to \mathcal{Y}$ is said to be \emph{injective} if for any arbitrary pair $x,x' \in \mathcal{X}$, the following properties are satisified: 1) for $x\neq x'$, then $f(x) \neq f(x')$, and 2) $f(x) = f(x')$ if and only if $x=x'$. For example, let $\mathcal{X}=\{0,1,2,3,4,5\}$ and $\mathcal{Y}=\{0,1,2,3,4,5,6,7,8,9,10\}$. Let $g$ be the association $g:\mathcal{X} \to \mathcal{Y}$, such that $g(x)=2x$ where $x \in \mathcal{X}$. Then the map $g$ is injective.
\par The map $f:\mathcal{X} \to \mathcal{Y}$ is said to be \emph{surjective} if $\mathcal{Y}$ is simply the set of all the images of $\mathcal{X}$ according to $f$. That is, $f(\mathcal{X})=\mathcal{Y}$. For example, consider the two sets $\mathcal{X}=\{0,\pm 1, \pm 2, \pm 3, \pm 4\}$, $\mathcal{Y}=\{0,1,4,9,16\}$ and the map $h:\mathcal{X} \to \mathcal{Y}$, such that $h(x) = x^2$, where $x \in \mathcal{X}$. Then the map $h$ is clearly surjective. 
\par Given the map $f: \mathcal{X} \to \mathcal{Y}$, $f$ is said to be \emph{bijective} if it is both injective and surjective. This means that: 1) for $x,x' \in \mathcal{X}$ and $x\neq x'$, then $f(x) \neq f(x')$, 2) $f(x) = f(x')$ if and only if $x=x'$ and 3) $\dim \mathcal{X} = \dim \mathcal{Y}$, \emph{i.e.} the two sets $\mathcal{X}$ and $\mathcal{Y}$ have the same dimension. If $f$ is bijective, it is also common to refer to it as a \emph{bijection} or a \emph{one-to-one correspondence}. Every bijection $f$ also has an inverse, denoted $f^{-1}$, such that $f^{-1}: \mathcal{Y} \to \mathcal{X}$. The map $f^{-1}$ is also bijective.
\par For more on bijective, injective and surjective maps, see for example \cite{book:Lang-1993}.

\section*{References}

\bibliography{biblio_fQMLS}

\end{document}